\documentclass[a4paper,11pt,twoside]{book}
\usepackage[style=verbose,bibstyle=numeric,sorting=none,maxnames=99,idemtracker=false,citestyle=numeric-comp,backend=biber,giveninits=true]{biblatex}
\usepackage{multicol}
\usepackage[utf8]{inputenc}
\usepackage{xspace}
\usepackage{amsmath,amsfonts,amssymb}
\usepackage[margin=1in]{geometry}
\usepackage{color}
\usepackage{csquotes}
\usepackage{hyperref}
\usepackage{graphics,graphicx}
\usepackage{enumitem}
\usepackage{tabularx}
\usepackage{subcaption}
\usepackage{isotope}
\usepackage{xcolor,epsfig,epstopdf}
\usepackage[normalem]{ulem} 
\usepackage{slashed}
\usepackage{bbold,bbm}
\usepackage{physics}
\usepackage{empheq}
\usepackage{fancyhdr} 
\usepackage{tgtermes}
\usepackage[american]{babel}
\usepackage{units}    
\usepackage{titlesec} 
\usepackage[font=footnotesize]{caption}  
\usepackage{soul}     
\setlist{nosep}  
\usepackage{wrapfig}
\usepackage{epsfig}
\usepackage{mathrsfs}
\usepackage{comment}
\usepackage{cancel}
\usepackage{hyperref}

\usepackage{tikz}
\usetikzlibrary{arrows.meta, automata, calc, positioning, quotes}
\usepackage{ragged2e}

\hypersetup{colorlinks, linkcolor = [rgb]{0,0.0,0.75}, citecolor = [rgb]{0,0.0,0.75}, urlcolor = [rgb]{0,0.0,0.75}}

\newcommand{\Abinitio}{\textit{Ab~initio}\xspace}
\newcommand{\abinitio}{\textit{ab~initio}\xspace}

\newcommand{\chieft}{$\chi$EFT\xspace}
\newcommand{\chipt}{$\chi$PT\xspace}
\newcommand{\pilesseft}{pionless EFT\xspace}
\newcommand{\LQCD}{LQCD\xspace}
\newcommand{\LO}{\ensuremath{{\rm LO}}\xspace}
\newcommand{\NLO}{\ensuremath{{\rm NLO}}\xspace}
\newcommand{\NNLO}{\ensuremath{{\rm NNLO}}\xspace}
\newcommand{\NNNLO}{\ensuremath{{\rm N}}$^3$\ensuremath{{\rm LO}}\xspace}
\newcommand{\NNNNLO}{\ensuremath{{\rm N}}$^4$\ensuremath{{\rm LO}}\xspace}
\newcommand{\NNLOsat}{\ensuremath{{\rm NNLO}_\text{sat}}\xspace}
\newcommand{\NN}{\ensuremath{NN}\xspace}
\newcommand{\NNN}{\ensuremath{3N}\xspace}
\newcommand{\NNNN}{\ensuremath{4N}\xspace}

\fancypagestyle{plain}{
  \fancyhf{} 
  \fancyfoot[LE,RO]{\thepage} 
  }

\titleformat{\chapter}[hang]{\vspace{-2.5cm}\Large\bfseries}{\thechapter}{10pt}{\Large\bfseries}
\titleformat*{\section}{\bfseries\fontsize{12pt}{12}} 
\titlespacing*{\section}{0pt}{9pt}{2.25pt}
\titlespacing*{\subsection}{0pt}{4.5pt}{4.5pt}

\newif\ifnewauthor
\newauthortrue

\newcounter{affils}

\makeatletter
\def\chapterauthor[#1]#2{\ifnewauthor\else, \fi
       #2\textsuperscript{{#1}}%
       \ifnewauthor\newauthorfalse\gdef\chapterauthors{{\it#2}}\else
               \g@addto@macro\chapterauthors{, {\it#2}}\fi
       \ignorespaces 
}
\makeatother

\def\chapteraffil[#1]{\item[#1 --]%
    \refstepcounter{affils}%
    \label{authaffil\the\value{chapter}.#1}%
}

\newenvironment{affils}
   {\addtocontents{toc}{\chapterauthors\string\par}\begin{enumerate}}
   {\end{enumerate}\global\newauthortrue\vspace{0.5cm}}

\DeclareBibliographyCategory{cited}
\AtEveryCitekey{\addtocategory{cited}{\thefield{entrykey}}}

\defbibenvironment{nolabelbib}
  {\list
     {}
     {\setlength{\leftmargin}{\bibhang}%
      \setlength{\itemindent}{-\leftmargin}%
      \setlength{\itemsep}{\bibitemsep}%
      \setlength{\parsep}{\bibparsep}}}
  {\endlist}
  {\item}

\begin{document}

\frontmatter
\begin{titlepage}
\vspace{1.5cm}
\begin{center}
    {\huge {\bf Nuclear Forces for Precision Nuclear Physics \\ -- a collection of perspectives --}}
\end{center}
\vspace{0.25cm}
\begin{center}
INT-PUB-22-002, LA-UR-22-20419, UMD-PP-022-02
\end{center}
\vspace{0.75cm}

\textbf{Editors:}

\noindent
Ingo Tews, Los Alamos National Laboratory
\vspace{-0.15 cm}

\noindent
Zohreh Davoudi, University of Maryland, College Park
\vspace{-0.15 cm}

\noindent
Andreas Ekström, Chalmers University of Technology
\vspace{-0.15 cm}

\noindent
Jason D. Holt, TRIUMF

\vspace{0.25cm}

\textbf{Abstract:} This is a collection of perspective pieces contributed by the participants of the Institute of Nuclear Theory's Program 
on {\it Nuclear Physics for Precision Nuclear Physics} which was held virtually from April 19 to May 7, 2021.

The collection represents the reflections of a vibrant and engaged community of researchers on the status of theoretical research in low-energy nuclear physics, the challenges ahead, and new ideas and strategies to make progress in nuclear structure and reaction physics, effective field theory, lattice QCD, quantum information, and quantum computing.

The contributed pieces solely reflect the perspectives of the respective authors and do not represent the viewpoints of the Institute for Nuclear theory or the organizers of the program.
\end{titlepage}
\tableofcontents
\chapter{Preface}
A sound theoretical description of nuclear forces is pivotal for understanding many important physical observables over a wide range of energy scales and densities: from few-nucleon physics to nuclear structure and reaction observables as well as astrophysical environments and associated phenomena.

Within the last three decades, significant progress in nuclear physics has been made possible, in part, thanks to the development of powerful \abinitio many-body methods for approximately solving the nuclear Schr\"odinger equation, and the development of nuclear forces using effective field theory (EFT), in particular chiral EFT (\chieft).
This progress means that it has now become increasingly important to quantify the theoretical uncertainties of the predictions, in particular the uncertainties stemming from the nuclear Hamiltonian itself because they often dominate the theoretical error budget. These uncertainties, primarily due to unknown or neglected physics, can lead to sizable errors when predicting nuclear observables of interest for next-generation experiments and astrophysical observations and, therefore, need to be managed and reliably quantified to enable precision nuclear physics. Indeed, theoretical predictions with quantified uncertainties facilitate the most meaningful comparisons with experimental and observational data. 

The Institute for Nuclear Theory at the University of Washington hosted a virtual three-week program to assess the state of low-energy nuclear physics and to evaluate pathways to further progress, with an emphasis on nuclear forces. The overarching questions addressed during the program were:
\begin{itemize}
    \item What are the current limitations of nuclear Hamiltonians? Which few- and many-body observables are ideal to constrain nuclear forces?
    \item How can novel computational and statistical tools be used to improve nuclear forces and their uncertainty estimates? What precision can be achieved by going to higher orders in \chieft?
    \item What is a suitable power counting for \chieft? What is the role of lattice quantum chromodynamics (\LQCD) studies of few-nucleon systems in constraining nuclear EFTs?
    \item What can be learned from quantum-information analyses of low-energy nuclear systems? Can quantum computing change the computational paradigm in nuclear physics in the upcoming decades?
\end{itemize}

The program brought together researchers with expertise in nuclear many-body techniques, EFT, and \LQCD for nuclear physics, to share recent advances and new developments, and to discuss shortcomings, generate new ideas, and identify pathways to address to the questions above.

To finish the program with a summary of outstanding problems and questions, possible benchmarks and solutions, and clearly stated tasks for the community, all participants were invited to contribute short {\it perspective pieces}.
These have been collected and merged into the present document. 
The wide range of topics covered by the contributed perspectives reflects the rich and stimulating developments that presently characterize a highly active nuclear-physics community. The various pieces touch upon a range of topics; renormalizability, power counting, unitarity, emulators, determination of low-energy constants, the complex nature of open-source computing in science, the three-nucleon continuum, collectivity, regulator dependencies, matching \LQCD to EFTs, variational \LQCD spectroscopy, quantum information and quantum entanglement, and quantum computing and its migration to nuclear physics. 

The hope is that this document will serve as an anthology for the community and help guide future developments, facilitate collaborative work between different sub-communities, and allow assessing the progress to be made in the next few years. 
\vspace{0.25cm}

\noindent
The program organizers and collection editors:
\\
Zohreh Davoudi, Andreas Ekstr\"om, Jason D. Holt, and Ingo Tews
\mainmatter
\chapter{Nuclear forces for precision nuclear physics: Status, challenges, and prospects}
\chapterauthor[1]{Zohreh Davoudi}
\chapterauthor[2]{Andreas Ekström}
\chapterauthor[3,4]{Jason D. Holt}
\chapterauthor[5]{Ingo Tews}
\begin{affils}
    \chapteraffil[1]{Maryland Center for Fundamental Physics and Department of Physics, University of Maryland, College Park, Maryland 20742, USA}
    \chapteraffil[2]{Department of Physics, Chalmers University of Technology, SE-412 96 G{\"o}teborg}
    \chapteraffil[3]{TRIUMF, Vancouver, BC V6T 2A3, Canada}
    \chapteraffil[4]{Department of Physics, McGill University, Montr\'eal, QC H3A 2T8, Canada}
    \chapteraffil[5]{Theoretical Division, Los Alamos National Laboratory, Los Alamos, New Mexico 87545, USA}
\end{affils}
\addtocontents{toc}{\protect\setcounter{tocdepth}{-2}}
This first contribution to this collection contains the perspective of the editors as well as a brief overview of the discussions during the program and the contributions to this collection. 
It therefore spans over a wide range of topics: current limitations of nuclear Hamiltonians and the calibration of nuclear forces, improved nuclear forces using novel computational and statistical methods, and improved power-counting schemes. It also enumerates ideas and questions related to  large-$N_c$ analysis for low-energy nuclear processes, \LQCD calculations for nuclear physics and their matching to the EFTs, and the role of quantum information sciences and quantum computing in theoretical nuclear physics.

\section{Current limitations of nuclear Hamiltonians and calibrating nuclear forces using data for few- and many-body observables}

A reoccurring 
question in the field is why some interactions derived in \chieft, even though adjusted to reproduce similar data, work better than others for particular observables across the nuclear chart. This question is related to several open challenges pertaining to the (chiral) Hamiltonians used in \textit{ab initio} many-body methods: uncertainty quantification, the regularization scheme and scale dependence, and the possibility of identifying an ideal set of observables to constrain Hamiltonians. 
In the coming years, it will be crucial to address these questions to identify which components of nuclear interactions are most important for accurately reproducing and predicting relevant nuclear observables. 

When talking about different interactions and their success in describing various nuclear observables, it is important to distinguish between the EFT itself and the individual model realizations of it. The latter are typically referred to as interactions and depend on choices for where, and how, to truncate the (asymptotic) EFT series, how to identify the low-energy constants (LECs) and their numerical values, and how to regularize the potential. These choices all contribute to the theoretical uncertainty of the interaction and the resulting predictions.

In addition, when comparing theoretical predictions for individual observables, additional uncertainties arise due to approximations pertaining to the employed many-body method used to approximately solve the Schr\"odinger equation. Of course, the underlying assumptions made when estimating theoretical uncertainties will also play a significant role.

It is crucial to estimate uncertainties in theory as well as experiment, without which one cannot identify relevant tensions/discrepancies among model predictions and experiments. 
Bayesian statistical inference is becoming the prevailing approach for uncertainty quantification, parameter estimation, and various statistical analyses of theoretical predictions and models. In recent years, Bayesian tools and prescriptions have become available to, e.g., estimate truncation errors in EFT, and it is very informative to specify such uncertainties in theoretical analyses of nuclear observables. Alongside any uncertainty estimation, it is key to specify the assumptions made and, if possible, enumerate any additional sources of uncertainty not accounted for.

In this context, a relevant question arises: how to best estimate the uncertainties due to approximations made when solving the many-body Schr\"odinger equation? This is sometimes referred to as the (many-body) method error. It will be very important for the community to find ways of better estimating these uncertainties, e.g., by comparing many-body methods at different levels of approximation against available benchmark data, and by comparing predictions between different \textit{ab initio} methods against each other and potentially against phenomenological models when data are not available. To facilitate such comparisons, it will be useful to more freely distribute relevant interaction matrix elements within the community and, if possible, make the many-body codes, as well as accurate emulators for many-body methods, available to other researchers. 
One way forward might be to create an online repository for such resources. While many obvious questions arise regarding storage space, documentation, and a recognition of scientific credit to the developers, it is nevertheless important to find ways to tackle these practical and logistical challenges.

Additionally, it is crucial to quantify the effects of different regulator schemes that might influence the performance of nuclear interactions by regulating different parts of the nuclear interaction differently. It might be that some problems with nuclear interactions are more persistent in some schemes compared to others. Can the community find arguments for or against certain schemes? For example, it is difficult to maintain relevant symmetries of the interactions with most regulators and nontrivial to consistently regulate currents and interactions. It is expected that regulator artifacts, i.e., systematic uncertainties due to the regulator choices, decrease at high orders in the EFT and for larger cutoffs. However, as was brought up in the program, if one needs very high orders in the calculations then one is likely working with the wrong expansion. Furthermore, high cutoffs are not accessible with most many-body methods, even though future method developments will enable the community to treat stiffer and bare \chieft interactions. 

Finally, it is important to investigate which observables are ideal to calibrate interactions. In principle, the LECs of a low-energy EFT, and any additional parameters necessary for uncertainty quantification, can be inferred from any set of low-energy data within the applicability domain of the EFT.
The challenge lies in identifying and combining a set of calibration data with sufficient information content to yield useful predictions. 
In addition to commonly used calibration data such as nucleon-nucleon scattering cross sections and bulk nuclear observables, a calibration data set could also include, e.g., nucleon-nucleus or nucleus-nucleus scattering, astrophysical observations of neutron stars or data on collective phenomena.
Hence, we can ask ourselves if it would be useful to come up with a minimal set of observables for validation of \textit{ab initio} approaches and interaction models.
Sensitivity studies might help to determine which observables are most useful to determine and test the various parts of nuclear interactions and should be included in such a set.
We stress however that such a set is only useful in combination with robust estimates of all uncertainties.

\section{Improving nuclear forces using novel computational methods and going to higher orders in EFT}

Since the introduction of nuclear EFTs, the EFT paradigm has proven itself as a useful principle for constructing high-precision interactions with the added benefit of systematic assessment of uncertainties. Going to higher orders in the EFT corresponds to including additional information on the short-range physics with the hope of improving the accuracy of the theoretical predictions. 
Predictions in the few-nucleon sector have now reached a high level of precision and accuracy when based on EFTs at sufficiently high order; even fifth-order calculations exist in some cases. 
One question is how to similarly improve the predictions for observables in heavier-mass systems? There are not yet any clear signs of systematic improvements in such systems when increasing the (chiral) order in EFT. 
More order-by-order comparisons of delta-full/delta-less interactions, constructed using the same methodology, are needed. Generally, it would most likely be useful for different groups to compare different schemes for constructing interactions in a more systematic way.

From a quantitative perspective, at least two complications arise when going to higher orders. 
First, with each additional order comes additional LECs and the numerical values of which need to be determined. 
Second, higher-order EFTs entail many-body interactions, e.g., three-nucleon forces, with associated unknown LECs that must be determined using data from three-nucleon systems, or beyond. 
It is computationally expensive to calibrate interactions using data from observables in few- and many-nucleon systems. 

In recent years, a large number of nuclear interactions have been constructed by the community. The question arises if this ``Skyrmification" of interactions is a positive or a negative trend.
Clearly, as long as the predictions from various interaction models agree within uncertainties, there is, in principle no problem. 
Indeed, a systematically developed family (or distribution) of interactions enables coherent model predictions and allows us to assess correlations.
In addition, operating with more than one interaction is a straightforward way of gauging theoretical uncertainties. 
As such, an ``antidote" to this ``Skyrmification" is a careful and honest uncertainty estimation. Theoretical predictions with relevant estimates of the underlying uncertainty will likely become standard practice in the coming years. 
It is important to note that the canonical $\chi^2$-per-datum measure does not account for e.g. model or method errors, but it is nevertheless a useful quantity for gauging the reproduction of, e.g., scattering data.

Emulators, i.e., computationally cheap, yet accurate, surrogate models for predicting the structure and reactions of few- and many-body systems, have emerged as powerful and useful tools since they provide access to an entirely new class of computational statistics methods for parameter estimation, sensitivity analysis, model comparison, and model mixing. 
In particular, emulators based on eigenvector continuation appear to be particularly efficient and accurate. This is an exciting development with the potential to facilitate new discoveries and to address several of the open problems mentioned before. 
Still, using emulators requires careful uncertainty quantification of the corresponding emulation error. 
Some methods, like Gaussian processes, yield uncertainties by design, but it remains to be established how to estimate the errors induced by eigenvector continuation emulators. 
Not all many-body methods lend themselves to emulation via eigenvector continuation, however, and the construction of emulators requires access to “split-format” interaction input. 
This again highlights the importance of a community repository for interaction codes and emulators.

\section{Improved power-counting schemes and constraining nuclear forces from lattice QCD}

To achieve renormalization group invariance it is of key importance to have the correct operators in place at the respective orders in the nuclear EFT expansion. There are, however, decades-long diverging viewpoints in the nuclear-theory community  about (non-perturbative) renormalization and power counting in \chieft. This was also a prominent topic of discussion during the program. 

In this context, the regulator cutoff plays a central role. It is an intermediate quantity necessary to regulate the interaction, and is often kept relatively small to converge present-day many-body calculations, but beyond this function it is not part of the underlying physical theory. It is clear that it is not meaningful to take the cutoff (much) smaller than the hard scale, or breakdown scale, of the EFT. 
In principle, the cutoff can be taken larger than the breakdown scale, but there are opposing viewpoints on how large it is meaningful to take it. This is intimately related to the question of inferring the importance of counterterms in the potential without understating or overstating their importance, as well as possible changes to the power counting in $A$-body systems.

To make progress, it is of interest to the community to find simple, or well-understood, benchmark systems to analyze renormalization, regularization, and power-counting strategies.
The present list of relevant, or realistic, benchmark systems appears to be rather short, and includes the zero-range limit at unitarity, systems described by \pilesseft, and the two-nucleon system.
In addition, when studying such systems at high values for the cutoff, spurious bound states might appear that could be difficult to treat in certain many-body methods.

An EFT does not dictate what the leading order (\LO) should contain beyond what is necessary to fulfill minimal symmetry and renormalization-group requirements. Studies of finite and infinite nuclear systems at \LO in \chieft point to deficiencies regarding saturation and spin-orbit interactions, two important properties observed in nuclear systems. 
Several questions related to the topic of constructing a \LO interaction emerged during discussions, such as: What is an ``optimal" convergence pattern for an EFT if you have to choose between ``smooth and steady but requiring more orders" or an ``irregular" start and then a rapid approach or “convergence” within fewer orders.

A standard avenue for constraining the LECs of the EFTs is to match to relevant experimental data. This may not be a straightforward endeavor when direct experimental measurements do not exist, necessitating the use of other related quantities and indirect phenomenological constraints.
Among various examples enumerated in this collection is a recent estimate on the \LO nucleon-nucleon (\NN) isotensor contact term in the neutrinoless double-$\beta$ decay within a minimal extension of the Standard Model. 
This was enabled by the application of a formalism similar to the Cottingham formula used in the study of the neutron-proton mass difference, with the result expressed in terms of a renormalized amplitude. 
This example highlights the need for a direct
matching of the EFT to calculations based in QCD for a variety of beyond-the-Standard-Model processes in the few-nucleon sector, from lepton-number non-conservation and CP violation to dark-matter-nucleus cross sections.

While \LQCD is the method of choice for constraining unknown LECs, its computational cost has hindered precise computations in the nuclear sector to date. In the absence of direct \LQCD constraints for the time being, large-$N_c$ considerations can provide valuable insights into the size and hence relative importance of interactions in the EFT, and may motivate prioritization of certain \LQCD calculations over the others. Among examples enumerated in this collection is the hadronic parity violation in the \NN sector where a combined large-$N_c$ and (pionless-)EFT analysis leads to only two independent \LO parity-violating operators. Importantly, there is an isotensor parity-violating LEC that contributes at this order. Such a guidance has motivated \LQCD calculations of the isotensor quantity that are computationally more accessible. Recent large-$N_c$ analyses have also revealed how questions regarding naturalness of the LECs and hence the size of contributions at given EFT orders may be impacted by the choice of basis.

Open questions to be studied in the coming years concern the expansion of nuclear binding energies in $1/N_c$, better understanding of the role of $\Delta$ in the large-$N_c$ analyses, and accidental cancellations that may ruin the large-$N_c$ countings. \LQCD can also play a role in the development of the large-$N_c$ studies in nuclear physics by providing constraints on higher partial waves in \NN scattering, parity-violating nuclear matrix elements, and three-nucleon observables for the organization of three-nucleon operators. Additionally, \LQCD calculations at $N_c \neq 3$ may provide insight into many of these questions, including into the role of $\Delta$ in single- and multi-nucleon sectors. 

The early and recent work in matching \LQCD results to EFTs has resulted in constraints on the two-body nuclear and hypernuclear interactions, revealing symmetries predicted by large-$N_c$ and entanglement considerations, albeit at unphysically-large quark masses. They have also enabled first QCD-based constraints on the \LO LECs in the pionless-EFT descriptions of deuteron's electromagnetic properties, and of the $np$ radiative capture, tritium $\beta$ decay, $pp$ fusion, and two-neutrino double-$\beta$ decay processes. A significant advance in this matching program involved making predictions for nuclei with atomic numbers larger than those obtained by \LQCD, hence demonstrating a full workflow involving \LQCD calculations, EFTs matching, and \textit{ab initio} many-body calculations based in the constrained EFTs, as described in this collection.

As the field moves forward, particularly once \LQCD computations of light nuclear systems will become a reality at the physical quark masses, more possibilities may be explored in this critical matching program. \LQCD in a finite volume matched to EFTs may help identify convergence issues in the EFTs, or help quantify the energy scale at which the nucleonic description of nuclei breaks down. Such a matching of \LQCD and EFT results in a finite volume can also facilitate constraints on few-nucleon operators without the need for complex, and generally not-yet-developed, matching formalisms to scattering amplitudes. A similar matching may be considered between \LQCD calculations at a given lattice spacing and the EFT-based many-body calculations at a corresponding UV scale. Additionally, phenomenological or EFT-inspired nuclear wavefunctions may lead to the construction of better interpolating operators for nuclear states in \LQCD calculations. To make progress, many-body methods that are set up fully perturbatively to eliminate the need for iterating the potential (hence preserving the strict renormalization-group invariance) may be preferred, nonetheless these methods need to overcome their present drawback in underbinding larger nuclei such as $^{16}$O.

Since the first \LQCD calculations of few-nucleon systems at unphysically large quark masses in early 2010s, the field has come a long way in pushing towards lighter quark masses and expanding the observables studied beyond lowest-lying energies, as described in this collection. A decade later, given advances in algorithms and methods and growth in computational resources, the field stands at a critical point where the first ground-breaking, but uncontrolled, calculations will give their place to a new generation of calculations that involve, for the first time, a more comprehensive set of two- and eventually multi-nucleon interpolating operators, enabling a systematic variational spectroscopy of nuclei with better control over excited-state effects. These will also involve ensembles with more than one lattice spacing such that the continuum limit of the lattice results can be taken systematically. Furthermore, the quark masses can be tuned at or near the physical values such that the results will correspond to those in nature.

The first variational studies of \NN systems have emerged in recent years, albeit still at large quark masses, with variational bounds on lowest-lying energies that are in tension with the previous non-variational estimates. These tensions may be attributed to one or more of the following: i) the variational basis of interpolating operators may be yet incomplete and while the upper bounds on energies are reliable, they may miss the presence of one or more lower-energy states if no operator in the set has significant overlap onto such states, ii) the previous non-variational ground-state results were dominated by excited-state effects at early times and misidentified the ground-state energies, or iii) as one study suggests, lattice-spacing effects may be significant, and comparing the results of two calculations at different input parameters may be ambiguous due to scale-setting inconsistencies. Investigating such possibilities will constitute a major endeavor in this field in the upcoming years, with promising directions already explored by various collaborations, as enumerated in this collection.

It is important to note that there are already a significant body of work, and related formal and numerical developments, in place in accessing phenomenologically interesting quantities in nuclear physics, from spectra and structure of light nuclei to nuclear scattering and reaction amplitudes. Therefore, once reliable and sufficient variational bases of operators are found and all systematic uncertainties are controlled, progressing toward the goal of matching QCD to EFT and many-body calculations will be within the reach.

\section{Prospects of quantum information sciences and quantum computing in nuclear physics}

The field of quantum information sciences (QIS) has grown to become a major area of scientific and technological developments in current times, benefiting from various partnerships between academia, government, and industry, as well as an ever growing workforce. Nuclear theorists, among other domain scientists, have recognized the potential of quantum computing in advancing many areas of nuclear physics that currently suffer from computationally intractable problems. 
These problems include accurate predictions for finite-density systems such as nuclei, phases and decomposition of dense matter (of relevance in neutron stars), real-time phenomena for description of reaction processes and evolution of matter after high-energy collisions (of relevance in collider experiments), as well as nuclear response functions (of relevance in the long-baseline neutrino experiments) and nuclear-structure quantities (of relevance in the upcoming Electron-Ion Collider).

In fact, the very first rudimentary (due to limited hardware technology) but ground-breaking (given their novel approach) calculations of small nuclear quantities have emerged, including quantum computing the binding energy of the deuteron and $^4$He, simulating models of nuclear response functions, and simple nuclear reactions, along with exploration of coherent neutrino propagation using simple models. 
Furthermore, with an eye on the grand challenge of obtaining nuclear dynamics from first-principle QCD-based studies, the field has witnessed a proliferation of concrete ideas, proposals, and algorithms for simulating quantum field theories, and illuminating hardware implementations in small systems. Additionally, by incorporating quantum entanglement and coherence in theoretical descriptions of nuclei, new understanding and insights have been reached in recent years. These research directions, which are still at an early stage, will form an exciting subfield of nuclear theory in the coming decade.

Many interesting open questions and underdeveloped areas will be studied by QIS-oriented nuclear theorists in the coming years: Can quantum entanglement provide a better organizational scheme for nuclear interactions, and a window into emergent symmetries beyond traditional considerations? Can the observed entanglement minimization in low-energy baryon-baryon interactions be understood from an \textit{ab initio} QCD analysis? Can the intricate balance between repulsion and attraction in nuclear media leading to complexities be characterized via quantum-information measures? Do the highly regular patterns in the shapes of nuclei (pointing to an emergent approximate symplectic symmetry), as verified by recent \textit{ab initio} studies based in chiral potentials, signal interesting quantum correlations and entanglement structure?  Can entanglement structure of nuclear wavefunctions and Hamiltonians in given nuclear many-body methods provide guidance on the most efficient bases for classical or quantum computation of certain nuclear processes?

Can customized gates provide a quicker and more scalable road to simulating nuclear dynamics than standard universal gates? What would be the role of hardware co-design, a process in which domain scientists, such as nuclear theorists, work in a feedback loop with quantum-hardware developers to impact the design of the next-generation quantum devices? Can nuclear theorists bring any benefit to the QIS field, given their long and advanced expertise in numerical Hamiltonian simulations, by providing state-of-the-art tools for simulating, hence optimizing, the quantum hardware? Have nuclear theorists identified concrete problems, i.e., the first realistic applications of quantum computing, so that quantum supremacy in the realm of nuclear physics can be claimed in the upcoming years?

What are the lessons to be learned from the integration of new theory and computing perspectives in nuclear theory over the past three decades (such as EFTs and \LQCD), such that we can effectively incorporate QIS tools and talent in nuclear theory too? What are the lessons to be learned, and projections to be made, from the course of developments of high-performance computing and its application in domain sciences, so that one can envision a high performance quantum-computing era in nuclear physics? Insights into these questions, and references to relevant recent progress, are presented in this collection.

{\bf Acknowledgements:} We are grateful to the Institute for Nuclear Theory, as well as an engaged, thoughtful, and critical nuclear-theory community, in particular the speakers, discussion leads, and participants of the INT program 21-1b on \textit{Nuclear Physics for Precision Nuclear Physics}. 

Z.D. acknowledges support from the the U.S.Department of Energy's (DOE's) Office of Science Early Career Award DE-SC0020271, Alfred P. Sloan foundation, and Maryland Center for Fundamental Physics at the University of Maryland, College Park. 
A.E. acknowledges support from the European Research Council (ERC) under the European Union’s Horizon 2020 research and innovation programme (Grant agreement No. 758027) and the Swedish Research Council project grant (Grant agreement No. 2020-05127). 
J.D.H. acknowledges support from the Natural Sciences and Engineering Research Council of Canada uunder grants SAPIN-2018-00027,
RGPAS-2018-522453, and the Arthur B. McDonald Canadian
Astroparticle Physics Research Institute.
I.T. was supported by the U.S. Department of Energy, Office of Science, Office of Nuclear Physics, under contract No.~DE-AC52-06NA25396, and by the U.S. Department of Energy, Office of Science, Office of Advanced Scientific Computing Research, Scientific Discovery through Advanced Computing (SciDAC) NUCLEI program. 

\addtocontents{toc}{\protect\setcounter{tocdepth}{1}}
\begin{refsection}
\chapter{Reflections on progress and challenges in low-energy nuclear physics}
\chapterauthor[1]{Dean Lee}
\chapterauthor[2]{Daniel R. Phillips}
\begin{affils}
    \chapteraffil[1]{Facility for Rare Isotope Beams and Department of Physics and Astronomy, Michigan State University, MI 48824, USA}
    \chapteraffil[2]{Department of Physics and Astronomy and Institute of Nuclear and Particle Physics, Ohio University, Athens, OH 45701, USA}
\end{affils}
\addtocontents{toc}{\protect\setcounter{tocdepth}{-2}}
During the program we enjoyed the many innovative talks and spirited discussions covering important and timely questions on nuclear forces, effective field theory, power counting, emulators, and lattice quantum chromodynamics.  Here we give some brief comments on a few of the topics. We hope our comments might have some value for others working in this field.

\section{Unitary limit and nuclear physics}

It is clear that the unitary limit for two-component fermions is immediately useful for describing the physics of dilute neutron matter.  Recent work also establishes that the unitary limit of four-component fermions is of relevance for understanding atomic nuclei \cite{Konig:2016utl}. In particular, the proximity of the (nucleon-nucleon) \NN system to the unitary limit results in the presence of universal correlations between few-nucleon observables~\cite{Bedaque:1998kg,Bedaque:1999ve,Platter:2004zs,Hammer:2011kg,Gattobigio:2019omi} and suggests that the Efimov effect may be nearly realized in the three-nucleon (\NNN) system~\cite{Efimov:1970zz,Rupak:2018gnc,Kievsky:2021ghz}.

If we consider quantum chromodynamics in the limit of large numbers of colors, then the most important \NN interactions have an underlying spin-flavor symmetry \cite{Kaplan:1995yg,Kaplan:1996rk,Cohen:2002qn,Cohen:2002im,CalleCordon:2009ps,Lee:2020esp} as do the dominant pieces of the three-nucleon force~\cite{Phillips:2013rsa}. For low momenta, the operators that are leading in large-$N_c$ are those permitted by Wigner-SU(4) symmetry~\cite{Wigner:1936dx}. The Wigner-SU(4) symmetry~\cite{Wigner:1936dx} also emerges as a symmetry of nuclear forces in the unitary limit~\cite{Mehen:1999qs}. The combined use of the momentum and large-$N_c$ expansion leads to useful insights into \NNN observables~\cite{Vanasse:2016umz} and the parity-violating NN force~\cite{Schindler:2015nga,Schindler:2018irz,Nguyen:2020quk}.

More recently, it has been suggested that the leading interactions in such a Wigner-SU(4) organization of the NN-force problem are also those that result in minimal entanglement in the \NN S-matrix~\cite{Beane:2018oxh,Low:2021ufv}. 

The fact that expansions around the unitary, chiral, and large-$N_c$ limit {\it all} provide insights into nuclear forces leads us to ask whether one
can systematically combine the around-unitarity and chiral expansions~\cite{Hammer:2019poc}? How about the chiral and large-$N_c$ expansions? Which limit is nuclear physics closer to: $m_q \rightarrow 0$ or $N_c \rightarrow \infty$? And what if the success of large-$N_c$ and/or the closeness of nuclear physics to the unitary limit are somehow a manifestation of a deeper quantum-information-theoretic phenomenon in QCD? If that were the case, how would it get built into an EFT for nuclear physics?

\section{Some questions about power counting in chiral EFT}
\begin{enumerate}
\item What order is the \NNN force in such a unified EFT? \LO---as suggested by Efimov physics~\cite{Kievsky:2016kzb}? \NLO---as it is in \chieft with explicit Deltas~\cite{vanKolck:1994yi}? \NNLO---as is currently practiced~\cite{Epelbaum:2002vt}? If it is \NLO, is that just the Fujita-Miyazawa piece of the 3N force? Or should it also include the short-distance operators with undetermined LECs?

\item  When we add higher-order corrections to a leading-order Hamiltonian in an \chieft do we intend those higher-order corrections to be treated in perturbation theory? How should we view the results if those higher-order corrections happen to make the Hamiltonian unbounded from below in certain parameter ranges?
\end{enumerate}

\section{Does a quantum phase transition make zero-range interactions a poor tool with which to describe nuclei?}

The parameters of nuclear physics appear to mean that nuclei sit near a quantum phase transition \cite{Elhatisari:2016owd}.  The phase boundary is between a Bose gas and a nuclear liquid. Which phase appears is controlled by the alpha-alpha scattering length. If the range of the nucleonic interactions are shorter than the size of the alpha particles, than the Pauli blocking between identical nucleons will cause the alpha-alpha interaction to be weakly attractive or even repulsive~\cite{Rokash:2016tqh,Kanada-Enyo:2020zzf}. Therefore, the alpha-particle size takes on a critical role in the structure of alpha-conjugate nuclei. 

The relationship between the alpha-particle size and the range of nucleonic interactions explains, for example, the instability of $^{16}$O against breakup into four alpha particles at leading order in pionless effective field theory \cite{Contessi:2017rww}.  The zero-range limit seems problematic for these systems, but a simple interaction near infinite scattering length with Wigner SU(4) symmetry, nonzero range, and significant local interactions seems to provide a useful starting point for studying atomic nuclei across the nuclear chart \cite{Lu:2018bat}.  

Similarly, if the nucleonic interactions have a significant range but the interactions are not local, then the alpha-alpha interaction may again not be sufficient to produce a nuclear liquid.\footnote{We use the term ``local'' to mean interactions whose interaction kernel is diagonal with respect to particle positions.} While the four-component unitary limit is relevant and useful for studying the physics of atomic nuclei with more than four nucleons, significant care must be taken regarding the range and locality of the nucleonic interactions relative to size of the alpha particles. 

\section{Some questions about eigenvector continuation}

Eigenvector continuation has recently emerged as a powerful tool that can reduce the computational load involved in solving the quantum many-body problem~\cite{Frame:2017fah,Konig:2019adq,Ekstrom:2019lss,Furnstahl:2020abp,Melendez:2021lyq}.

\begin{enumerate}
\item A meta-question is whether this method should really even be called eigenvector continuation. Recently the subspace-emulation strategy proposed in Ref.~\cite{Frame:2017fah} has been applied without using any eigenvectors~\cite{Melendez:2021lyq}. This raises the question of what the minimal conditions are for this very successful strategy to be used.

\item How do we estimate the errors/convergence of eigenvector continuation? Some work in this direction is reported in Ref.~\cite{Sarkar:2020mad}.

\item What does the workflow for improving nuclear forces with novel fitting strategies and eigenvector continuation look like in practice? A recent work on constraining \NNN-force parameters by the BUQEYE collaboration~\cite{Wesolowski:2021cni}, uses eigenvector-continuation emulation of \NNN and \NNNN bound-state calculations to facilitate simultaneous calibration of both \chieft parameters and the parameters of the statistical model that encoded the impact of the \chieft truncation on observables. 

\item Can we use eigenvector continuation to extend a power counting scheme where perturbation theory is not converging? For example, what happens if we do continuation in $g_A$ (or $g_A^2/f_\pi^2$) to control the strength of the one-pion-exchange potential? 
\end{enumerate}
\printbibliography[heading=subbibliography]
\end{refsection}
\addtocontents{toc}{\protect\setcounter{tocdepth}{1}}
\begin{refsection}
\chapter{Dependence of nuclear \textit{\textbf{ab initio}} calculations for medium mass nuclei on the form of the nuclear force regulator}
\chapterauthor[1]{Petr Navr{\'a}til}
\begin{affils}
\chapteraffil[1]{TRIUMF, 4004 Wesbrook Mall, Vancouver, British Columbia V6T 2A3, Canada}
\end{affils}

\addtocontents{toc}{\protect\setcounter{tocdepth}{-2}}
\Abinitio calculations for medium mass nuclei with chiral nuclear interactions as input show that there is a substantial dependence of binding energies, radii and other observables on the regulator used in particular in the three-nucleon interaction. This dependence is rather weak in light nuclei and therefore it was overlooked at first. The chiral effective field theory (\chieft) suggests that the regulator effects should be of higher order than the used perturbation expansion. However, in practical calculations for medium mass nuclei the effect is large, bigger than what one would anticipate from the chiral perturbation theory (\chipt).

The sensitivity is in particular significant to the functional form of the regulator, local vs. non-local or semi-local. The use of local regulators in the chiral 3N interaction, i.e., depending on the transferred momentum,  was widespread in the past as technically it is easier to implement the most complicated chiral three-nucleon (\NNN) terms with this type of regulator because the resulting interaction is local in the coordinate space~\cite{Navratil:2007zn}. However, obtained Hamiltonian with low-energy constants (LECs) typically determined in mass $A$=2-4 systems overbinds medium mass nuclei and underestimates nuclear radii~\cite{Binder:2013xaa,Lapoux:2016exf}. On the other hand, applications of non-local regulators in the \NNN interaction that depend on the relative nucleon momenta, gives much better results in medium mass nuclei for both binding energies and radii~\cite{Huther:2019ont}. It should be noted that most chiral nucleon-nucleon (\NN) interactions used in \abinitio calculations include non-local regulators~\cite{Entem:2003ft,Machleidt:2011zz,Entem:2017gor}, i.e., it appears that the use of a non-local regulator in the \NNN interaction is more consistent. Still, it has been argued recently that theoretically best justified is the application of semilocal regulators that preserve chiral symmetry~\cite{Epelbaum:2014sza,Reinert:2017usi}. However, the corresponding consistently regularized \NNN interactions have not been developed yet beyond \NNLO and results available so far show a similar overbinding problem as calculations with the local \NNN forces~\cite{Maris:2020qne}.

Recently, \NNN interactions that combine the use of both local and non-local regulators have been introduced~\cite{Soma:2019bso}. These interactions provide a good description of binding energies of light and medium mass nuclei including $^{132}$Sn\cite{Miyagi:2021pdc}. At the same time, the calculated radii are typically underestimated compared to experiment~\cite{Soma:2020dyc} and compared to the most successful interaction for the description of radii, the \NNLOsat~\cite{Ekstrom:2015rta}.

The strong dependence of binding energies and radii of medium mass nuclei on the regulator type needs to be further investigated and understood as it appears contrary to expectations from \chipt.

\section{Inclusion of the \NNN contact interaction at \NNNNLO } Contact terms contributing to the chiral \NNN interaction at \NNNNLO order has been derived recently~\cite{Girlanda:2011fh}. These fourteen terms accompanied by LECs impact the spin-orbit strength and the isospin dependence of the nuclear force among other other effects. It has been demonstrated that analysing power puzzle in the d-p data can be resolved considering even a subset of these terms~\cite{Kievsky:1999nw,Girlanda:2016bze}. Having 14 additional LECs at disposal, a high-precision fit to three-nucleon data should be possible, i.e., a partial wave analysis similar to that performed for the nucleon-nucleon data should be feasible. Such an analysis would undoubtedly result in a much better quality 3N interaction for applications in nuclei across the nuclear chart.

Recently, the spin-orbit ($E_7$) \NNNNLO contact term has been tested in calculations of $^7$Be(p,$\gamma$)$^8$B radiative capture. This term was shown to considerably improve the the analyzing power in the p-d scattering. 
Its inclusion improved the structure description of $^7$Be and $^8$B nuclei as well as of the $^7$Be(p,$\gamma$)$^8$B S factor when compared to experiment~\cite{Kravvaris:2022xx}.

The \NNN interaction \NNNNLO contact terms include a $T{=}3/2$ contribution. It might help to improve the description of nuclei far from stability with a large neutron or proton excess. Applications of this $T{=}3/2$ 3N contribution is worth exploring.

\section{Importance of calculations that include continuum effects} Most of {\it ab initio} calculations used to test nuclear forces involve bound-state observables such as binding energies, excitation energies, radii, and electroweak transitions between bound states. It should be noted that considering nuclear properties affected by continuum provide a complementary and often more comprehensive and strict test of nuclear forces. The reason is that even a straightforward experimental information such as elastic scattering cross section comprises information from bound states of scattering nuclei, resonances of the composite system, background phase shifts etc.~\cite{Kumar:2017yem} Even for describing an isolated resonance, one needs to calculate and compare not only its energy but also its width. Methods capable of including continuum effects should be applied more broadly in tests of nuclear forces. This is obviously done in few-body systems, e.g., \NN scattering and nucleon-deuteron scattering~\cite{Witala:2013ioa,Skibinski:2017vqs,Urbanevych:2020sjs}. However, continuum calculations for light and medium mass nuclei should also be considered~\cite{Navratil:2016ycn, Michel:2021jkx}.

\printbibliography[heading=subbibliography]
\end{refsection}
\addtocontents{toc}{\protect\setcounter{tocdepth}{1}}
\begin{refsection}
\chapter{Collective observables for nuclear interaction benchmarks}
\chapterauthor[1]{Kristina D. Launey}
\chapterauthor[1]{Grigor H. Sargsyan}
\chapterauthor[1]{Kevin Becker}
\chapterauthor[1]{David Kekejian}
\begin{affils}
    \chapteraffil[1]{Department of Physics and Astronomy, Louisiana State University, Baton Rouge, LA 70803}
\end{affils}

\newcommand{\SpR}[1]{\ensuremath{\mathrm{Sp}( #1,\mathbb{R} )}}
\newcommand{\braketop}[3]{\ensuremath{\left\langle #1 | #2 | #3 \right\rangle}}

\begin{figure*}[th]
\begin{center}
\includegraphics[width=\textwidth]{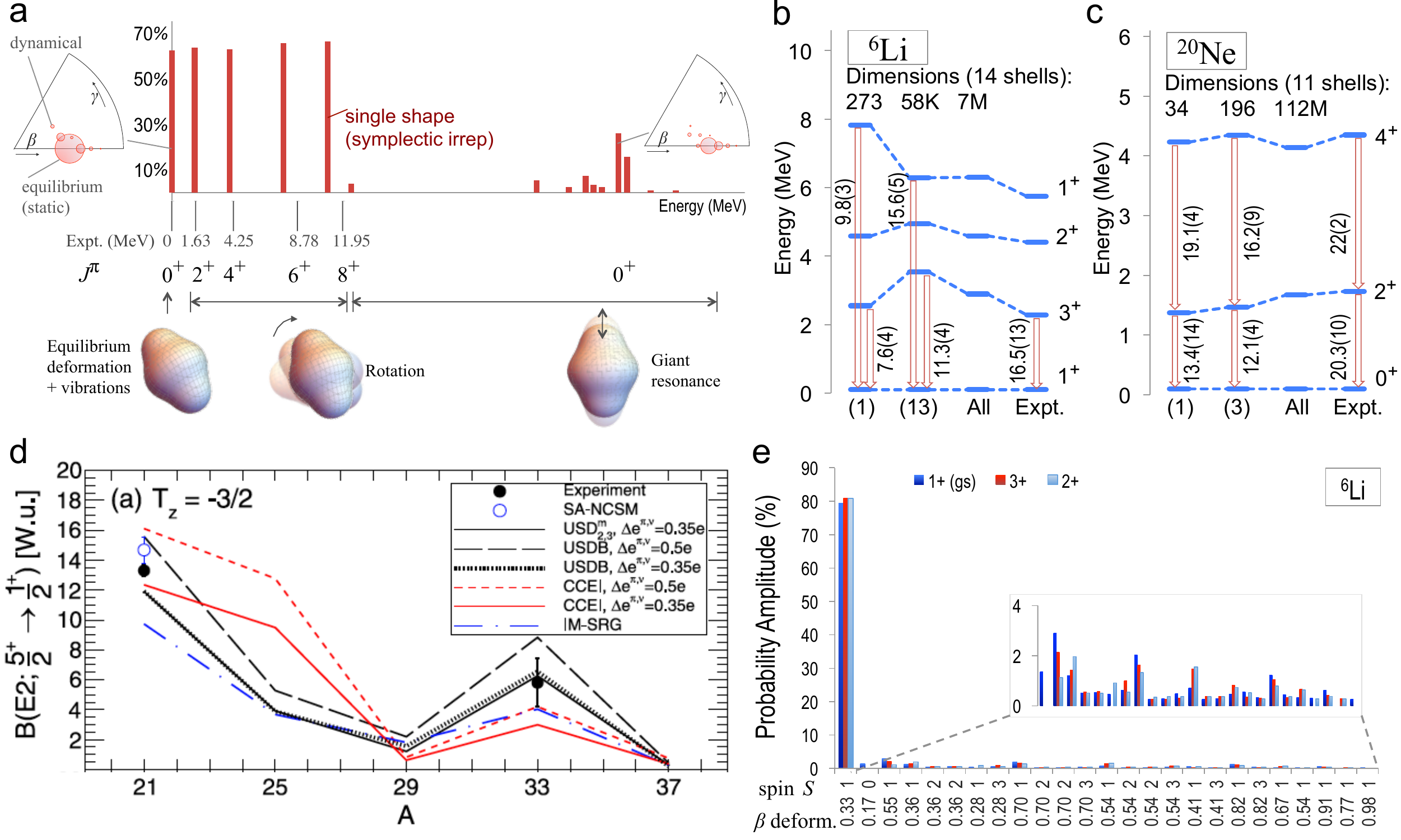}\\
\caption{\label{fig}  
{\small 
{\bf a.} Emergence of almost perfect symplectic symmetry in nuclei from first principles \cite{Launey:2021sua,Launey:2016fvy}, enabling  {\it ab initio} descriptions of clustering and collectivity in terms of nuclear shapes. {\bf b.} Observables for $^{6}$Li   calculated   in the SA-NCSM using only a small number of nuclear shapes (specified in the $x$-axis labels) and compared to experiment (``Expt."); dimensions of the largest model spaces used  are also shown. {\bf c.} The same, but for  $^{20}$Ne; for comparison, the corresponding complete dimension  is $3.8\times 10^{10}$. {\bf d.} Experimental and theoretical $B(E2; {5\over 2}^+ \rightarrow  {1\over 2}^+)$ values
for $T_z = -{3 \over 2}$; for $A = 21$,  the {\it ab initio}  SA-NCSM calculation is shown without the use of effective charges (figure adapted from Ref. \cite{Ruotsalainen:2018clb}).
{\bf e.}  Symplectic  \SpR{3} irreps  or nuclear shapes that compose the rotational band  states of $^6$Li; each irrep is specified by its equilibrium shape, labeled by the deformation $\beta$ and total intrinsic spin $S$. Figures adapted from Ref.~\cite{Dytrych:2018vkl}, unless otherwise stated.}}
\end{center}
\end{figure*}

\addtocontents{toc}{\protect\setcounter{tocdepth}{-2}}
Wave functions are not observables, however, they can provide critical information about the nuclear correlations and spin mixing in nuclei. Correlations drive properties of nuclei beyond the mean field and define important observables such as transitions, excitation spectra, and reaction observables (see, e.g., Ref.~\cite{Launey:2021sua}). Collective correlations are responsible for deformation of nuclei, and the \abinitio symmetry-adapted no-core shell-model (SA-NCSM) approach \cite{Launey:2016fvy,Dytrych:2018vkl} has  unveiled the ubiquitous presence of deformation in light to medium-mass nuclei. The importance of deformation is anticipated to hold even more strongly in heavy nuclei \cite{Rowe_1985,Castanos:1991mig,Jarrio:1991alv,Bahri:1999vm,Heyde_2011}. 
Specifically, Refs. \cite{Dytrych:2018vkl,Launey:2020rkj} show that nuclei and their excitations are dominated by only a few collective shapes that rotate (Fig. \ref{fig}a), which naturally emerge from first principles. Typically, low-lying states have a predominant shape that is realized by the most deformed configuration in the valence shell and particle-hole excitations above it. However, the probability amplitudes of these shapes vary to some extent from one parametrization of the chiral potentials to another, and this has a significant effect on reproducing collective observables.

\section{Nearly perfect symplectic symmetry in nuclei---radii and quadrupole moment operators}

First-principle nuclear structure calculations with various chiral potentials show that the special nature of the strong nuclear force determines highly regular patterns in nuclei that can be tied to an emergent approximate symmetry, the \SpR{3} symplectic symmetry \cite{Rosensteel_1977,Launey:2016fvy,Dytrych:2018vkl}. Since this symmetry does not mix nuclear shapes and is only slightly broken in nuclei, nuclear states are readily described by only a few subspaces that respect this symmetry, or a few nuclear shapes (Fig. \ref{fig}a-c). These subspaces extend to higher-lying harmonic oscillator (HO) shells and are imperative for reproducing collective observables -- see, e.g., Fig. \ref{fig}d for $B(E2)$ in $^{21}$F; see also the outcomes of a many-particle modeling \cite{Dreyfuss:2012us,Tobin:2013dya} that utilizes interactions inspired by the symplectic effective field theory  \cite{kekejian_2020}; see also Refs. \cite{Henderson:2017dqc,Williams:2019ezy} for $E2$ transitions in Mg isotopes.  
Besides the predominant shape(s), there is a manageable number of shapes, each of which contributes at a level that is typically at least an order of magnitude  smaller, as shown in  Fig.  \ref{fig}e.
Furthermore, practically the same symplectic content observed for the low-lying states in  $^{20}$Ne, Fig. \ref{fig}a, and for those in $^6$Li, Fig. \ref{fig}e,  is a rigorous signature of rotations of a shape and can be used to identify members of a rotational band. 
A notable outcome is that excitation energies and transition rates for a few nuclear shapes closely reproduce the experimental data, Fig. \ref{fig}c and d.  

This has important implications:

\begin{description}
\item[Deformation.] The nuclear deformation is calculated by the quadrupole moment operator $$Q_{2}=\sqrt{16\pi/5 }\sum_{k=1}^A r_k^2Y_{2}(\hat r_k).$$
This operator is a symplectic generator, which means that it has zero matrix elements between two shapes $\ket{\sigma_i}$ and $\ket{\sigma_j}$: $\braketop{\sigma_i}{Q_2}{\sigma_j}=0$, $\forall i \ne j$. This is important, since the largest fraction of the quadrupole moments and $E2$ transitions strengths,  and, hence nuclear collectivity,  necessarily emerges within the predominant symplectic irrep(s) or nuclear shape(s). The more the mixing of shapes, the smaller these collective observables. 
\item[Radii.] The nuclear size is calculated by the monopole moment operator $r^2=\sum_{k=1}^A{\vec r_k \cdot \vec r_k}$. This operator is a symplectic generator, which means that it has zero matrix elements between two shapes $\ket{\sigma_i}$ and $\ket{\sigma_j}$: $\braketop{\sigma_i}{r^2}{\sigma_j}=0$, $\forall i \ne j$. Different from the quadrupole moment, the rms radius provides the average radius of a shape that describes the size of a shape regardless if it is deformed or not, thereby converging at comparatively smaller model spaces compared to deformation-related observables.
\item[Spin mixing.] The total intrinsic spin is a good quantum number for each nuclear shape. Hence, more spin mixing implies more mixing of shapes, thereby reducing collective observables.
\item[Shape vibrations.] A nuclear shape is composed of an equilibrium shape (typically, a configuration in the valence shell) and vibrations. An important point is that a nuclear shape becomes energetically favored only when vibrations are allowed to develop within a model space. In limited model spaces, collective observables are highly reduced for two reasons: (1) shapes of enhanced deformation are suppressed, while other less deformed shapes enter the eigenfunctions, and (2) vibrations are largely suppressed, thus for example, for $^{20}$Ne's predominant shape, $B(E2; 2^+ \rightarrow 0^+_{\rm gs})= 13.4(14)$ W.u. in 11 shells  (Fig. \ref{fig}c), whereas it reduces to 4.2 W.u. for the  equilibrium shape only (valence shell).
\end{description}

\noindent
Note the critical difference between rms radii, $\braketop{\Psi}{r^2}{\Psi}=\sum_{i}c_i^2\braketop{\sigma_i}{r^2}{\sigma_i}$, and quadrupole moments, $\braketop{\Psi}{Q_2}{\Psi}=\sum_{i}c_i^2\braketop{\sigma_i}{Q_2}{\sigma_i}$: radii of individual shapes always add, whereas quadrupole moments can add (for oblate shapes), subtract (prolate) or be zero (spherical shapes). The quadrupole moments, and, hence, $E2$ transitions further decrease if largely deformed shapes are suppressed. This exposes the role of the  symplectic symmetry, established as a remarkably good symmetry of the strong nuclear force in the low-energy regime,  in guiding toward and calculating precise nuclear observables.

\section{Recommendations} 
Collective observables are essential for parametrizing nuclear interactions and for benchmark studies to ensure the proper account of the physics of nuclear dynamics in \textit{ab initio} modeling. First, it is important to monitor collective observables for various potential models and, in particular, perform global sensitivity analysis that probes the sensitivity of  collective observables  \cite{Becker_2020} to the low-energy constants that enter the chiral potentials,  similarly to that of bulk properties  \cite{Ekstrom:2019lss}. Second, for benchmark studies, it is imperative to include quadrupole moments and/or $E2$ transitions. An ideal test case is the $B(E2; 2^+ \rightarrow 0^+_{\rm gs})$ in $^{12}$C, which is well measured 4.65(26) W.u.; alternatively, the quadrupole moment of the first excited $2^+$  state in $^{12}$C could be used (requires a calculation of a single state), although there are still large uncertainties in the recommended value, 6(3) $e$ fm$^2$. A particularly interesting case is the  $B(E2; 3^+ \rightarrow 1^+_{\rm gs})$ transition rate in $^{6}$Li that remains a challenge to realistic interactions. For such calculations, it is beneficial if a many-body approach is used, such as the SA-NCSM, that does not require renormalization of the chiral interactions in the nuclear medium, does not use effective charges, and admits any type of the nuclear interaction including non-local interactions.

{\bf Acknowledgements:}
This work was supported by the U.S. National Science Foundation (PHY-1913728) and benefitted from computing resources provided by the National Energy Research Scientific
Computing Center NERSC (under Contract No. DE-AC02-05CH11231), Frontera computing project at the Texas Advanced Computing Center (under National Science Foundation award OAC-1818253) and LSU ({\tt www.hpc.lsu.edu}).
\sloppy
\printbibliography[heading=subbibliography]
\end{refsection}
\addtocontents{toc}{\protect\setcounter{tocdepth}{1}}
\begin{refsection}
\chapter{Role of the continuum couplings in testing the nuclear interactions}
\chapterauthor[1]{Marek P{\l}oszajczak}
\begin{affils}
\chapteraffil[1]{Grand Acc\'el\'erateur National d'Ions Lourds (GANIL), CEA/DSM - CNRS/IN2P3, BP 55027, F-14076 Caen Cedex, France}
\end{affils}

\addtocontents{toc}{\protect\setcounter{tocdepth}{-2}}
The development of the nuclear forces for precision nuclear physics demands reliable methods to improve interactions by analyzing the discrepancies between results of chosen many-body approach and experimental data. Hence, the appropriate many-body approach and the relevant choice of studied observables are the two essential ingredients of testing of the interactions. 

The many-body calculation of nuclear observables brings new aspects to the problem of testing the interactions, such as the dependence of interaction in a given nucleus on nucleon number, or the role of couplings to reaction channels and scattering continuum. These two aspects are a consequence of the attempts to improve interactions by calculating various observables in heavier nuclei.   

One aspect concerns the changing role of the many-body forces with increasing number of nucleons because the interplay between two- and higher-body forces depends on the number of possible 2-body, 3-body, etc. couplings (talk of C.-J. Yang) in many-particle systems. This induces the dependence of EFT power counting on particle number \cite{Yang:2020pgi}. On the other hand, the relative number of the different k-tuple couplings depends on the nucleon density distribution, i.e., the relative number of nucleons in the surface region and in the interior of nucleus. These many-body effects could be absorbed in the effective parameters of the EFT interaction fitted locally to the individual nuclei or their small sets.

Another aspect is related to the role of the continuum coupling which leads to the appearance of new energy scale(s) in the EFT power counting related to the distance from the threshold of reaction channel(s). An optimal way to test interactions in the vicinity of different particle emission channels is to employ the many-body approach which preserve unitarity at each opening of new reaction channel.

 Weak binding brings another aspect which is related to the coupling to the scattering states and various particle decay channels. Indeed, experimental data even at low excitation energies contain states which are either weakly bound or unbound like, for example, $A$=5 nuclei which are all unbound even in the ground state. Calculation of the binding energy in long isotopic chains brings the issue of dependence on the asymmetry of proton $S_p$ and neutron $S_n$ separation energies \cite{Gade:2004zz,Gade:2008zz}, such as the weakening of interaction between unlike nucleons and the asymmetry of neutron-neutron and proton-proton interactions. 
 
 The continuum couplings can be included either in the complex-energy continuum shell model using Berggren ensemble of single-particle states \cite{Michel:2002tm,Michel:2021jkx}, the so-called Gamow shell model, or in the real-energy continuum shell model in the projected subspaces \cite{Okolowicz_2003}. Gamow shell model can be formulated either in core + valence particle model space \cite{Michel:2002tm,Hu:2020pxl} or in the no-core basis \cite{Papadimitriou:2013ix}. Berggren representation has been also used in the coupled-cluster approach \cite{Hagen:2006pq}. Another approach which has been used extensively, is the no-core shell model, including couplings to the reaction channels via the $R$-matrix approach \cite{Baroni:2013fe}. 
 
The problem of deriving a reliable inter-nucleon interaction (\NN, \NNN, \NNNN,$\ldots$) for precision nuclear physics calculations cannot be separated from the choice of the model space and the many-body approach. Unitarity, which is the fundamental property of Quantum Mechanics, is violated by most microscopic nuclear theories used. Coupling to the environment of scattering continuum and decay channels does not reduce to refitting parameters of the interaction which have been adjusted to observables in well-bound states. It is necessary that new families of the interactions are tested on a broader set of data including resonances and low-energy scattering observables, using the many-body frameworks respecting the unitarity. 

\printbibliography[heading=subbibliography]
\end{refsection}
\addtocontents{toc}{\protect\setcounter{tocdepth}{1}}
\begin{refsection}
\chapter{Comments on the $\boldsymbol{\chi}^2$ values in the three-nucleon sector}
\chapterauthor[1]{Alejandro Kievsky}
\begin{affils}
\chapteraffil[1]{Istituto Nazionale di Fisica Nucleare, Largo Pontecorvo 3, I-56100 Pisa, Italy}
\end{affils}
\addtocontents{toc}{\protect\setcounter{tocdepth}{-2}}
In recent years a substantial progress has been achieved in the development of 
accurate descriptions of the interaction between nucleons using the systematic framework
provided by chiral perturbation theory. In the nucleon-nucleon (\NN) sector, potentials obtained up to 4th and 5th
order in the nuclear EFT expansion have provided an extremely accurate description 
of the \NN world data with a value of $\chi^2$ per datum close to one.
Most of the data that these potentials
describe are proton-proton and neutron-proton scattering cross-sections and analyzing powers. The same class of data exist
in other sectors, as the three-nucleon (\NNN) or four-nucleon scattering (\NNNN).
Focusing in the \NNN system, it is a fact that all realistic
\NN potentials describe poorly some of these observables. In particular,
vector and tensor analyzing powers in proton-deuteron scattering are underpredicted by
almost all those \NN potentials even if they are supplemented by a
\NNN force. However at present the \NN and \NNN
interactions are considered at different orders. As example, in table~\ref{tab:chisq}, we show the $\chi^2$
per datum obtained after solving proton-deuteron scattering at low energies~\cite{Kievsky:2001fq, Marcucci:2009xf}.
Two different interactions are considered, the widely used Argonne V18 interaction (AV18),
without and with the inclusion of the Urbana IX (UR) \NNN force,
and the chiral based interaction by Entem and Machleidt (Idaho-\NNNLO), with and without 
the inclusion of the \NNN force at \NNLO. The strength
of this force depends on two low-energy constants (LECs) that have been determined by
fixing the triton binding energy and the doublet neutron-deuteron scattering length.
In all the cases examined the $\chi^2$ per datum of the analyzing powers could
exceed the value of one hundred, in particular the vector analyzing powers $A_y$ and $iT_{11}$.
The tensor analyzing powers are slightly better described, however the $\chi^2$ value of $T_{21}$
is very high in some cases.

\begin{table}[h]
\caption{$\chi^2$ per datum obtained in the description of the proton-deuteron vector and
tensor analyzing powers \label{tab:chisq}}
\centering
\begin{tabular}{ll|ccccc}
\hline
Energy   & potential & $A_y$&$iT_{11}$&$T_{20}$&$T_{21}$&$T_{22}$ \\
\hline
1 MeV    & AV18      & 276  & 112     & 3.5    & 4.5    &  2.8    \\
         & AV18+UR   & 190  & 61      & 1.0    & 2.5    &  0.7    \\
         & Idaho-N3LO& 197  & 68.7    & 4.0    & 2.5    &  1.5    \\
         & N3LO+N2LO & 139.9& 49.5    & 2.7    & 2.5    &  0.9    \\
\hline
3 MeV    & AV18      & 313  & 205     & 4.8    & 6.7    &  12     \\
         & AV18+UR   & 271  & 144     & 5.4    & 11     &  2.4    \\
         & N3LO      & 186  & 108.3   & 1.9    & 2.8    &  4.4    \\
         & N3LO+N2LO & 114  & 85.8    & 3.6    & 8.3    &  1.6    \\
\hline
5 MeV    & AV18      & 211  & 99      & 6.8    & 12     &  7.8    \\
         & AV18+UR   & 186  & 59      & 26     & 36     &  1.5    \\
\hline
7 MeV    & AV18      & 303  &  90     & 19     & 38     &  1.9    \\
         & AV18+UR   & 239  &  56     & 40     & 81     &  4.2    \\
\hline
9 MeV    & AV18      & 292  & 165     & 42     & 70     &  38     \\
         & AV18+UR   & 218  & 134     & 63     & 86     &  7.2    \\
\hline
10 MeV   & AV18      & 288  & 29      & 10     & 6.2    &  24     \\
         & AV18+UR   & 224  & 23      & 13     & 6.1    &  7.6    \\
\hline
\end{tabular}
\end{table}

These results suggest that the complicated structure of the \NNN force has
to be further analysed. Recently, the contact three-nucleon interaction 
at \NNNNLO has been worked out showing that there are thirteen new LECs to be determined. Moreover, the spin structure is sufficiently flexible
to guarantee a better description of the polarization observables at low energies.
The results of Ref.~\cite{Girlanda:2018xrw} show that using the proton-deuteron data as input, it is possible
to fit these \NNN LECs and obtain values of the $\chi^2$ per datum similar to those
obtained in the \NN sector.

\printbibliography[heading=subbibliography]
\end{refsection}
\addtocontents{toc}{\protect\setcounter{tocdepth}{1}}
\begin{refsection}
\chapter{Perspectives on an open-source toolchain for \abinitio nuclear physics}
\chapterauthor[1,2]{Matthias~Heinz}
\chapterauthor[3,4]{Thomas Duguet}
\chapterauthor[5]{Harald W. Grie{\ss}hammer}
\chapterauthor[6,7]{Heiko Hergert}
\chapterauthor[1,2]{Alexander Tichai}
\begin{affils}
\chapteraffil[1]{Technische Universit\"at Darmstadt, Department of Physics, 64289 Darmstadt, Germany}
\chapteraffil[2]{ExtreMe Matter Institute EMMI, GSI Helmholtzzentrum für Schwerionenforschung GmbH, 64291 Darmstadt, Germany}
\chapteraffil[3]{IRFU, CEA, Universit\'e Paris-Saclay, 91191 Gif-sur-Yvette, France}
\chapteraffil[4]{KU Leuven, Instituut voor Kern- en Stralingsfysica, 3001 Leuven, Belgium}
\chapteraffil[5]{Institute for Nuclear Studies, Department of Physics,
The George Washington University, Washington DC 20052, USA}
\chapteraffil[6]{Facility for Rare Isotope Beams, Michigan State University, East Lansing, MI 48824, USA}
\chapteraffil[7]{Department of Physics \& Astronomy, Michigan State University, East Lansing, MI 48824, USA}
\end{affils}

\addtocontents{toc}{\protect\setcounter{tocdepth}{-2}}

One topic of discussion at the workshop was the open-sourcing of codes necessary to evaluate and transform nuclear interaction matrix elements for use in many-body codes.
We offer here some perspectives on what this could look like and how it could be achieved.

\section{Introduction}
The long-term benefits of open-sourcing computational tools are obvious:
a standard set of validated, time-tested tools would provide a strong base for any further interaction and many-body developments.
Even in the short term, the open availability of nuclear interaction matrix elements in formats used for many-body methods should allow many-body practitioners to experiment with different interactions and encourage greater cross-talk between many-body specialists and interaction specialists.
Another consequence would be that the development effort required to begin research on many-body methods would be reduced, opening access to the field for more researchers without the resources to obtain or develop all of the required matrix-element-processing tools.
Combined with the continuous open-sourcing and/or publication of many-body codes, this would provide an excellent environment for nuclear theorists with various backgrounds to perform different studies of nuclear interactions in medium-mass systems.

However, there exists a natural concern regarding how more cutting-edge developments that are open-sourced are appropriately acknowledged and how the people responsible for these developments obtain the scientific credit they are due. If implemented with proper usage guidelines, the open availability of interaction matrix elements can provide an enforceable standard for how to deal with citations and acknowledgments when using interactions and tools provided by others.

\section{Sharing and distribution of interaction matrix elements}
A simple starting point would be to make matrix elements available in formats used by configuration-space many-body methods, i.e., (angular-momentum- and isospin-coupled) single-particle harmonic-oscillator (HO) matrix elements, through a public read-only repository.
These could also be made available in the form of individual terms proportional to different low-energy constants (LECs) to allow for fits or sensitivity studies in realistic applications.
The matrix elements should be provided with some attached metadata on the code(s) and version(s) used to generate the matrix elements and who actually generated them.
Ideally, this metadata should have all the information needed to reproduce the matrix elements using standard tools available in the repository, which we discuss next.
For this reason, the metadata should support updates that help to clarify different strategies or approximations that went into generating the matrix elements to make them more reproducible.
The starting interaction matrix elements can also be provided in code form (or as data with a read-in library), with standard interfaces to obtain coordinate-space/momentum-space matrix elements.

\section{Tools for matrix element usage}
To make the generation of single-particle HO matrix elements fully reproducible, one would make available the tools to do the individual transformations required to generate these matrix elements (starting from the center-of-mass (COM) to lab frame transformation).
A full set of these tools would be nucleon-nucleon (\NN) and three-nucleon (\NNN) COM-to-lab transformation codes, \NN and \NNN coordinate-space/momentum-space to Jacobi HO transformation codes, and possibly momentum-space and Jacobi HO similarity renormalization group codes.
Any matrix elements made available in a data format should provide a link to some documentation that defines how the data format is structured.
Additionally, matrix element formats should provide routines to read in matrix elements and access specific matrix elements as a small library (to be employed by the end user).
A set of conversion tools to convert between equivalent matrix element formats may also prove useful.

\section{Explorations of nuclear forces in medium-mass systems}
Providing \NN and \NNN matrix elements broken up into terms proportional to individual LECs in a single-particle HO format would immediately allow many-body theorists to experiment with LEC choices and study their effects in medium-mass nuclei.
For interaction specialists to do the same, one would need open-source versions of (at least) standard many-body solvers for closed-shell systems.
We believe that the effort to open-source matrix elements outlined here should be paralleled by the practice of open-sourcing many-body codes.
The ideal candidates for open-sourcing are stable older versions of codes where the original developers have completed the studies for which the code was intended.
This way many-body practitioners may still exclusively profit off of their cutting-edge developments, but their publication of established versions opens their code to more people, leading to citations and collaboration opportunities with other interested nuclear theorists.
In nuclear theory, there are some examples of open-source many-body codes that have helped open the field of many-body theory to more people: the self-consistent Green's function code by Barbieri~\cite{Dickhoff:2004xx,Barbieri:2007zy,Barbieri:2009nx,Soma:2013ona,Barbieri_2021}, the in-medium similarity renormalization group code by Stroberg~\cite{Stroberg:2016ung,Stroberg_2021}, and the many shell-model codes made available by their developers, such as \textsc{NuShellX}~\cite{Brown:2014bhl,Rae_2021}, \textsc{KShell}~\cite{Shimizu:2019xcd,Shimizu_2021}, \textsc{BIGSTICK}~\cite{Johnson:2018hrx,Johnson_2021}, and \textsc{Antoine}~\cite{Caurier:2004gf,Caurier_2021}.

\section{Ensuring proper credit and acknowledgment}
By standardizing access to and use of interaction matrix elements, there exists the opportunity to set usage guidelines that would ensure ``good behavior'' by users.
These should be detailed in a general usage license for the tools and data made available on this public repository.
Additionally, publishers of interaction matrix elements (including those of single-particle HO matrix elements that employ advanced strategies to handle 3N forces) should be allowed to extend the general usage license to include required citations and/or acknowledgment.
There is also the opportunity to require procedures (e.g., convergence checks) to be taken care of and discussed.
However, these requirements beyond general usage and acknowledgment would be additional restrictions and should be added with caution so they do not inadvertently restrict the usage of matrix elements too much.
``Good behavior'' would encompass abiding by the license and using the publicly available codes where possible to standardize the matrix elements,
as small undocumented changes in the matrix element transformations can lead to unexpectedly large changes in observables, which could hurt reproducibility.

\section{Considerations for publishing codes}
For codes made publicly available, ``publication'' would generally mean pushing a version of the source to a public repository (for example, on GitHub or GitLab).
The code in the repository would be available for all to use (provided they comply with the license, which should be generally compatible with the usage license for the matrix element repository outlined above) and would allow for community inspection and contributions (for example, optimizations or to add support for other file formats).
The matrix element repository would specify ``pinned'' standard versions of relevant codes that are the currently best supported versions and whose usage to generate matrix elements for the repository would be considered ``good behavior.''
For released versions associated with recent developments, it may make sense to associate a released version of the code with a journal publication.
This should be included in the usage license and would make appropriate acknowledgment unambiguous.

\section{Closing thoughts}
We have outlined what an open-source repository of matrix elements and transformation tools might look like, along with some principles to guide behavior and encourage people to contribute.
This is only the very beginning, and next steps would be to secure buy-in from more of the community and to form a ``collaboration'' to secure computational and organizational resources to manage this.
On the latter point, there is the possibility to draw on prior experience from other fields where these practices are more established, like lattice quantum chromodynamics~\cite{USQCD}, quantum chemistry~\cite{Psi4}, and astrophysics.
We acknowledge that some institutions (for example, national laboratories) may make it challenging to contribute to open-source projects; we hope the establishment of an organization for these efforts could aid in the streamlining of contributions to these efforts by researchers that would otherwise be hindered by bureaucratic overhead.
Buy-in from key matrix element producers (broadly speaking, including those who generate single-particle matrix elements for use in their many-body codes) and an inclusive global partnership would make open-source matrix elements the new standard and allow nuclear theory to profit from the unprecedented open accessibility of information and computational tools for \abinitio nuclear-theory calculations.

\textbf{Acknowledgments:} We thank Benjamin Bally and Zohreh Davoudi for useful discussions on open-source collaborations. This work was supported in part by the US Department of Energy under contracts DE-SC0015393 (H.W.G.), DE-SC0017887 (H.H.), and DE-SC0018083 (NUCLEI SciDAC-4 Collaboration, H.H.), by the Deutsche Forschungsgemeinschaft (DFG, German Research Foundation) -- Projektnummer 279384907 -- SFB (M.H., A.T.), and by the BMBF Contract No.\ 05P18RDFN1 (M.H., A.T.).

\printbibliography[heading=subbibliography]
\end{refsection}
\addtocontents{toc}{\protect\setcounter{tocdepth}{1}}
\begin{refsection}
\chapter{Few-body emulators based on eigenvector continuation}
\chapterauthor[1]{Christian Drischler}
\chapterauthor[1,2]{Xilin Zhang}
\begin{affils}
\chapteraffil[1]{Facility for Rare Isotope Beams, Michigan State University, MI 48824, USA}
\chapteraffil[2]{Department of Physics, The Ohio State University, Columbus, OH 43210, USA}
\end{affils}

\newcommand{\eg}{\textit{e.g.}\xspace}
\newcommand{\ie}{\textit{i.e.}\xspace}

\newcommand{\etal}{\textit{et~al.}\xspace}

\addtocontents{toc}{\protect\setcounter{tocdepth}{-2}}

In this contribution we briefly recapitulate the progress made in constructing fast and accurate emulators for few-body scattering and reaction observables based on eigenvector continuation. 
Emulators have been game changers and we envision them to play a key role in future workflows in nuclear physics and beyond. 
They have the potential to push the frontier of precision nuclear physics even further by enabling full Bayesian analyses of nuclear structure, scattering, and reaction observables, as well as by facilitating constraints for chiral interactions from (lattice) quantum chromodynamics (QCD).
The future will show what other exciting applications are within reach.

\section{Making the impossible possible: emulators} \label{sec:intro}

The power of emulators lies in trading an exact solution with a highly accurate approximation obtained using only a (small) fraction of the computational resources.
One needs to train an emulator {\it only once} on a small number of exact solutions (\eg, to the many-body Schr{\"o}dinger equation) in the model's phase space and can then efficiently make approximate predictions at all other points instead of  evaluating the model exactly. 
Trained emulators also allow one to make intricate model calculations publicly available as self-contained mini-applications. 
This enables users to make fast and accurate model predictions without having the detailed knowledge and computational resources otherwise necessary to build and run an application from complex (and sometimes closed source) code bases.
We consider this an important feature for future workflows in nuclear physics and beyond.

Emulators have been game changers in nuclear physics, where Bayesian methods have become standard tools for rigorously quantifying uncertainties in model predictions~\cite{Furnstahl:2014xsa,Zhang:2015ajn,King:2019sax}; \eg, for low-energy observables derived from chiral effective field theory (\chieft)~\cite{Epelbaum:2008ga,Machleidt:2011zz,Hammer:2019poc,Epelbaum:2019kcf}. 
Bayesian parameter estimation~\cite{Zhang:2015ajn,Wesolowski:2018lzj}, model comparison~\cite{Phillips:2020dmw}, and sensitivity analysis~\cite{Ekstrom:2019lss} can provide important insights to validate and improve model predictions. But their application in statistical analyses typically requires the model's parameter space to be repeatedly evaluated for (large-scale) Monte Carlo sampling, which was in most cases prohibitively slow due to the computationally expensive nature of nuclear structure,  scattering, and reaction calculations. Emulators have recently changed this practical limitation and made (what was thought) the impossible possible~\cite{Konig:2019adq,Ekstrom:2019lss}, by significantly reducing computing cost for evaluating models.\footnote{Applications of emulators are not limited to statistical analyses. For instance, they have also been used as interpolants and extrapolants in solving chaotic three-body problems~\cite{Breen_2020} and in quantum molecular calculations~\cite{Bogojeski_2020}.}

Implementations of emulators include Gaussian processes~\cite{rasmussen2006gaussian}, neural networks (see \eg, Refs.~\cite{Breen_2020,Bogojeski_2020}), and eigenvector continuation (EC)~\cite{Frame:2017fah, Sarkar:2020mad}. In low-energy nuclear physics, the number of EC-driven emulators applied to few- and many-body bound state calculations (\ie, subspace projection methods) is increasing~\cite{Konig:2019adq, Ekstrom:2019lss, Demol:2019yjt, Bai:2021xok, Yoshida:2021jbl}. 
As discussed in Sarah Wesolowski's talk~\cite{SarahTalk} and her recent work~\cite{Wesolowski:2021cni}, such a few-body bound state emulator enabled the construction of the first set of order-by-order chiral interactions with theoretical uncertainties fully quantified. The challenge is now to extend these efficient emulators for bound state calculations to scattering and reactions.

In this contribution to the ``perspective pieces'' we focus on the extension of EC-driven emulators to two-body (Section~\ref{sec:two_body}) and higher-body scattering (Section~\ref{sec:three_body}), as it was presented in our talks~\cite{ChristianTalk,XilinTalk} at the 2021 INT program ``Nuclear Forces for Precision Nuclear Physics''\footnote{\url{https://sites.google.com/uw.edu/int/programs/21-1b}}. We then provide an outlook for new EC-driven emulators for few-body systems in periodic boxes and external potential traps as well as their interesting applications (Section~\ref{sec:few_body_discrete_levels}).

\section{Setting the stage: Two-body scattering and reactions}
\label{sec:two_body}

Furnstahl~\etal~\cite{Furnstahl:2020abp} have recently demonstrated that EC can be used to construct extremely effective trial wave functions for fast and accurate variational calculations of two-body scattering observables, as discussed in Christian Drischler's talk~\cite{ChristianTalk}.
Specifically, Ref.~\cite{Furnstahl:2020abp} applied the Kohn variational principle (KVP) for the $K$-matrix to a range of test potentials, including nucleon-nucleon (\NN) and optical potentials. 
More recently, Melendez~\etal~\cite{Melendez:2021lyq} studied Newton's variational method using EC-motivated trial $K$-matrices (instead of trial wave functions) and emulated, \eg, neutron-proton cross sections based on a modern chiral \NN potential.
The emulator's high accuracy and speedup compared to exact scattering calculations are remarkable.
Following a different approach, Miller~\etal~\cite{Miller:2021vby} implemented the wave-packet continuum discretization (WPCD) method with GPU acceleration, which is capable of approximating scattering solutions fast at several energies simultaneously. 

Comparing the efficacies of the available emulators for scattering observables quantitatively is an important task for future work. 
Such a comparison requires a scattering scenario with matching (real and complex) interactions and a common definition of the term {\it exact solution} for reference calculations. 
Important benchmarks then include the emulator's accuracy and speedup relative to the exact scattering solution, and their susceptibility to numerical noise as well as spurious singularities known as Kohn (or Schwartz) anomalies~\cite{PhysRev.124.1468}. 
We also envision studies of a wide range of modern interactions (\eg, different resolution scales and regularization schemes) for both proof-of-principle calculations and uncertainty quantification. Applications of statistical methods using emulators will provide important insights into the idiosyncrasies of nuclear potentials to help address known issues. For issues in \chieft, see, \eg, Ref.~\cite{Furnstahl:2021rfk}.

Also important are studies aimed at quantifying the emulator's intrinsic errors and their dependence on the underlying training set such as the position and number of the training points. 
In particular, the rate of convergence of EC for scattering calculations needs to be investigated further. 
Progress along those lines has already been made for bound-state calculations~\cite{Sarkar:2020mad}. 
A machine learning algorithm that positions and/or adapts a given number of training points such that the emulator's intrinsic errors are minimized would be extremely useful for future applications of EC-driven emulators. 

Variational calculations of scattering observables are known to be prone to Kohn anomalies (see Refs.~\cite{Melendez:2021lyq, Drischler:2021xx} for recent discussions). 
For a given set of model parameters, the anomalies occur at energies where no (unique) stationary approximation due to the variational functional exits. 
While Kohn anomalies can be straightforwardly spotted in proof-of-principle calculations (in which the exact solution is also computed), their presence can limit in practice the applicability of, \eg, Monte Carlo sampling of the model's parameter space. 
As discussed in Christian Drischler's talk~\cite{ChristianTalk, Drischler:2021xx}, the generalized KVP has been used to efficiently detect and mitigate those anomalies (see also Ref.~\cite{Kievsky:1997zf}). 
To this end, the method in Ref.~\cite{Drischler:2021xx} assesses the consistency of stationary approximations obtained from a family of functionals with different scattering boundary conditions. 
This strategy is applicable to other variational calculations. 
Although Kohn anomalies were not an issue in the proof-of-principle calculations in Refs.~\cite{Furnstahl:2021rfk,Melendez:2021lyq}, it is important to study their emergence in more detail and implement efficient detection algorithms, especially, for Monte Carlo sampling. The rich literature in this field is an excellent starting point for future work along those lines~\cite{nesbet1980variational,PhysRevA.40.6879,doi:10.1063/1.454462}. 

These advances in developing fast and accurate emulators for two-body scattering are promising for future extensions to scattering problems where emulators are essential such as three-body scattering (see Section~\ref{sec:three_body}). 

Further, the fast convergence observed by Melendez~\etal~\cite{Melendez:2021lyq} using EC-motivated trial matrices (rather than trial wave functions) might indicate that the EC concept could also be applied to other stationary calculations. This would open exciting possibilities for future applications of emulators, some of which might not even be in sight at the moment.

\section{Rising action: Three and higher-body scattering and reactions}
\label{sec:three_body}
The two-body EC scattering and reaction emulator~\cite{Furnstahl:2020abp} has been generalized to three-boson elastic $s$-wave scattering, as reported in Xilin Zhang's talk~\cite{XilinTalk}. The first results are encouraging: the emulators have similar accuracy and computing speed compared to those in the two-body sector. The speed up can reach an order of $10^6$:  directly solving a three-body nuclear scattering problem takes about $10^3$ seconds while the emulator's cost on a laptop is milliseconds if the interactions' parametric dependencies are factorized from the operators. The paper summarizing this work is in preparation~\cite{Zhang:2021xx}.  

This work opens up the possibility of fitting chiral three-nucleon interactions efficiently to proton-deuteron scattering and reaction data, which has been difficult due to the expensive cost of solving three-body Faddeev equations. However, further progress needs to be made before achieving this goal, including generalizations to arbitrary partial-wave channels\footnote{The variational approaches (\eg, Refs.~\cite{Kohn:1948zz,newton2002scattering}) also work for fixed scattering angles. This suggests emulators can be constructed without using a partial-wave decomposition.} and spin statistics, and nontrivial generalizations above the deuteron break-up threshold,\footnote{There exists work on Kohn-type variational approaches in this kinematic region, see \eg, Refs.~\cite{Lieber:1972uk,Viviani:2001sy}} as well as implementations of these emulators for fitting chiral interactions. 

Three-body emulators will also be useful in analyzing 
deuteron-nucleus scattering and reactions (and in fact any process that can be described in terms of three-cluster dynamics). Full exploration of an effective Hamiltonian's parameter space (including testing new interaction operators) will become possible thanks to the emulators' speed.  However, a new challenge\footnote{One difficulty in full three-body calculations is the implementation of the Coulomb interaction with large Sommerfeld parameters (see, \eg, Ref.~\cite{Hlophe:2019wwg}). It will be interesting to study the manifestation of this difficulty in the EC emulators.} in these applications comes from the fact that some parametric dependence cannot be factorized from the associated operator, such as the range of the interactions. An immediate solution~\cite{Zhang:2021xx} is to decompose the interaction potential as a linear combination of a series of potentials (\eg, those constructed using orthogonal polynomials), but its feasibility, depending on the required number of basis, needs to be studied case by case. In Ref.~\cite{Zhang:2021xx}, the method of emulator-in-emulator is also explored and demonstrated to solve this issue, which uses Gaussian process within the EC-driven emulators.     

In the longer term, four-nucleon interactions will become relevant.  
With that in mind, it will be necessary to further extend the EC scattering and reaction emulators to four-body systems. The emulators will not only enable parameter estimation but also provide users effortless applications of these expensive calculations in their own research.

\section{(Not really the) Final act: Few-body systems with discrete energy levels}
\label{sec:few_body_discrete_levels}
Besides experimental constraints, there will be valuable information on NN interactions from lattice QCD calculations at and around the physical pion mass. However, the hadronic systems studied in LQCD live in periodic boxes instead of the free space. This requires extra steps to connect the LQCD results (without infinite-volume extrapolation) with the free-space observables we are interested in. For two-hadron systems, the L{\"u}scher formula~\cite{Luscher:1990ux} is often used to extract free-space scattering phase shifts from discrete eigenenergies at various box sizes. Its generalization to three-hadron systems is being studied intensively by multiple groups (see \eg, Ref.~\cite{Jackura:2019bmu}). 

For constraining NN interactions using LQCD results, we suggest an alternative approach to the three-body L{\"u}scher's method, by treating the LQCD simulations as computer experiments. The chiral potentials can be solved in the same periodic boxes as used in the LQCD's calculations in order to construct the mapping between the low-energy couplings in the nuclear interactions and the discrete energy levels. Again, the EC-driven emulators---for few-nucleon systems in periodic boxes here---will be the key to fitting those couplings to the eigenenergies, as they enable needed rapid solutions of the corresponding Faddeev equations.  The computer-experiment strategy was used by Barnea~\etal~\cite{Eliyahu:2019nkz} to analyze the LQCD's results (at an unphysical pion mass) in few-nucleon systems, as discussed in Nir Barnea's talk at this INT program. However, instead of using an emulator, they employed the so-called Stochastic Variational Method to speed up solving the few-body Schr{\"o}dinger equation. As pointed out during this INT program, a method different from  the generalized L{\"u}scher method is desirable. It will be interesting to compare different methods of extracting constraints from LQCD on NN interactions in the future.

In the area of nuclear calculations, Xilin Zhang and collaborators~\cite{Zhang:2019cai,Zhang:2020rhz} have been developing analysis tools to extract two-cluster scattering phase shifts from the system's discrete eigenenergies in harmonic potential traps computed using nuclear many-body methods. The goal is to take advantage of the progress in many-body calculations for medium-mass nuclei to enable \abinitio calculations of scattering and reactions in the same mass region. 

For two-cluster systems in external potential traps, the so-called BERW formula akin to the L{\"u}scher formula in LQCD were studied for some time and lately improved (see Ref.~\cite{Zhang:2019cai} and references therein). For three-cluster systems, both three-hadron-L{\"u}scher-method and EC-emulator approach are worth pursuing here. Therefore, the EC-driven emulators for few particle systems in external potential traps need to be developed as well. 

From a broader perspective, few-body systems in periodic boxes and harmonic traps have discrete energy levels as self-bound systems have, while they also have sub-systems' scattering information encoded in the energy levels thanks to the  L{\"u}scher and BERW formula. Developing EC-driven emulators for these systems will reveal the elusive connection between the bound~\cite{Konig:2019adq, Ekstrom:2019lss} and scattering state emulators, paving the way for a unified understanding of the two as well as the associated variational principles.   


{\bf Acknowledgements:}
We thank R.~J. Furnstahl, A.~J. Garcia, P. Giuliani, A.~E. Lovell, J.~A. Melendez, F.~M. Nunes, and M. Quinonez for sharing their invaluable insights with us. 
We are also grateful to the organizers of the (virtual) INT program ``Nuclear Forces for Precision Nuclear Physics'' (INT--21--1b) for creating a stimulating environment to discuss eigenvector continuation and variational principles.
This material is based upon work supported by the U.S. Department of Energy, Office of Science, Office of Nuclear Physics, under the FRIB Theory Alliance award DE-SC0013617. 
Xilin Zhang was also supported by the National Science Foundation under Grant No.~PHY--1913069 and by the NUCLEI SciDAC Collaboration under Department of Energy MSU subcontract RC107839-OSU.
\sloppy
\printbibliography[heading=subbibliography]
\end{refsection}
\addtocontents{toc}{\protect\setcounter{tocdepth}{1}}
\begin{refsection}
\chapter{The relevance of the unitarity limit and the size of many-body forces in Chiral Effective Field Theory}
\chapterauthor[1]{Harald W. Grie{\ss}hammer}
\chapterauthor[2]{Sebastian K{\"o}nig}
\chapterauthor[3]{Daniel Phillips}
\chapterauthor[4,5]{Ubirajara van Kolck}

\begin{affils}
\chapteraffil[1]{Institute for Nuclear Studies, Department of Physics,
The George Washington University, Washington DC 20052, USA}
\chapteraffil[2]{Department of Physics, North Carolina State University, Raleigh, NC 27695, USA}
\chapteraffil[3]{Department of Physics and Astronomy and Institute of Nuclear and Particle Physics, Ohio University, Athens, OH 45701, USA}
\chapteraffil[4]{Department of Physics, University of Arizona,
Tucson, AZ 85721, USA}
\chapteraffil[5]{Universit{\'e} Paris-Saclay, CNRS/IN2P3, IJCLab,
91405 Orsay, France}
\end{affils}

\newcommand{\mpi}{m_\pi}
\newcommand{\Mhi}{M_{\text{hi}}}
\newcommand{\OO}{\mathcal{O}}

\addtocontents{toc}{\protect\setcounter{tocdepth}{-2}}

We collect in this document some thoughts
which we believe are relevant for systematically incorporating the closeness of
nuclear physics to the unitarity limit into chiral effective field theory (\chieft).
Our discussion contains no quantitative insights beyond those already made in the literature. But the excellent talks and stimulating discussions at the recent INT program gave us an opportunity to take stock of the role of the
unitarity limit in the precision description of nuclei. We thank the organizers for their efforts to overcome the constraints of the online format and foster an interactive and collaborative atmosphere---and for the invitation to contribute this piece to the various ``Perspectives'' coming out of the program.

The first section focuses on many-body forces in \chieft and the interplay of the
chiral and unitarity limits in nuclear systems with $A \leq 4$; a second section discusses the particular (peculiar?) role of the ${}^1$S$_0$ channel; a final section then looks at how the
issues discussed in the previous two play out in nuclei with $A=6$ and beyond.

\section{Chiral limit or unitarity limit: which one is more relevant for nuclei with $A \leq 4$?}

Both the \pilesseft expansion around the unitarity limit and the \chieft expansion
around the chiral limit yield good descriptions of nuclei with $A\leq 4$ nucleons.
Pionless EFT and \chieft are distinctively different since they are constructed as expansions around entirely different limits: the former
embraces the large nucleon-nucleon (\NN) scattering lengths as an emergent low-momentum scale and 
treats the pion mass $\mpi$ as a high scale, while the latter treats $\mpi$ as a low
scale.  In light of this contrast, what is the connection between the two expansions?

\begin{itemize}

\item Recent
  work~\cite{Song:2016ale,Yang:2020pgi,Peng:2021xx,Koenig:2021int} has shown that a
  renormalizable, perturbative approach 
  to \chieft with reasonable convergence \emph{is} possible for light
  nuclei.
  Results following the EFT power counting that converge to the experimental values are
  obtained not only for binding energies, but also for charge radii, with the caveat that
  so far only the lowest two orders have been studied.

 \item The unitarity expansion in \pilesseft exhibits more rapid convergence for
  nuclei with $A \leq 4$ \cite{Konig:2016utl,Konig:2019xxk} 
  than \chieft, a success that stems largely from
  having a three-nucleon (\NNN) force at leading order (\LO)~\cite{Bedaque:1998kg}.
  This force can be used to put the \NNN bound state at the right energy,
  and few-body
  universality then guarantees that the four-body binding energy~\cite{Platter:2004zs}
  and the three-body radius~\cite{Platter:2005sj} will be approximately 
  reproduced. This makes for an excellent starting point for the description of $A \leq 4$ (bound) systems. 
  The presence of a four-nucleon (\NNNN) force at 
  next-to-leading order (\NLO)~\cite{Bazak:2018qnu}
  spoils the predictivity a bit.
  However, only a single input parameter is needed to fix this force,
  and fitting it to the \isotope[4]{He} binding energy, one can still predict the
  radius and other properties. 
  The results that this produces for few-body observables will
  be studied in future work.

\end{itemize}

\subsection{Connections}

\begin{itemize}

  \item Recent studies of correlations between few-body observables in \chieft show that
  $A=3,4$ binding energies and radii provide highly degenerate information  on the \NNN force~\cite{Wesolowski:2021cni}.
  This observation is consistent with light nuclei being within the lower-energy regime of
  \pilesseft 
  Based
  on this \emph{phenomenological} effect, it has been argued
  before~\cite{Kievsky:2016kzb} that \chieft should feature a (pure contact)
  \NNN force at \LO as well, but a clear justification for this
  conjecture is not provided by that work.  

 \item It is remarkable how close the Tjon lines come out to each other in
  \chieft and the unitarity expansion~\cite{Koenig:2021int}.
  Assuming convergence of both expansions to the physical point means that
  they ultimately have to intersect there, and we already know that
  the Tjon band from \pilesseft captures well all the points from
  phenomenological potentials.
  But is there anything interesting to be learned
  from the fact that the slope of the \LO chiral Tjon line (generated by residual cutoff dependence) agrees well with
  the relation between three- and four-body binding energies $B_4 = 4.610(1) B_3$  that
  prevails in the unitarity limit~\cite{Deltuva:2012ig}?

 \item It is instructive to consider the importance of many-body forces from the
  perspective of varying the resolution of the interaction.
  This can be done by Similarity Renormalization Group (SRG) transformations.
  It is well
  known that maintaining unitarity of these transformations induces higher-body
  forces, even if the interaction initially is given only at the two-body level.
  Moving from the regime of \chieft (typical momentum $Q
  \sim \mpi$) to \pilesseft ($Q \ll \mpi$), where pions are ``integrated
  out,'' corresponds to lowering the effective resolution.
  In light of this it
  makes sense to consider \pilesseft as the ultimate limit of a \chieft
  interaction transformed to very low resolution, and in this limit there is a
  \NNN contact force at \LO and a \NNNN force at \NLO.
  In the two-body sector, SRG-induced operator structures have been found to resemble
  simple contact terms~\cite{Anderson:2010aq,More:2017syr}, so it is not
  unreasonable to assume the dominant induced three-body forces have a similar
  form.
  Based on this picture, one might imagine that considering the low-resolution limit will inform, and potentially adjust, the \chieft
  power counting.
\end{itemize}

\subsection{Conjectures}
\begin{itemize}

\item What would be the best, most rigorous, arguments to determine the order 
  at which a contact \NNN force first appears in \chieft?
  We know such a force
  is not \emph{required} for RG invariance at \LO~\cite{Song:2016ale,Yang:2020pgi}, but is there a possible
  numerical signature that would indicate a departure from naive dimensional analysis?
  It could 
  be informative to study how the coupling constant of such a force runs
  with the cutoff when it is included at \LO and fixed to reproduce exactly the triton
  binding energy---as done in \pilesseft~\cite{Bedaque:1998kg}.
  For recent efforts in this direction in the context of ${}^4$He trimers and the
  long-range van der Waal's force see Ref.~\cite{Odell:2021ryo}.
  Alternatively, analytic results for the \LO three-body wave function at short
  distances would enable an analysis of the anomalous dimensions induced by the strong \LO
  interactions.
  The analog of the analysis in Ref.~\cite{PavonValderrama:2014zeq} for
  electroweak operators in the two-body sector would then reveal the extent to which those
  \LO interactions alter the naive dimensional analysis (NDA) result for the three-body contact.

 \item For the intermediate-range \NNN force in Deltaless \chieft ($\sim c_D$), one
  can consider matrix elements between correlated ${}^1$S$_0$ \NN states
  plus a third spectator nucleon.
  The large scattering length in this channel
  means one should promote this force by one order compared to
  its NDA order~\cite{Kaplan:1998tg,vanKolck:1998bw,Birse:1998dk,PavonValderrama:2014zeq}.
  This moves
  the $c_D$ term from $\OO(Q^3/\Mhi^3)$ to $\OO(Q^2/\Mhi^2)$~\footnote{These orders are computed counting powers of $4\pi$ as Weinberg did~\cite{Weinberg:1991um}, cf.\ the final bullet in this section.}, where $\Mhi\sim m_\rho$ (the rho mass) is
  the assumed breakdown scale of \chieft.
  RG invariance would then seem to require that the pure contact \NNN
  force ($\sim c_E$) is promoted to the same degree, since matrix elements of $c_D$
  alone will be regulator dependent. 

 \item If \NNN forces appear earlier than Weinberg's $\OO(Q^3/\Mhi^3)$, then at what
 order in \chieft do \NNNN forces enter?
 The argument of the previous bullet also implies a promotion of the particular piece of
 the \NNNN force that is proportional to $c_D^2$.
 In this context, it is interesting to note that calculations of nuclear matter that employ
 \NN and \NNN forces adjusted to reproduce \NN data and the triton binding energy
 still generate at least a portion of the Coester line~\cite{Drischler:2017wtt}. 

\item Adjusting the counting of factors of $4\pi$ associated with
  (nonrelativistic) reducible loops, as suggested by Friar~\cite{Friar:1996zw} and inferred
  from \pilesseft~\cite{vanKolck:2020llt}, gives a promotion over Weinberg's power counting by one order for the \NNN force and by two orders for the \NNNN force. This is in addition to any other enhancement.
\end{itemize}

\section{The perpetually vexatious ${}^1$S$_0$ channel}

What features of the ${}^1$S$_0$ channel matter for finite nuclei?
For $A \leq 4$, it seems to be sufficient to formulate a LO that has the ${}^1$S$_0$
amplitude close to the unitarity limit at low energies.
This is one of the ingredients of the success of \pilesseft, and the unitarity limit
(and nothing else) delivers the promising \LO results in Ref.~\cite{Konig:2016utl}. 

But as the \NN energy increases away from threshold, the \LO ${}^1$S$_0$ phase shift
predictions in such an approach rapidly deviate from data.
Attempts to formulate an EFT that reproduces higher-energy features of the phase shift in
this channel at LO have a more-than-two-decade history now~\cite{Lutz:1999yr}.
The idea is to change the \chieft power counting in order to generate more energy
dependence in the ${}^1$S$_0$ phase shift at lower orders.
However, only recently have the implications of such an approach for finite nuclei been elucidated in a systematic way. 

\begin{itemize}
\item One can consider a promotion of the short-distance operator that
contributes to the \NN effective range~\cite{Beane:2001bc,Long:2012ve}. Unfortunately that can only be done in an RG-invariant fashion with the energy dependence stemming from a dibaryon field.

\item
The ${}^1$S$_0$ phase shift goes through zero at at a
momentum of $\approx 340$ MeV. If the zero is not present at \LO, higher orders must overcome \LO at larger momenta, which implies poor convergence in the vicinity of this zero. The analysis of Ref.~\cite{SanchezSanchez:2017tws} formulates an EFT that at \LO respects
the unitarity limit and also reproduces the
zero.
With the (potentially significant) caveat that the interaction is transformed from an
energy-dependent form to a momentum-dependent form, leaving only on-shell
results invariant, Ref.~\cite{SanchezSanchez:2020kbx} reports that including this
higher-energy feature of the ${}^1$S$_0$ phase shift in the Weinberg \LO EFT description of \NN
physics improves the description of finite nuclei.

\item Alternatively one can consider a promotion of 
correlated two-pion exchange contributions~\cite{Ordonez:1995rz,Kaiser:1998wa} to lower
orders. Promising results for finite nuclei and nuclear matter have recently been obtained when these diagrams with $N\Delta$ intermediate states are included at $\OO(Q^2/\Mhi^2)$~\cite{Piarulli:2016vel,Jiang:2020the}.
Promotion of ``sub-leading" Deltaless two-pion-exchange contributions proportional to $c_1$, $c_3$, and $c_4$ is supported by the large-$N_c$ limit, where two-pion-exchange mechanisms involving the $\Delta(1232)$ are the $SU(4)$ partners of the iterated one-pion-exchange diagrams that appear at \LO in chiral EFT in the ${}^3$S$_1$ channel.
But large-$N_c$ presumably demands that such mechanisms be included at \LO if the
EFT is to respect both chiral symmetry and the large-$N_c$ limit.
If two-pion-exchange graphs with $\Delta(1232)$ intermediate states are \LO in
the \NN system then would that not mean the Fujita-Miyazawa 3NF is leading order in the
three-body system?
And is there really enough scale separation between the momentum
associated with $N\Delta$ intermediate states and, say, $m_\rho$ to justify treating the
former as a low-energy excitation and the latter as a high-energy one?
  \end{itemize}

\section{What about ``real'' nuclei?}

For $A\geq 5$, both \pilesseft~\cite{Stetcu:2006ey,Contessi:2017rww,Bansal:2017pwn,Dawkins:2019vcr,Gattobigio:2019omi}
and \chieft~\cite{Yang:2020pgi} tend to yield unstable nuclei at \LO.
What is the significance of this additional similarity between the two EFTs?

\begin{itemize}
 \item While instability is appropriate for $A=5,8$, it raises the question, how can stability
 arise in other nuclei, for example $A=16$?
 Can it be obtained in (distorted-wave) perturbation theory?

 \item In Ref.~\cite{Bansal:2017pwn}, \NLO \pilesseft corrections were iterated.
 This effectively includes the interaction range at \LO, but destroys RG
 invariance~\cite{Phillips:1996ae}.
 Is this the only practical calculational scheme to produce $p$-shell nuclear stability in \pilesseft? 
    
 \item In \chieft the interaction range is included at \LO via one-pion exchange,
 and still it is not sufficient in a renormalized approach to ensure
 stability, at least at accessible cutoff values~\cite{Yang:2020pgi}.
 Should one trade the range by other formally higher-order interactions, such as the
\NNN force in \chieft or the \NNNN force in Pionless EFT?
    
 \item Jerry Yang has suggested in his talk during the program~\cite{Yang:2021int} that
 few-body forces
 might be enhanced by combinatorial factors of $A$. This might be sufficient to promote
 few-body forces to \LO in \chieft. Preliminary results indicate that this is sufficient
 for $A=16$ stability thanks to the \NNN force. But then \NNNN forces might
 become important for heavier nuclei, complicating the description of nuclei significantly.
    
 \item With the original Weinberg prescription~\cite{Weinberg:1991um}, there is no
 saturation in symmetric nuclear matter until one reaches the order in the expansion where \NNN forces appear~\cite{Drischler:2017wtt,Sammarruca:2018bqh}. Renormalized \chieft leads, without
 \NNN forces in LO, to saturation with significant
 underbinding~\cite{Machleidt:2009bh}. Can a nuclear EFT that at \LO (and \NLO in the Weinberg prescription) does not produce realistic nuclear matter be a convergent description of nuclei? Ref.~\cite{Drischler:2020yad} argues that the \LO and \NLO uncertainties at the canonical saturation density of $n_0=0.16~{\rm fm}^{-3}$ are too large to say whether or where saturation occurs. However, results computed at \NNLO (and \NNNLO) (within the Weinberg prescription) fall within the \LO and \NLO error bands and show saturation occurring at a density and binding energy per nucleon consistent with the ``empirical" value. In contrast, adding a \NNN force to Weinberg's LO does yield nuclear
 saturation~\cite{Kievsky:2018xsl}.
 What do these results tell us about the ordering of few-body forces and the organization of nuclear EFT? 
\end{itemize}

\printbibliography[heading=subbibliography]
\end{refsection}

\addtocontents{toc}{\protect\setcounter{tocdepth}{1}}
\begin{refsection}
\chapter{Nuclear forces in a manifestly Lorentz-invariant formulation of chiral effective field theory}
\chapterauthor[1]{Xiu-Lei Ren}
\chapterauthor[2]{Evgeny Epelbaum}
\chapterauthor[3]{Jambul Gegelia}
\begin{affils}
\chapteraffil[1]{Institut f\"ur Kernphysik \&  Cluster of Excellence PRISMA$^+$, Johannes 
Gutenberg-Universit\"at  Mainz,  D-55128 Mainz, Germany}
\chapteraffil[2]{Ruhr-Universit\"at Bochum, Fakult\"at f\"ur Physik und Astronomie, Institut f\"ur Theoretische Physik II, D-44780 Bochum, Germany}
\chapteraffil[3]{High Energy Physics Institute, Tbilisi State  University,  0186 Tbilisi, Georgia}
\end{affils}

\addtocontents{toc}{\protect\setcounter{tocdepth}{-2}}

We outline the advantages and disadvantages of manifestly Lorentz-invariant formulation of chiral effective field theory (\chieft)for the nuclear forces compared to the non-relativistic formalism. 

\section{Introductory remarks \label{Sec:intro}}

Chiral perturbation theory is an effective field theory (EFT) of the
strong interaction applicable at low energies. It shares all
symmetries of the underlying fundamental theory---the quantum
chromodynamics (QCD).  
While Lorentz-invariance is a cornerstone of quantum field
theories in general, a systematic non-relativistic expansion can be
made for physical quantities if 
particle velocities are much smaller than the speed of the light.  
This expansion can also be done at the Lagrangian level 
leading to a non-relativistic EFT. The two approaches yield exactly
the same non-relativistic expansions for physical quantities, provided
that 
one takes special care to address the non-commutativity of the
expansion of the effective Lagrangian and calculations of quantum
corrections. It is important to keep in mind  that the ultraviolet (UV) behavior of
loop integrals of quantum corrections is completely different in the two
approaches. In the few-nucleon sector, one deals with effective potentials,
whose low-energy behavior is systematically calculable order-by order
in \chieft. On the other hand, the (infrared) power counting has no
status in the UV region. As we do
not know the short-range behavior of few-body potentials, one
might argue that all UV extensions of the effective potential are
equally good/bad.  While the effective potentials are not even
uniquely defined, we do know the physical spectrum of QCD
(assuming that it indeed describes the nature).  
Considering, e.g., nucleon-nucleon (\NN) scattering, the actual short-distance behavior of
the nuclear force can certainly not be singular since no corresponding
deeply bound states are observed in nature. At the conceptual level,
all complications caused by singular EFT interactions result
from a naive extension of non-relativistic potentials from large
distances to short ones, where the infrared ordering of various
contributions is invalid.  
Therefore, approximating a non-singular potential with a singular leading order
(\LO) contribution, supplemented by a finite number of contact interactions, 
is only appropriate if the cutoff is kept of the order of the hard
scale of the problem. On the other hand, if one employs a non-singular
extension of the one-pion exchange potential from long to short
distances, then the short-range details
of the \LO approximation indeed do not matter after removing the
regulator, since a finite number of counter terms are required to
renormalize the amplitude. 
The formalism based on the manifestly Lorentz-invariant 
formulation is well suited for this purpose. While being
equivalent to the non-relativistic formulation in the infrared region,
performing a resummation of  a certain class
   of $1/m$-corrections in a way consistent with the underlying Lorentz symmetry leads to
the effective potentials that admit a better UV behavior.   

\section{Chiral nuclear forces from the Lorentz-invariant Lagrangian using time-ordered perturbation theory
\label{Sec:2}} 

These ideas have been taken up in Ref.~\cite{Epelbaum:2012ua} to
formulate a \emph{renormalizable} framework for \NN scattering
based on the manifestly Lorentz-invariant effective Lagrangian.  
In the resulting modified Weinberg approach, the \LO amplitude is
obtained by solving the Kadyshevsky equation while higher-order
corrections are treated perturbatively.  
Symmetry-preserving regularization within this formalism has been considered in Ref.~\cite{Behrendt:2016nql}.
A fully Lorentz-covariant form of the effective potential 
based on a new power counting has been suggested in
Refs.~\cite{Ren:2016jna,Li:2016mln}. A systematic approach relying on
the Lorentz-invariant Lagrangian and time-ordered perturbation theory
has been further developed in Ref.~\cite{Baru:2019ndr},  where the
effective potential and the scattering equation (Kadyshevsky equation)
are obtained within the same framework. 
Restricting the non-perturbative treatment to the (non-singular) \LO
potential and assuming the validity of perturbation theory for
higher-order interactions, one can systematically remove \emph{all}
divergences from the amplitude and, therefore, employ arbitrarily large 
values of the cutoff. Alternatively, the full effective potential can
be treated non-perturbatively. The milder UV behavior then offers a larger
flexibility regarding admissible cutoff values, which generally need
to be kept of the order of the breakdown scale. 
Therefore, we expect that this approach should lead to better description of systems with larger numbers of nucleons.
However, one should
keep in mind that the derivation of corrections to the interaction beyond \LO is
computationally more demanding as compared to its non-relativistic
counterpart. Work is in progress towards extending the analysis of \NN
scattering within the modified Weinberg approach beyond
\LO~\cite{Ren:2021progress}. Notice further that the actual solution of the  Kadyshevsky
integral equation is facilitated by the fact that it can be rewritten
in the form of the  standard Lippmann-Schwinger equation for a modified
 potential.
Last but not least, the relativistic formulation of \chieft can
also be merged with the Dirac-Brueckner-Hartree-Fock theory. It would
further be interesting to perform \abinitio studies of finite nuclei
and nuclear matter using the relativistic version of chiral nuclear
interactions. 

{\bf Acknowledgements:}
This work was supported in part by BMBF (Grant No. 05P18PCFP1), by DFG and NSFC through funds provided to the
Sino-German CRC 110 ``Symmetries and the Emergence of Structure in QCD" (NSFC
Grant No.~12070131001, Project-ID 196253076 - TRR 110), by Collaborative Research Center ``The Low-Energy
Frontier of the Standard Model'' (DFG, Project No. 204404729 - SFB 1044), by the Cluster of
Excellence ``Precision Physics, Fundamental Interactions, and Structure of Matter'' (PRISMA$^+$, EXC 2118/1)
within the German Excellence Strategy (Project ID 39083149), by the  Georgian Shota Rustaveli National Science Foundation (Grant No. FR17-354).

\printbibliography[heading=subbibliography]
\end{refsection}

\newcommand{\ve}{\varepsilon}
\newcommand{\no}{\nonumber}

\addtocontents{toc}{\protect\setcounter{tocdepth}{1}}
\begin{refsection}
\chapter{Nuclear forces for precision nuclear physics: Some thoughts on the status, controversies and challenges}
\chapterauthor[1]{Evgeny Epelbaum}
\chapterauthor[1,2]{Ashot~Gasparyan}
\chapterauthor[1,3]{Jambul Gegelia}
\chapterauthor[1]{Hermann Krebs}
\begin{affils}
\chapteraffil[1]{Ruhr-Universit\"at
  Bochum, Fakult\"at f\"ur Physik und Astronomie, Institut f\"ur
  Theoretische Physik II, D-44780 Bochum, Germany}
  \chapteraffil[2]{NRC ``Kurchatov Institute'' - ITEP, B. Cheremushkinskaya 25, 117218 Moscow, Russia}
  \chapteraffil[3]{High Energy Physics Institute, Tbilisi State  University,  0186 Tbilisi, Georgia}
\end{affils}

\addtocontents{toc}{\protect\setcounter{tocdepth}{-2}}

We outline the status of chiral effective field theory (\chieft) for nuclear systems, summarize our understanding of renormalization group invariance in this context and discuss some of the most pressing challenges and opportunities in the field.  

\section{Introductory remarks and disclaimer} 

This paper is a summary of the contributions and opinions of the Bochum
participants in the INT program on Nuclear Forces for Precision Nuclear
Physics, held at the INT, Seattle, April 19 - May 7, 2021, regarding
some of the topics addressed during this meeting.
It is not intended to provide a review of the field, and we also
made no attempt to be exhaustive in the references. A more 
complete and detailed discussion of (most of) the considered topics
can be found in the recent review article by some of us
\cite{Epelbaum:2019kcf} and in references
therein, see also the earlier review \cite{Epelbaum:2008ga}.
\begin{figure}[b!] 
\includegraphics[width=\textwidth]{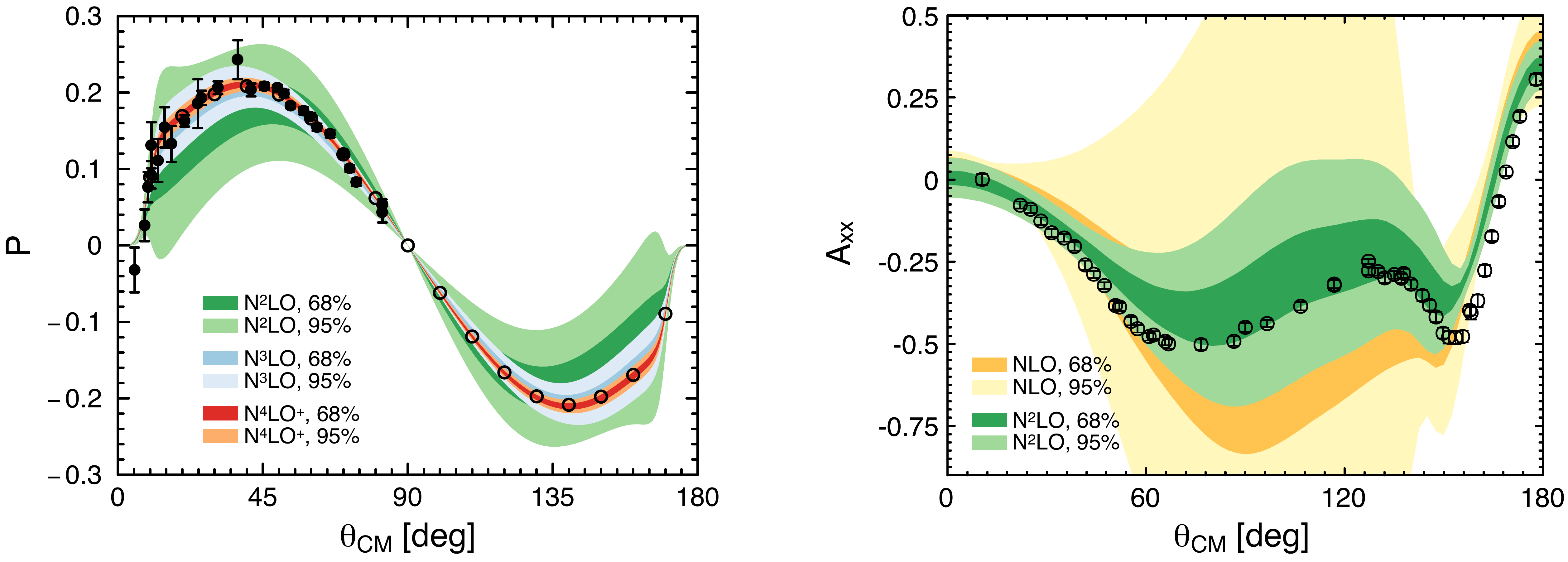}
\caption{\chieft predictions for the analyzing power $P$ in
  proton-proton scattering at $E_{\rm lab} = 142$~MeV (left panel) and
  for tensor analyzing power $A_{xx}$ in nucleon-deuteron elastic
  scattering at $E_{\rm N} = 135$~MeV. The light- (dark-) shaded
  bands depict the $68\%$ ($95\%$) degree-of-belief truncation errors
  at the corresponding order. Solid and open circles with error bars
  are experimental data from Refs.~\cite{Taylor:1960} and \cite{Sekiguchi:2002sf} respectively. Open
  circles without error bars show the predictions of the Nijmegen
  partial-wave analysis \cite{Stoks:1993tb}. For more details see Ref.~\cite{Epelbaum:2019kcf}.} 
\label{Fig:1}
\end{figure}

Our paper is organized as follows. In section \ref{Sec:2_epel}, we discuss
the status of \chieft for nuclear
systems. We limit ourselves to its standard, finite-cutoff formulation
based on the Weinberg approach \cite{Weinberg:1990rz,Weinberg:1991um}
and summarize
the most pressing open issues 
in the field. Section \ref{Sec:3} is devoted to the
ongoing debate concerning a proper renormalization of nuclear
chiral EFT. We provide a brief account of the renormalization program
in the meson and single-baryon sectors of chiral perturbation theory
(\chipt) before critically addressing the so-called renormalization group (RG)
invariant approach in the few-nucleon sector. Formal
aspects of renormalization in the finite-cutoff formulation of \chieft are
discussed in section \ref{Sec:4}, while some concluding remarks are
made in section  \ref{Sec:5}.

\section{Nuclear interactions from \chieft: Current status and open
questions} \label{Sec:2_epel}

\subsection{The nucleon-nucleon sector}
Starting from the pioneering work by
Weinberg in the early
nineties~\cite{Weinberg:1990rz,Weinberg:1991um},
see also \cite{Ordonez:1993tn,Ordonez:1995rz},
the nucleon-nucleon (\NN) interactions have been worked out
up to fifth order (\NNNNLO) in
\chieft \cite{Reinert:2017usi,Entem:2017gor}, see~\cite{Epelbaum:2019kcf} for a recent review and
references therein. At the highest EFT order, the interactions from
Ref.~\cite{Reinert:2017usi} lead to 
an excellent description of \NN scattering data below pion production
threshold with $\sim 40\%$ less adjustable parameters as
compared to high-precision phenomenological potentials. They also show
a clear evidence of the (parameter-free) chiral two-pion exchange
potential. In Ref.~\cite{Reinert:2020mcu},
a full-fledged partial-wave analysis of \NN data up to the pion
production threshold has been performed in the framework of 
\chieft (including a selection of mutually consistent data
and a complete treatment of isospin-breaking interactions),
thereby achieving a statistically perfect description of experimental data,
see the left panel of Fig.~\ref{Fig:1}
as a representative example. In this sense, one may claim ``mission
accomplished'' in the \NN sector of \chieft.  

\subsection{The three-nucleon force challenge}

The situation with three-nucleon (\NNN) forces is more intricate.
Although they have been worked out completely to
fourth chiral order (\NNNLO) a long time ago
\cite{Bernard:2007sp,Bernard:2011zr}, and partially even 
to \NNNNLO \cite{Girlanda:2011fh,Krebs:2012yv,Krebs:2013kha}, their
implementation in few-/many-body 
calculations is more demanding. The issue is rooted in the
regularization of the \NNN force.  Loop contributions to the \NNN force starting
from \NNNLO have been derived using dimensional
regularization (DR). On the other hand,
the $A$-body Schr\"odinger equation is regularized  using a
cutoff. Differently to the \NN sector, this
mismatch in the regularization can \emph{not} be
compensated by counterterms from the effective chiral
Lagrangian \cite{Epelbaum:2019kcf}. Curing this conceptual problem
will require a
re-derivation of the three- and four-nucleon (\NNNN) forces starting from
\NNNLO
using a cutoff regulator which (i) maintains the chiral
symmetry and (ii) is consistent with the one employed in the \NN sector.  
The higher derivative regularization method proposed in
Ref.~\cite{Slavnov:1971aw}  offers one possible approach to
tackle this challenge by introducing the cutoff at the level of the
effective Lagrangian.
Work along this line is in progress.  \\
A precise description of nucleon-deuteron scattering data remains a
major unsolved challenge in nuclear physics as reflected by very large values of the
$\chi^2/{\rm datum}$ in the \NNN sector \cite{Gloeckle:1995jg,KalantarNayestanaki:2011wz} as compared to
$\chi^2/{\rm datum} \sim 1$ for the reproduction of \NN scattering data
\cite{Reinert:2017usi,Reinert:2020mcu}. \chieft
predictions up to  \NNLO generally agree with the data, see the
right panel of Fig.~\ref{Fig:1} for an example, but the accuracy at
this order is fairly low. Based on the experience in
the \NN sector, the solution of the \NNN force challenge will likely require pushing
the EFT expansion of the \NNN force (at least) to \NNNNLO.  Apart from the
regularization issue mentioned above, one will then face the
computational challenge associated with the determination of
low-energy constants (LECs) entering the \NNN force from $A \ge 3$-nucleon
data, see Ref.~\cite{Girlanda:2018xrw} for pioneering steps along this line. The eigenvector continuation technique
\cite{Frame:2017fah,Ekstrom:2019lss,Wesolowski:2021cni} may be the key
technology to make such an
analysis feasible in the near future.

\subsection{Nuclear currents}

Nuclear vector, axial-vector, pseudoscalar
and scalar currents have also been extensively studied in \chieft, see Ref.~\cite{Krebs:2020pii} and references therein. 
The vector and axial-vector currents have been worked
out using two different techniques, namely the method of a unitary
transformation by the Bochum-Bonn group
\cite{Kolling:2009iq,Kolling:2011mt,Krebs:2016rqz,Krebs:2019aka,Krebs:2020plh}
and time-ordered perturbation 
theory by the JLab-Pisa group
\cite{Pastore:2008ui,Pastore:2009is,Pastore:2011ip,Baroni:2015uza,Baroni:2016xll}.
The results of these calculations
disagree with each other, with the differences being most pronounced
for the axial-vector currents. As an attempt to shed light on this
issue, the Bochum-Bonn group has tried to reproduce the results of the JLab-Pisa group
for the axial-vector currents using their 
method~\cite{Krebs:2020rms}, thereby confirming their own
expressions. To exclude the possibility that
the approach of the JLab-Pisa group has been misinterpreted,
a comparison could be carried out 
at the level of Hilbert-space operators \cite{Krebs:2020rms}. \\
Independently of this discrepancy, the implementation of the nuclear currents
starting from  \NNNLO is also affected by the already mentioned
issue with mixing up two different regularization methods \cite{Krebs:2020pii}.  
Even though no results for the 
current operators using the higher-derivative regularization are available
yet, a high-accuracy
calculation of the deuteron charge and quadrupole form factors
has been performed recently at \NNNNLO
\cite{Filin:2019eoe,Filin:2020tcs}. This was possible since
the loop corrections to the two-body charge density, whose consistent
regularization is not yet available, do not contribute to the deuteron
form factors thanks to the isospin selection rules. 

\subsection{On the role of the $\Delta$(1232) resonance}

Nuclear interactions discussed so far have been derived using
pions and nucleons as the only explicit degrees of freedom. 
Given the low excitation energy of the $\Delta(1232)$ resonance and
its strong coupling to the $\pi$N system, one may expect its explicit
inclusion in the effective Lagrangian to yield 
a more efficient EFT framework. Indeed, clear
evidence of the improved convergence of the $\Delta$-full formulation of
\chieft for pion-nucleon scattering was found in recent studies 
\cite{Siemens:2016hdi,Yao:2016vbz,Siemens:2016jwj}. On the
other hand, the available results for the \NN system in the $\Delta$-less
framework already show a generally good convergence pattern, with the
estimated breakdown scale of the order of $\Lambda_b \sim 600 - 650$~MeV
\cite{Epelbaum:2014efa,Furnstahl:2015rha,Epelbaum:2019wvf}. Moreover,
the LECs accompanying the \NN contact interactions come
out of a natural size \cite{Reinert:2017usi,Epelbaum:2019kcf} with no signs of enhancement from the
$m_\Delta - m_N \sim 2 M_\pi$ scale. This may indicate that the
longest-range contributions of the $\Delta$-resonance to the \NN force
are efficiently mimicked via the saturation of $\pi$N
LECs.\footnote{Naively, integrating out $\Delta$-resonance
  contributions to the two-pion exchange potential is expected to
  induce large contributions to \NN contact interactions governed by
  the scale $m_\Delta - m_N$. However, a close inspection of the
  corresponding analytical expressions in Refs.~\cite{Kaiser:1998wa,Krebs:2007rh} suggests that
  the scale is actually given by $2(m_\Delta - m_N)$, which is
  numerically already close to $\Lambda_b$.} It is thus
not \textit{a priori} clear, that the explicit inclusion of the $\Delta$-resonance
would result in a significantly larger value of
$\Lambda_b$, leading to a better convergence of \chieft in the few-nucleon sector. \\
$\Delta$-contributions to the nuclear forces have so far been
worked out to third order (\NNLO) \cite{Kaiser:1998wa,Krebs:2007rh},
and the results up to this order do seem to indicate a superior 
performance of the $\Delta$-full approach \cite{Ekstrom:2017koy}, see also
Refs.~\cite{Piarulli:2014bda,Piarulli:2016vel} and references therein. Clearly, to 
assess the role of the $\Delta$-resonance in quantitative terms, the calculations within the
$\Delta$-full approach will have to be pushed beyond \NNLO.
As a first step along this line, some of us
have recently worked out the \NNLO contributions of the $\Delta(1232)$
to the two-pion exchange \NNN force topology using DR \cite{Krebs:2018jkc}.
Notice, however, that the already mentioned issue with the inconsistent regularizations is also
relevant for the $\Delta$-full formulation of \chieft, and all loop
contributions will need to be
(re-) derived using e.g.~the higher derivative regularization. 

\section{Renormalization group invariance and power counting} \label{Sec:3}

\subsection{Renormalization in chiral perturbation theory}

In the strictly perturbative domain of chiral perturbation theory
(\chipt) comprising its mesonic and the heavy-baryon formulations, a
finite number of counterterms is needed to remove divergences at
every order in the chiral expansion. When using DR,
all the required counterterms are generated by bare LECs from the effective
Lagrangian of that corresponding order. Consequently, renormalized
amplitudes calculated up to any finite order are independent of the DR
scale $\mu$.\footnote{The exact independence on the choice of
  renormalization conditions is a (nice) artifact of DR. It does not
  hold when using more general regularization schemes that keep track
  of power-law divergences.} 
  On the other hand, the inclusion of an infinite number of counterterms is required (or 
  implicitly assumed) e.g.~in the infrared regularized formulation of
  manifestly Lorentz-invariant baryon \chipt \cite{Becher:1999he} that
  allows one to non-perturbatively resum $1/m$-corrections within the
  heavy-baryon approach. This leads to a residual dependence of
  renormalized amplitudes on the renomalization scale.  
 This feature is, however, perfectly acceptable from the EFT point of
 view since scale-dependent terms are of a higher chiral order, both formally and numerically.  
     
\subsection{Residual renormalization-scheme dependence in \pilesseft}   

In the few-nucleon sector of \chieft, certain types of diagrams
must be resummed non-perturbatively to accommodate for the appearance of shallow bound and
virtual states. Residual renormalization scheme dependence of the
calculated observables does not pose a conceptual problem in this case
either. In particular, resonant P-wave
systems have been studied recently in \pilesseft without auxiliary dimer fields
\cite{Epelbaum:2021sns}. This analytically solvable example
provides an explicit 
demonstration that
    the exact RG invariance at order $\mathcal{O} (p^n)$ (i.e.,
    $\partial T^{(n)}/\partial \mu_i = 0$ for $\forall \mu_i$)
    is \emph{not} necessary to claim consistent EFT. Rather, it is sufficient
    to have   $\partial T^{(n)}/\partial \mu_i = \mathcal{O} (p^{n+1})$.

\subsection{\chieft for nuclear systems and the cutoff choice}

    In contrast to the previously considered cases, 
      renormalization is carried out \emph{implicitly} in the
      non-perturbative domain of \chieft and  in
      \pilesseft for systems with more than two nucleons. 
      This is achieved by numerically
      expressing bare LECs $C_i (\Lambda )$ in terms of
      observables, in other words, by fitting them to experimental data. \\
      Removing the UV divergences in each order of the loop expansion
      of the scattering amplitude $T(\Lambda ) =
      \sum_{n=0}^{\infty} \hbar^n T_n (\Lambda )$ requires the
      inclusion of an infinite
      number of counterterms in the case of the one-pion exchange (OPE) potential (see, e.g.,
      Ref.~\cite{Savage:1998vh}), which is usually not possible in
      practical calculations.  
      Notice that the operations of summation in $T(\Lambda ) = \sum_{n} \hbar^n
      T_n$
      and $\lim_{\Lambda \to \infty} T(\Lambda )$ do not
      commute unless \emph{all}  counterterms needed to render all
      terms in the series finite are taken into account.
     Without including all the necessary counterterms, the resulting partially
      renormalized \cite{Epelbaum:2018zli} amplitude $T(\Lambda )$
      for $\Lambda \gg \Lambda_b$ is actually determined by the ambiguous behaviour of $V
      (r)$ at $r \ll \Lambda_b^{-1}$, which is outside of the EFT validity range.  
      Within such treatment, we see no \textit{a priori} reason for
      $T(\Lambda )\big|_{\Lambda \gg \Lambda_b}$ to represent a
      meaningful result (in the EFT sense). Under such circumstances,
      the cutoff $\Lambda$ should be kept of the order of the expected
      breakdown scale $\Lambda_{b}$ \cite{Lepage:1997cs,Gegelia:1998iu,Epelbaum:2018zli}. We also see no immediate relation
      between $T(\Lambda )\big|_{\Lambda \gg
        \Lambda_b}$ and the power counting for \emph{renormalized} short-range
      interactions. 

    
\subsection{The RG-invariant formulation of \chieft}

    As stated in Ref.~\cite{Hammer:2019poc}, the approximate RG invariance
      of the truncated amplitude $T^{(\nu)} (p, \Lambda)$ requires the condition 
      \begin{equation}
      \label{RGinv}
\frac{\Lambda}{T^{(\nu)} (p, \Lambda)} \frac{d T^{(\nu)} (p,
  \Lambda)}{d \Lambda} = \mathcal{O} \left( \frac{p^{\nu +
      1}}{\Lambda_b^\nu \Lambda} \right)
      \end{equation}
      to be satisfied. 
      As argued in that paper, in the absence of
      analytical results, \emph{``varying the regulator parameter widely
      above the breakdown scale is usually the only tool available to
      check RG invariance''}. Moreover, RG invariance of the
      truncated amplitude is often identified with the requirement that
      \emph{``the result converges with respect to $\Lambda$, i.e., the
      observable can only depend on negative power of $\Lambda$ after
      renormalization''} \cite{Yang:2019hkn}. The final result for the
      calculated observables is then understood as the corresponding
      $\Lambda \to \infty$ limits, see e.g. Refs.~\cite{Song:2016ale,deVries:2020loy}.

\subsection{A practical implementation of the RG-invariant approach: Lessons from a
   toy-model example}
   
While we agree with Eq.~(\ref{RGinv}), we object to the
suggested method of its numerical verification by varying $\Lambda$ widely
above $\Lambda_b$. To demonstrate possible issues  
with taking $\Lambda$ very large without subtracting all divergences,
consider a toy-model example of the scattering amplitude of two
heavy particles in $4+1$ space-time
dimensions\footnote{The unrealistic choice of the toy model is
  dictated by our wish to have an analytically solvable
  example. The realistic case of \NN scattering by the OPE potential in
chiral EFT is conceptually similar to the considered example but
requires numerical treatment.} 
\begin{equation}
T\left( \vec p,\vec q \, \right)=V\left( \vec p,\vec q\, \right)+m \int
\frac{d^4 k }{(2\pi)^4}\,V( \vec p,\vec
k)\,\frac{1}{m E -k^2+i\,0^+}\,T( \vec k,\vec q \, )\,,
\nonumber
\end{equation}
with the effective potential given by 
\begin{equation}
 V(\vec p,\vec q \, )\; =\; \frac{\alpha}{\big[ \left(\vec p-\vec
q \, \right)^2+M^2 \big]^2} + V_C \; \equiv \;  V_L(\vec p,\vec q \, ) \,
+\, C \, +\,  \cdots .
\nonumber
\end{equation}
Here, $V_C$ is a series of contact interactions with $C$ being the LO
term. Further, $E$ and $m$  refer to the energy and mass of the scattered
particles, respectively, while $M$ denotes the light mass of the exchanged
meson that sets the soft scale in the problem.
The coupling constant $\alpha$ is chosen such that the long-range
potential $V_L$ contributes at LO for momenta of the order of $M$. 
The solution to the LO integral equation can be written as
\begin{equation}
T(\vec p,\vec q \, ) \; =\;  T_L (\vec p,\vec q \, ) \, +\, \frac{\Psi_L (\vec
p \, ) \Psi_L (\vec q \, )}{1/C-G_E}\,, \label{solT}
\end{equation}
where the finite quantities $T_L$ and $\Psi_L $ are given by
\begin{eqnarray}
T_L \left( \vec p,\vec q \, \right) & = & V_L \left( \vec p,\vec
q \, \right) \, + \, m \int \frac{d^4 k }{(2\pi)^4}\, V_L( \vec p,\vec
k) \frac{1}{m E -k^2+i\,0^+}  T_L ( \vec k,\vec q \,)\,,
\nonumber\\
 \Psi_L (\vec q \, ) & = & 1 + m  \int \frac{d^4 k
}{(2\pi)^4}\,\frac{1}{m E-k^2+i\,0^+}\,T_L( \vec k,\vec q \, )\,. \label{blocks}
\end{eqnarray}
On the other hand, $G_E$ contains divergences and is given by
\begin{eqnarray}
G_E \; = \;   m \int \frac{d^4 k
}{(2\pi)^4}\,\frac{1}{mE-k^2+i\,0^+}  
  \, + \, m^2 \int \frac{d^4 k_1 }{(2\pi)^4}\,\frac{d^4 k_2
}{(2\pi)^4}\,\frac{T_L( \vec k_1,\vec
k_2)}{\left(m E-k_1^2+i\,0^+ \right)\left( m E -k_2^2+i\,0^+ \right)}\,.  \nonumber
\end{eqnarray}
Using cutoff regularization, $G_E$ can be written as 
\begin{equation}
G_E \; =\; a_1 \Lambda^2 \, +\, a_2 \ln^2\frac{\Lambda}{M} \,
+\,  (a_3+ a_4  E)\ln\frac{\Lambda}{m} \, + \, G_E^f \, + \, {\cal
O}\left(\frac{1}{\Lambda^2}\right), \label{GE}
\end{equation}
where $a_i$ are some constant factors
and $G_E^f$ is a finite $\Lambda$-independent part.

Coming back to the amplitude in Eq.~(\ref{solT}), we perform implicit
renormalization by fixing the bare LEC $C(\Lambda )$ from the
requirement to reproduce the on-shell amplitude at some kinematical
point $E_0$.  For the sake of definiteness, suppose that the
denominator of the last term in Eq.~(\ref{solT}) takes some value $d$
for $E=E_0$. This finally yields the scattering amplitude
\begin{equation}
T(\vec p,\vec q \, ) \; =\;  T_L (\vec p,\vec q \; )\, +\, \frac{\Psi_L (\vec
p\, ) \Psi_L (\vec q\, )}{d -a_4 (E - E_0)  \ln (\Lambda/m)   - (
G_E^f - G_{E_0}^f )}\,, \label{solTR}
\end{equation}
where  we have dropped the irrelevant ${\cal  O} (\Lambda^{-2})$-terms. The
survival  of the $ \ln (\Lambda/m) $-term in this result is a
consequence of the amplitude being only \emph{partially} renormalized 
by the counterterms generated by the LEC $C$.  

On the other hand, the amplitude in Eq.~(\ref{solT})  can be {\it fully} 
renormalized using the BPHZ procedure, i.e.~by subtracting {\it all}
divergences and subsequently taking the $\Lambda\to\infty$
limit. Choosing the subtraction points of the order of the hard scale
$\Lambda_b \sim m$ in order to avoid distortion of the long-range
interaction and fixing the renormalized LEC $C_R$ at the same
kinematical point, the subtractively renormalized amplitude takes the form
\begin{equation}
T(\vec p,\vec q \, ) \; =\;  T_L (\vec p,\vec q \; )\, +\, \frac{\Psi_L (\vec
p\, ) \Psi_L (\vec q\, )}{d -a_4 (E - E_0)  \ln (\Lambda_b/m)   - (
G_E^f - G_{E_0}^f )}\,. \label{TSR}
\end{equation}
This result agrees, to a good accuracy, with the
partially renormalized one in Eq.~(\ref{solTR}) as long as $\Lambda \sim
\Lambda_b$. On the other hand, choosing $\Lambda\gg m $, one 
approaches  for $E \neq E_0$ the expression
\begin{equation}
 T(\vec p,\vec q \, ) \approx T_L(\vec p, \vec q \, )\,. \label{Twrong}
\end{equation}
While finite in the $\Lambda \to \infty $ limit, this result cannot be
correct in general.\footnote{To see this, one can regard the
  underlying interaction to be just $V_L(\vec p,\vec q \, ) + C$,
  regularized with a sharp cutoff $\Lambda = \Lambda_b$. Then, the
  amplitude in Eq.~(\ref{TSR}) coincides with the exact result for this underlying
  model up to $\mathcal{O} (\Lambda_b^{-2})$ terms in the
  denominator.}
Another analytic example along this line with a long-range
interaction of a separable type is presented in Ref.~\cite{Epelbaum:2009sd}. 

The above considerations illustrate the issues that can arise by 
attempting to numerically verify the validity of  Eq.~(\ref{RGinv})
from the variation of $\Lambda$ in a wide range above $\Lambda_b$. For the
considered model, Eq.~(\ref{RGinv}) is fulfilled both for $\Lambda \sim \Lambda_b$ and
for $\Lambda \gg \Lambda_b$. However, if the scattering amplitude
would only be available numerically, following the approach advocated by the
practitioners of the RG-invariant method as done e.g.~in Refs.~\cite{Hammer:2019poc,Yang:2019hkn,Song:2016ale,deVries:2020loy}
would lead one to choose the approximately
$\Lambda$-independent but incorrect solution given in Eq.~(\ref{Twrong}).   

In our view, a valid alternative approach to verify the approximate RG invariance
as defined in Eq.~(\ref{RGinv}) is by comparing the residual
$\Lambda$-dependence for the available cutoff range, $\Lambda
\sim \Lambda_b$, against the expected truncation
uncertainty, which can be estimated using Bayesian methods \cite{Furnstahl:2015rha,Melendez:2019izc}. Such
self-consistency checks are, in fact, already being routinely performed in
chiral EFT calculations, see e.g.~Refs.~\cite{Epelbaum:2014efa,Epelbaum:2014sza,Reinert:2017usi,Epelbaum:2019zqc,Filin:2020tcs,Acharya:2018qzk}.

\subsection{The RG-invariant formulation of nuclear \chieft: Open issues}

In light of the above considerations, we encourage the supporters of
the large-cutoff RG-invariant approach as defined above to take a position on the
following issues:

\begin{itemize}
\item
  The large-cutoff (i.e., $\Lambda \gg
  \Lambda_b$) behavior of the scattering amplitude for singular LO
  interactions $V_{\rm LO}(r)$, like e.g.~the OPE
  potential, is controlled by the (ambiguous) behavior of $V_{\rm LO}(r)$ at short
  distances $r \ll
  \Lambda_b^{-1}$ that governs the terms with positive powers and/or
  logarithms of $\Lambda$
  in  the loop
  expansion  $T_{\rm LO}= \sum_n \hbar^n T_n$ (unless one succeeds to completely renormalize the
  amplitude by subtracting all UV divergent terms in $T_{\rm LO}$, which requires the inclusion of an
  infinite number of counterterms). If only a finite number of
  counterterms are included, what is the rationale behind
  expecting $T_{\rm LO} \big|_{\Lambda \gg  \Lambda_b}$ to represent
  a valid/meaningful EFT prediction? 
\item
In Ref.~\cite{Epelbaum:2021sns}, some of us have considered resonant P-wave
systems using the formulation of \pilesseft without auxiliary dimer
fields. For P-wave systems with an enhanced scattering volume,
expressing the leading order (\LO) LECs $C_2(\Lambda )$, $C_4(\Lambda )$ in terms of
the scattering volume and effective ``range'' $r$ is only possible for
$\Lambda \sim \Lambda_b \sim r$ as a consequence of the Wigner bound.
How is this feature to be interpreted from the point of view of the RG-invariant EFT approach?
\item
  Given the apparent non-uniqueness of the approximate RG-invariance criterion
  in Eq.~(\ref{RGinv}) when varying the cutoff from $\Lambda \sim
  \Lambda_b$ to  $\Lambda \gg \Lambda_b$ as demonstrated in the
  toy-model example, how can one make sure to avoid running into a UV
  stable but unphysical solution if no analytical 
  results are available?
\end{itemize}

\section{Renormalizability in the EFT sense: A formal proof} \label{Sec:4}

Formal aspects of renormalizability of the finite-cutoff formulation
of \chieft for \NN scattering were discussed in the talk by Ashot Gasparyan, see Ref.~\cite{Gasparyan:upcoming}
for more details. In the considered framework, the cutoff
$\Lambda\sim\Lambda_b$ is
chosen in such a way that no spurious bound states are generated at (\LO) 
in the \NN spin-triplet channels, but its form is not restricted
otherwise (in particular, it can be of a local or
non-local type).  Following the Weinberg power counting,
the LO interaction is resummed up to an infinite order by iterating the Lippmann-Schwinger equation,
whereas the subleading (\NLO) terms are iterated only once.

The standard requirement for a theory to be renormalizable is the
ability to absorb all UV divergences appearing in the $S$-matrix into a redefinition
(renormalization) of parameters in the underlying Lagrangian.
Clearly, introducing a finite cutoff automatically tames all
infinities in the scattering amplitude,
and the problem is shifted to the appearance of positive powers of
$\Lambda$ in place of the soft scales such as
$M_\pi$ or external three-momenta as dictated by the 
power counting. Such power-counting violating contributions originate from the
integration regions with momenta of the order of the cutoff.
It is thus natural to extend the notion of renormalizability by
demanding that all  power-counting breaking terms are
absorbable into shifts 
of the LECs at lower orders.

For the \LO \NN amplitude, one usually assumes that no power-counting
breaking contributions appear since positive powers of the
cutoff in the iterations of the Lippmann-Schwinger equation 
are compensated by the corresponding inverse powers of the hard scale
that appears as a prefactor in the \LO potential.
This conjecture, along with the renormalizability of the NLO scattering
amplitude, is rigorously proven in Ref.~\cite{Gasparyan:upcoming} to all orders
in the loop expansion. 
An extension of the proof to the purely non-perturbative case with a
non-convergent LO series is in progress.

To accomplish the
proof, it was essential to
introduce the regulator at the level of the Lagrangian without actually
affecting it. This is achieved by adding the regulator terms to the LO
interaction while systematically subtracting them from the
perturbative \NLO interaction. The resulting approach allows one to strongly
reduce the cutoff dependence of observables, the feature that has been verified by considering
several examples of the NN phase shifts \cite{Gasparyan:upcoming}. 

\section{Concluding remarks} \label{Sec:5}

\chieft offers a model independent and
systematically improvable approach to low-energy nuclear dynamics,
which---if pushed to sufficiently high orders in the EFT expansion---should be capable of making reliable and accurate predictions. It
thus is expected to shed light on the long-standing problems in
nuclear physics such as e.g.~the \NNN force challenge. Today, 30 years after Weinberg's
seminal papers \cite{Weinberg:1990rz,Weinberg:1991um} that laid out the foundations of the method,
the term ``Precision Nuclear Physics'' is not merely
a dream anymore. Indeed, modern \NN interactions derived in \chieft
have reached the precision of the most sophisticated phenomenological
potentials.  Further recent examples of precision nuclear physics studies
in \chieft
include the determination of the pion-nucleon coupling constants from
\NN data at
the $\sim 1\%$ accuracy level \cite{Reinert:2020mcu} and the
calculation of the
deuteron structure radius at
the $\sim 0.1\%$ accuracy level \cite{Filin:2019eoe,Filin:2020tcs}, both carried out at
\NNNNLO. To push the precision frontier beyond the \NN sector, 
it will be necessary  to develop \emph{consistently
  regularized} high-precision many-body forces and currents up
through \NNNNLO. Work on this ambitious goal is in progress, and it will hopefully help to mature low-energy nuclear physics into
precision science.

{\bf Acknowledgements:}
It is a pleasure to thank Zohreh Davoudi, Andreas Ekstr\"om, Jason Holt
and Ingo Tews for making this wonderful INT program possible. We are
also grateful to our long-standing collaborator Ulf-G.~Mei{\ss}ner
as well as to Patrick Reinert, Xiu-Lei Ren and the LENPIC Collaboration for sharing their
insights into the discussed topics.  
This work was supported in part by BMBF (Grant No.~05P18PCFP1), by DFG and NSFC through funds provided to the
Sino-German CRC 110 ``Symmetries and the Emergence of Structure in QCD" (NSFC
Grant No.~12070131001, Project-ID 196253076 - TRR 110), by DFG (Grant
No.~426661267)
and  by the Georgian Shota Rustaveli National Science Foundation (Grant No. FR17-354).

\sloppy
\printbibliography[heading=subbibliography]
\end{refsection}

\addtocontents{toc}{\protect\setcounter{tocdepth}{1}}
\begin{refsection}
\chapter{Challenges and progress in computational and theoretical low-energy nuclear physics}
\chapterauthor[1]{Chieh-Jen Yang}
\begin{affils}
\chapteraffil[1]{Department of Physics, Chalmers University of Technology, SE-412 96 G{\"o}teborg, Sweden}
\end{affils}

\addtocontents{toc}{\protect\setcounter{tocdepth}{1}}

\vspace{0.5cm}

I consider the challenges we have been facing in theoretical nuclear physics can be mainly categorized into two aspects: (i) The challenge in computing complex systems. (ii) The challenge in search for a better theoretical foundation. Breakthrough in these two directions are both important and should complement each other in order to make true progress. Fortunately, there are several important achievements in both directions presented in this workshop. In the following I highlight two breakthroughs (one in each direction) and one problem which requires further investigations in this workshop.  


\begin{itemize}
\item Breakthroughs in computational aspect

   Eigenvector continuation is a powerful tool, which allows fast simulations and testings in ab-initio calculations\cite{Frame:2017fah}. It has been applied to no-core-shell-model and coupled-cluster methods\cite{Ekstrom:2019lss} and is crucial for optimizing the low-energy-constants (LECs) in order to obtain a better global fit. So far this technique has been applied mainly to the bound-state problems. The effort to extend it to 3-body scattering, as presented in this workshop\cite{Zhang:2021jmi}, is therefore very interesting.

\item Breakthroughs in theoretical foundation

   Many-body forces play an important role in complex systems. They emerge naturally when the degrees of freedom are reduced from elementary particles to composite ones. Regardless how they are derived, most of the existing calculations performed today treat them as (parts of) the potential on top of two-body interactions without additional considerations. One of the most intriguing recent discoveries, as presented in this workshop, is that this could be very wrong. Due to a combinatorics argument\cite{Yang:2019hkn}, the importance of three-nucleon forces is estimated to be as important as the leading two-nucleon forces for nuclei with number of nucleons A=10-20. This means, under chiral effective field theory (\chieft), the leading three-nucleon forces---which are conventionally regarded as next-to-next-to-leading order (\NNLO)---should be promoted to leading order (\LO) in the calculation of $^{16}$O. This is confirmed by explicit calculations, where a physical $^{16}$O is obtained for the first time under a consistent \chieft at \LO~\cite{Yang:2021vxa}. Further investigations in this direction, e.g., the importance of four-nucleon forces and high-order corrections, are highly desirable.

\item One problem requires further investigations

    It is shown in this workshop that, at least under the widely adopted Weinberg power counting (WPC), there is a limitation in optimizing the LECs in \chieft potentials. In particular, one faces the choices to either sacrifice the description of nucleon-nucleon (\NN) and few-body observables in order to describe saturation-related properties, or the other way around\cite{Jiang:2020the,Nosyk:2021pxb}. Since the potentials been tested are of considerable high-order (\NNLO), such large discrepancy/uncertainty is not acceptable. A call for a rearrangement of EFT power counting---with the number of nucleons taken into account, as suggested in Ref.\cite{Yang:2021vxa}---might be necessary. Naively, this would partially release the burden of the LECs presented in the three-nucleon forces so that they do not have to fit all systems (from light to heavy mass nuclei) at the same time. 
\end{itemize}    
\sloppy
\printbibliography[heading=subbibliography]
\end{refsection}

\addtocontents{toc}{\protect\setcounter{tocdepth}{1}}
\begin{refsection}
\chapter{On the determination of $\boldsymbol{\pi N}$ and $\boldsymbol{NN}$ low-energy constants}
\chapterauthor[1]{Martin Hoferichter}
\begin{affils}
\chapteraffil[1]{Albert Einstein Center for Fundamental Physics, Institute for Theoretical Physics, University of Bern, Sidlerstrasse 5, 3012 Bern, Switzerland}
\end{affils}

\addtocontents{toc}{\protect\setcounter{tocdepth}{-2}}

Constructing precision nuclear forces from chiral effective field theory (\chieft) requires good control over subleading orders in the chiral expansion, in particular, of the low-energy constants (LECs) that parameterize degrees of freedom beyond the range of validity of the EFT. While some of the LECs can be determined from other observables, there are many cases in which this is not possible, leaving ultimately lattice QCD (\LQCD) as the tool of choice. Here, we describe some of the recent developments and benchmarks of this program. 

\begin{enumerate}

 \item The long-range part of the nucleon-nucleon (\NN) and three-nucleon forces is related to $\pi N$ physics, encoded in the LECs $c_i$, $d_i$, and $e_i$ at the respective order. At a given order, these LECs can be determined precisely by matching to the subthreshold parameters of $\pi N$ scattering via the solution of Roy--Steiner equations~\cite{Hoferichter:2015tha,Hoferichter:2015hva} in combination with experimental input from pionic atoms~\cite{Baru:2010xn,Baru:2011bw,Strauch:2010vu,Hennebach:2014lsa,Hirtl:2021zqf}, leaving the convergence of the chiral expansion as the dominant uncertainty. 
 
 \item These issues in the chiral convergence become apparent when comparing the expansion in the subthreshold and threshold regions---with the former being most relevant for \NN kinematics---as the heavy-baryon expansion fails to simultaneously describe these two kinematic regions. The convergence improves with a covariant formulation and when including explicit $\Delta$ degrees of freedom~\cite{Siemens:2016jwj}, but in the latter case at the expense of introducing additional LECs. Only the leading one, the $\pi N \Delta$ coupling $h_A$, can be determined from phenomenology, while the subleading coefficients, $g_1$, $b_{4,5}$, are only constrained by large-$N_c$ arguments~\cite{Siemens:2016jwj}.
 
 \item Not all subleading $\pi N$ LECs can be directly extracted from $\pi N$ scattering, as the LEC $c_5$, which appears as an isospin-breaking contribution, is determined from the strong part of the proton--neutron mass difference, which is not directly observable. A phenomenological determination is possible via the Cottingham formula~\cite{Cottingham:1963zz}, which relates the elastic contribution to nucleon form factors and the inelastic ones to nucleon structure functions when assuming a suitable high-energy behavior. The resulting separation of the nucleon mass into strong and electromagnetic contributions~\cite{Gasser:1974wd,Gasser:2015dwa,Gasser:2020mzy,Gasser:2020hzn} differs from \LQCD~\cite{Borsanyi:2014jba,Brantley:2016our,Horsley:2019wha} by $2.3\sigma$.
 
 \item A formalism similar to the Cottingham approach was used in~\cite{Cirigliano:2020dmx,Cirigliano:2021qko} to estimate the leading-order contact term~\cite{Cirigliano:2018hja,Cirigliano:2019vdj} in neutrino-less double-$\beta$ decay, based on the elastic contribution, which gives the dominant effect in the case of the nucleon mass difference. This defines a benchmark for future calculations in \LQCD~\cite{Davoudi:2020gxs}. Moreover, the calculation in~\cite{Cirigliano:2020dmx,Cirigliano:2021qko} is performed in dimensional regularization, but the result presented in terms of a renormalized, physical amplitude, which can then be matched to schemes applied in nuclear-structure calculations~\cite{Wirth:2021pij,Jokiniemi:2021qqv}. This strategy may prove useful for future \LQCD calculations as well. 
 
 \item Before turning to the \NN sector, benchmark quantities for simpler $\pi N$ matrix elements include the axial coupling $g_A$ and the $\sigma$-term $\sigma_{\pi N}$. While for the former \LQCD calculations have reached few-percent accuracy~\cite{Chang:2018uxx,Gupta:2018qil}, the situation for the latter remains unresolved, with \LQCD~\cite{Aoki:2021kgd,Durr:2015dna,Yang:2015uis,Yamanaka:2018uud,Alexandrou:2019brg,Borsanyi:2020bpd} favoring values significantly smaller than phenomenological determinations~\cite{Hoferichter:2015dsa,Hoferichter:2016ocj,RuizdeElvira:2017stg}. Recently, it was suggested that the origin could lie in larger-than-expected excited-state contamination~\cite{Gupta:2021ahb}, which may be of relevance for \LQCD calculations of other $\pi N$ and \NN LECs.  
 
 \item Given its relation to phenomenology via the Cheng--Dashen low-energy theorem~\cite{Cheng:1970mx}, the $\sigma$-term also serves as an important benchmark for matrix elements required for searches for physics beyond the Standard Model. Another such indirect relation that allows one to determine LECs for non-standard currents proceeds via a unitarity argument~\cite{Cirigliano:2017tqn,VonDetten:2021rax}, connecting the energy dependence of vector and antisymmetric tensor matrix elements~\cite{Hoferichter:2016duk,Hoferichter:2018zwu}.
\end{enumerate}

{\bf Acknowledgements:}
Support by the Swiss National Science Foundation (Project No.\ PCEFP2\_181117)
is gratefully acknowledged.
\sloppy
\printbibliography[heading=subbibliography]
\end{refsection}

\newcommand{\Nc}{\ensuremath{N_c}\xspace}
\newcommand{\LONc}{\ensuremath{\text{LO-in-\Nc}}\xspace}
\newcommand{\NLONc}{\ensuremath{\text{NLO-in-\Nc}}\xspace}
\newcommand{\NNLONc}{\ensuremath{\text{N${}^2$LO-in-\Nc}}\xspace}
\newcommand{\neutrinoless}{\ensuremath{0 \nu \beta \beta}\xspace}
\newcommand{\oneS}{{{}^{1}\!S_0}}
\newcommand{\threeS}{{{}^{3}\!S_1}}
\newcommand{\onePone}{{{}^{1}\!P_1}}
\newcommand{\Ptwo}{{{}^{3}\!P_2}}
\newcommand{\Pone}{{{}^{3}\!P_1}}
\newcommand{\Pzero}{{{}^{3}\!P_0}}
\newcommand{\PJ}{{{}^{3}\!P_J}}
\newcommand{\oneP}{{{}^{1}\!P_1}}
\newcommand{\threeD}{{{}^{3}\!D_1}}
\newcommand{\Galilean}{\overset{\leftrightarrow}{\nabla}^i}
\newcommand{\LRC}{\overset{\leftrightarrow}{C}{}}
\newcommand{\del}{\nabla}
\newcommand{\calO}{\ensuremath{\mathcal{O}}}
\newcommand{\calQ}{\ensuremath{\mathcal{Q}}}
\newcommand{\calI}{\ensuremath{\mathcal{I}}}
\newcommand{\calG}{\ensuremath{\mathcal{G}}}
\newcommand{\calC}{\ensuremath{\mathcal{C}}}
\newcommand{\calA}{\ensuremath{\mathcal{A}}}
\newcommand{\calL}{\ensuremath{\mathcal{L}}}
\newcommand{\calM}{\ensuremath{\mathcal{M}}}
\newcommand{\Lagr}{\mathcal{L}}
\newcommand{\CovDer}{\mathcal{D}}
\newcommand{\1}{\mathbbm{1}}
\newcommand{\Langle}{\Big\langle}
\newcommand{\Rangle}{\Big\rangle}
\renewcommand\vector{\mathbf}
\newcommand{\pplus}{\vector{p}_+}
\newcommand{\pminus}{\vector{p}_-}
\newcommand{\vsig}{\vec{\sigma}}
\newcommand{\vtau}{\vec{\tau}}
\newcommand{\Czerotrip}{\ensuremath{C_0^{(^3 \! S_1)}}\xspace}
\newcommand{\Czerosing}{\ensuremath{C_0^{(^1 \! S_0)}}\xspace}
\newcommand{\Ctrip}{\ensuremath{C_2^{(^3 \! S_1)}}\xspace}
\newcommand{\Csing}{\ensuremath{C_2^{(^1 \! S_0)}}\xspace}
\newcommand{\Csd}{\ensuremath{C^{(SD)}}\xspace}
\newcommand{\ConeP}{\ensuremath{C^{(^1 \! P_1)}}\xspace}
\newcommand{\CPzero}{\ensuremath{C^{(^3 \! P_0)}}\xspace}
\newcommand{\CPone}{\ensuremath{C^{(^3 \! P_1)}}\xspace}
\newcommand{\CPtwo}{\ensuremath{C^{(^3 \! P_2)}}\xspace}
\newcommand{\CtripLO}{\ensuremath{C_{2,{\LONc}}^{(^3 \! S_1)}}\xspace}
\newcommand{\CsingLO}{\ensuremath{C_{2,{\LONc}}^{(^1 \! S_0)}}\xspace}
\newcommand{\CtripNNLO}{\ensuremath{C_{2,{\NNLONc}}^{(^3 \! S_1)}}\xspace}
\newcommand{\CsingNNLO}{\ensuremath{C_{2,{\NNLONc}}^{(^1 \! S_0)}}\xspace}
\newcommand{\asing}{\ensuremath{a^{(^1 \! S_0)}}\xspace}
\newcommand{\rsing}{\ensuremath{r^{(^1 \! S_0)}}\xspace}
\newcommand{\atrip}{\ensuremath{a^{(^3 \! S_1)}}\xspace}
\newcommand{\rtrip}{\ensuremath{r^{(^3 \! S_1)}}\xspace}
\newcommand{\Lone}{\ensuremath{{}^{\nopi} L_1}\xspace}
\newcommand{\Ltwo}{\ensuremath{{}^{\nopi} L_2}\xspace}
\newcommand{\CMs}{\ensuremath{C^{(M)}_s}\xspace}
\newcommand{\CMv}{\ensuremath{C^{(M)}_v}\xspace}
\newcommand{\gnu}{\ensuremath{g_\nu^{\NN}}\xspace}
\newcommand{\fpi}{\ensuremath{f_\pi}}
\newcommand{\GeV}{\ensuremath{\mathrm{GeV}}}
\newcommand{\nopi}{\ensuremath{\pi\hskip-0.40em /}}
\newcommand{\eftnopi}{EFT$_{\nopi}$\xspace}
\newcommand{\LambdaNoPion}{\Lambda_{\nopi}\xspace}
\newcommand{\Qorder}{Q/\LambdaNoPion}
\newcommand{\chiPT}{$\chi$PT\xspace}
\newcommand{\Ndag}{N^\dagger}

\newcommand\redsout{\bgroup\markoverwith{\textcolor{red}{\rule[0.5ex]{2pt}{1.4pt}}}\ULon}
\newcommand{\replace}[2]{\redsout{\protect#1}\textbf{\color{purple}#2}}
\newcommand{\add}[1]{\textbf{\color{purple}#1}}
\newcommand{\remove}[1]{\redsout{#1}}
\renewcommand{\emph}[1]{\textit{#1}}
\newcommand{\rps}[1]{{\sffamily\bfseries{\color{brown}[rps: #1]}}}
\newcommand{\mrs}[1]{{\bf \color{blue} \small [mrs: #1]}}
\newcommand{\trr}[1]{{\bf \color{red} \small [TRR: #1]}}
\newcommand{\saori}[1]{{\bf \color{orange} \small [saori: #1]}}

\addtocontents{toc}{\protect\setcounter{tocdepth}{1}}
\begin{refsection}
\chapter{The possible role of the large-$\boldsymbol{\Nc}$ limit in understanding nuclear forces from QCD}
\chapterauthor[1]{Thomas R. Richardson}
\chapterauthor[1]{Matthias R.~Schindler}
\chapterauthor[2]{Roxanne P. Springer}
\begin{affils}
\chapteraffil[1]{Department of Physics and Astronomy, University of South Carolina, Columbia, SC 29208, USA}
\chapteraffil[2]{Department  of  Physics,  Duke  University,  Durham,  NC  27708,  USA}
\end{affils}
\addtocontents{toc}{\protect\setcounter{tocdepth}{-2}}

The community has developed a procedure for attempting to understand nuclear physics starting from QCD: lattice QCD (\LQCD) is used to calculate the nonperturbative physics that determines the low energy constants (LECs) of an effective field theory (EFT) possessing the symmetries of QCD (and/or beyond-the-standard-model physics), which is then input into many-body calculations to address heavier nuclei.  
The large-\Nc limit of QCD \cite{tHooft:1973alw}, where \Nc is the number of colors, can play a role in this procedure.
One- and two-nucleon matrix elements can be expanded in powers of $1/\Nc$. When combined with an EFT expansion, either pionless or chiral, the number of independent LECs at a given order in the combined expansion may be reduced.
These constraints can be used to prioritize \LQCD calculations and also provide some simplifications to the input needed for many-body calculations.

The large-\Nc limit of QCD has been used to provide theoretical constraints for a variety of applications in the two- and three-nucleon (\NNN) sectors, see, e.g., Refs.~\cite{Kaplan:1995yg,Kaplan:1996rk,Banerjee:2001js,Belitsky:2002ni,Cohen:2002im,Riska:2002vn,Phillips:2013rsa,Phillips:2014kna,Epelbaum:2014sea,Schindler:2015nga,Samart:2016ufg,Schindler:2018irz,Vanasse:2019fzl,Richardson:2020iqi, Richardson:2021xiu}.
In the large-\Nc limit, Wigner's SU(4) symmetry emerges, which is also manifest in the beta decays of some medium-mass nuclei \cite{Kaplan:1995yg, Kaplan:1996rk}.
In the SU(3) sector, an SU(6) symmetry among baryon-baryon interactions is predicted in the large-\Nc limit, with an accidental SU(16) emerging for certain values of LECs. These patterns have been observed in \LQCD calculations with larger-than-physical values of the quark masses~\cite{Wagman:2017tmp,Illa:2020nsi}.
In the \NNN sector, LECs in \chieft also broadly align with the large-\Nc hierarchy \cite{Phillips:2013rsa,Epelbaum:2014sea}.
The large-\Nc analysis in the parity-violating sector demonstrates that the number of leading-order couplings in \pilesseft is reduced from five to two \cite{Schindler:2015nga}. This also highlights the need for a determination of the isotensor parity-violating LEC. 
The isotensor LEC in particular is an opportunity for \LQCD to make a prediction in the absence of experimental data.
Reference \cite{Vanasse:2019fzl} considered the impact of the dual expansion on T-violating interactions.
The application of the large-\Nc approach to external magnetic and axial vector fields \cite{Richardson:2020iqi} offers a partial explanation for the disparate sizes of the isoscalar and isovector magnetic LECs despite these terms occurring at the same order in the \pilesseft power counting. 
These results also indicate that naturalness, i.e., the concept that LECs at the same order in the power counting should be the same size, may be hidden depending on the choice of basis; therefore, caution should be taken when attempting to quantify naturalness.
Finally, the large-\Nc analysis of charge-independence-breaking (CIB) two-nucleon interactions~\cite{Richardson:2021xiu} provides a justification for the assumptions of Refs.~\cite{Cirigliano:2018hja,cirigliano_renormalized_2019} relating a new lepton-number-violating LEC to an experimentally determined combination of CIB LECs. This result was recently corroborated using a different method \cite{Cirigliano:2020dmx, Cirigliano:2021qko}. 

These examples demonstrate the utility of combining the large-\Nc and EFT expansions. 
Further, this dual expansion could be used to estimate the relative sizes of the couplings, which could potentially reduce the number of contributions required at any given order for many-body calculations.
Additionally, large-\Nc constraints 
may help prioritize calculations for the lattice community.
Lastly, a large-\Nc analysis of new beyond-the-standard-model couplings can provide constraints to potentially guide the interpretation of experimental results, e.g., for dark matter direct detection. We think that the procedure of understanding nuclear phenomena via the combination of \LQCD, EFTs, and many-body techniques may benefit from implementing large-\Nc constraints.

{\bf Acknowledgements:}
We are grateful to Saori Pastore for useful discussions.  
We thank the Institute for Nuclear Theory at the University of Washington for its stimulating research environment during the INT-21-1b program ``Nuclear Forces for Precision Nuclear Physics,'' which was supported in part by the INT's U.S.~Department of Energy grant No.~DE-FG02-00ER41132.
This material is based upon work supported by the U.S.~Department of Energy, Office of Science, Office of Nuclear Physics,
under Award Numbers DE-SC0019647 (TRR and MRS) and DE-FG02-05ER41368 (RPS). 

\printbibliography[heading=subbibliography]
\end{refsection}

\addtocontents{toc}{\protect\setcounter{tocdepth}{1}}
\begin{refsection}
\chapter{Towards robustly grounding nuclear physics in the Standard Model}

\chapterauthor[1]{Zohreh Davoudi}
\chapterauthor[2,3]{William Detmold}
\chapterauthor[4]{Marc Illa}
\chapterauthor[5]{Assumpta Parre\~no}
\chapterauthor[2,3]{Phiala E. Shanahan}
\chapterauthor[2,6]{Michael L. Wagman}

\begin{affils}
\chapteraffil[1]{Maryland Center for Fundamental Physics and Department of Physics, University of Maryland, College Park, MD 20742, USA}
\chapteraffil[2]{Center for Theoretical Physics, Massachusetts Institute of Technology, Cambridge, MA 02139, USA}
\chapteraffil[3]{The NSF AI Institute for Artificial Intelligence and Fundamental Interactions}
\chapteraffil[4]{InQubator for Quantum Simulation (IQuS), Department of Physics, University of Washington, Seattle, WA 98195, USA}
\chapteraffil[5]{Departament de F\'{\i}sica Qu\`{a}ntica i Astrof\'{\i}sica and Institut de Ci\`{e}ncies del Cosmos,	Universitat de Barcelona, Mart\'{\i} i Franqu\`es 1, E08028, Spain}
\chapteraffil[6]{Fermi National Accelerator Laboratory, Batavia, IL 60510, USA}
\end{affils}
\addtocontents{toc}{\protect\setcounter{tocdepth}{-2}}

Nuclear physics is entering an exciting era in which aspects of nuclear structure and reactions can be directly computed from the Standard Model of particle physics. 
Lattice quantum chromodynamics (\LQCD) will play a vital role in this era by providing a systematically improvable route through which to obtain nonperturbative quantum chromodynamics (QCD) predictions for few-nucleon systems.
In particular, robust QCD predictions with quantified uncertainties for observables, including the energy spectra of multi-nucleon systems and matrix elements of electroweak and beyond-Standard-Model (BSM) currents, will provide valuable information about nuclear structure and interactions complementary to that obtained from experimental measurements. Such predictions can be used to constrain the parameters of low-energy effective field theories (EFTs), as well as to validate and inform phenomenological models of nuclei based on nucleon degrees of freedom. 

Both physical and computational challenges will restrict LQCD studies to few-nucleon systems for the foreseeable future; exponential degradation of signal versus noise at large Euclidean times~\cite{Parisi:1983ae,Lepage:1989hd} arising from sign problems~\cite{Wagman:2016bam} and tensor contraction complexity \cite{Detmold:2012eu} make LQCD calculations of (multi-)baryon correlation functions computationally demanding, while the smallness of finite volume (FV) energy gaps between  states in such systems complicate their analysis.
Fortunately, the most relevant inputs to EFTs of nuclei are two- and three-baryon interactions as well as one- and two-baryon electroweak and BSM currents.
It is in this relatively computationally accessible few-baryon sector that LQCD calculations will have the largest impact on nuclear EFTs.
Pioneering LQCD calculations of few-nucleon systems performed over the last two decades have been used to motivate, develop, and test different strategies for using the immediate results of LQCD calculations -- FV Euclidean correlation functions formed from particular sets of composite operators designed to interpolate to the desired states -- to obtain FV energy spectra and matrix elements and to constrain the inputs of nuclear effective theories.
The first dynamical LQCD calculations of two-nucleon correlation functions, performed by the NPLQCD Collaboration in 2006~\cite{Beane:2006mx}, used asymmetric correlation functions with localized sources and non-local sinks to constrain the contact operators describing two-nucleon interactions with both Weinberg~\cite{Weinberg:1990rz} and Beane-Bedaque-Savage-van-Kolck~\cite{Beane:2001bc} power counting in the two $s$-wave scattering channels.
Calculations of analogous two-baryon correlation functions with non-zero strangeness by the NPLQCD Collaboration in 2012 were used to constrain contact interactions in an EFT for hyperon-nucleon systems that was then used to predict hyperon-nucleon phase shifts as well as in-medium energy shifts of hyperons relevant for the neutron-star equation of state~\cite{Beane:2012ey}. 
Instead of obtaining scattering amplitudes via EFTs that are constrained by LQCD, L{\"u}scher's quantization condition~\cite{Luscher:1986pf} and its generalizations (reviewed in Refs.~\cite{Briceno:2017max,Hansen:2019nir}) have also been explored as a complementary strategy for relating the immediate results of LQCD calculations for two-baryon systems to infinite-volume quantities such as scattering phase shifts.
Constraints on $s$-wave scattering at particular values of the quark masses have been made using this method by the NPLQCD Collaboration~\cite{Beane:2006mx,Beane:2013br}.Constraints on higher-partial-wave scattering were first made by the CalLat Collaboration~\cite{Berkowitz:2015eaa} in 2015 by applying these methods to asymmetric LQCD correlation functions with displaced as well as local sources.

LQCD and EFT have advanced together over the last decade and been applied to study increasingly complex systems. 
In 2013, calculations of baryon-number $A\in\{2,3,4\}$ nuclear (and $A\in\{2,3,4,5\}$ hypernuclear) correlation functions by the NPLQCD Collaboration with unphysically large quark masses corresponding to $m_\pi = 806$ MeV~\cite{Beane:2012vq} were used to constrain two- and three-body contact interactions in pionless EFT by Barnea {\it et al.}~\cite{Barnea:2013uqa}, who went on to predict binding energies of $A\in\{5,6\}$ nuclei at these quark masses. More refined EFT matching directly to FV energies has been recently pursued in Ref.~\cite{Eliyahu:2019nkz}.
Calculations of larger nuclei in pionless EFT including ${}^{16}$O and ${}^{40}$Ca at both $m_\pi = 806$ MeV (matched to the aforementioned LQCD results) and with physical quark masses were performed by multiple groups in 2017~\cite{Contessi:2017rww,Bansal:2017pwn}.
Calculations of additional hyperon-nucleon and hyperon-hyperon scattering channels by the NPLQCD Collaboration in 2017 suggested new emergent symmetries in baryon-baryon interactions~\cite{Wagman:2017tmp}. The appearance of these symmetries at lighter quark masses has been tested by recent calculations constraining SU(3)$_f$-breaking hypernuclear interactions~\cite{Illa:2020nsi}.
The structure of light nuclei with $m_\pi = 806$ MeV and $m_\pi = 450$ MeV has been probed by calculations of scalar, axial, tensor, and vector nuclear matrix elements by the NPLQCD Collaboration over the last several years~\cite{Beane:2013kca,Beane:2014ora,Beane:2015yha,Chang:2015qxa,Detmold:2015daa,Savage:2016kon,Shanahan:2017bgi,Chang:2017eiq} that have revealed shell-model-like structure at unphysically large quark masses.
The first nuclear-reaction studies from LQCD, albeit at large quark masses, were reported in Refs.~\cite{Beane:2015yha, Savage:2016kon, Tiburzi:2017iux, Shanahan:2017bgi}, paving the way to constraining short-distance LECs of the EFTs in $pp$ fusion, and single- and double-$\beta$ decay processes in light nuclei \cite{Davoudi:2020ngi}.
Techniques for matching FV results for few-nucleon systems between LQCD and EFT have been further developed and in the last year have been used to enable a quark-mass extrapolation of the Gamow-Teller matrix element governing triton $\beta$ decay~\cite{Detmold:2021oro,Parreno:2021ovq} as well as first constraints on the quark momentum fractions of ${}^3$He~\cite{Detmold:2020snb}.

Enabled by the early development of efficient algorithms~\cite{Detmold:2012eu,Doi:2012xd}, all the LQCD calculations of multi-baryon systems described above used local, or sometimes displaced, sources and non-local sinks built from products of momentum-projected baryons.
An alternative approach developed by the HALQCD Collaboration is based on determining nuclear potentials from Bethe-Salpeter wavefunctions of multi-baryon systems~\cite{Ishii:2006ec, Inoue:2010es} and is argued to avoid systematic uncertainties from excited-state effects involving unbound elastic scattering states~\cite{Aoki:2020bew} (inelastic states still contaminate the correlation functions used in this approach).
However, short-distance features of the potentials determined using these methods depend on the sink interpolating operator choice, making it very challenging to quantitatively assign systematic uncertainties to predictions that depend on these short-distance features~\cite{Beane:2010em, Birse:2012ph, Yamazaki:2017gjl, Iritani:2018zbt, Drischler:2019xuo}.

In the last few years, there has been exciting progress in enlarging the scope of interpolating operators that can be practically included in multi-baryon LQCD calculations. Increased computing power and new algorithmic approaches have allowed rigorous variational constraints on finite-volume energies of two-baryon systems.
The first variational calculations of two-nucleon systems with symmetric correlation functions with multi-baryon sources and sinks were performed by Francis \emph{et al}.~\cite{Francis:2018qch} and were enabled by the Laplacian-Heaviside method~\cite{Peardon:2009gh} for computing approximate all-to-all quark propagators.
A follow-up to this calculation~\cite{Green:2021qol} found significant discretization effects in multi-nucleon FV energy shifts and, perhaps relatedly, interesting tensions in comparison with previous results from the NPLQCD and CalLat Collaborations using asymmetric correlation functions.
A further variational calculation~\cite{Horz:2020zvv} using the stochastic Laplacian-Heaviside method~\cite{Morningstar:2011ka} included two-nucleon interpolating operators with zero and one unit of relative momentum in correlation-function matrices with several values of center-of-mass momentum that were diagonalized to construct orthogonal approximations to the ground state and first unbound excited state.
These results using different discretizations and interpolating operators again show tensions with earlier results.
The most recent variational study of two-nucleon systems as of this writing was performed by the NPLQCD collaboration ~\cite{Amarasinghe:2021lqa} and used sparsened timeslice-to-all quark propagators~\cite{Detmold:2019fbk} and  included a considerably larger set of hexaquark operators, quasi-local operators with exponential nucleon wavefunctions inspired by EFT bound-state wavefunctions, and  scattering operators with up to $\sqrt{6}$ units of relative momentum together  (in the center-of-mass frame).
Direct comparisons between asymmetric correlation functions and variational results on the same gauge-field ensemble in this study indicate that estimates of the FV energy spectrum depend significantly on the interpolating-operator set such that it is difficult to achieve systematically-controlled results at the available level of statistics. Similarly, comparison of variational results from different choices of interpolating-operator sets leads to different bounds on the ground-state energy, although the upper bounds on energy levels provided by variational methods are robust in all cases.

There are multiple possible explanations of these results. On one hand, asymmetric correlation functions could appear to be exhibiting single-state dominance due to delicate cancellations between ground and excited-state contributions and the true ground-state energy could be larger than the value determined by previous asymmetric calculations~\cite{Amarasinghe:2021lqa}.
If such delicate cancellation is in place, the observed volume insensitivity of asymmetric correlation functions associated with the obtained ground states in previous two-nucleon studies~\cite{Wagman:2017tmp}, which signals the bound nature of the state in the volume, will be a surprising coincidence. On the other hand, it is straightforward to construct interpolating-operator overlap models for which asymmetric correlation functions would reveal the true ground-state energy while variational methods provide an upper bound that is dominated at realistic statistical precision by contributions from a higher-energy state that has larger overlap with all of the interpolating operators used in the study~\cite{Amarasinghe:2021lqa}.
A simple toy example of this is given by a pair of interpolating operators $A$ and $B$ that are used to probe a three-state system with true energy levels
\begin{equation}
  E_0^{(AB)} = \eta - \Delta, \hspace{20pt} E_1^{(AB)} = \eta, \hspace{20pt} E_2^{(AB)} = \eta + \delta.
\end{equation}
Define normalized overlap factors for operators $A$ and $B$ onto these states by
\begin{equation}
  \mathcal{Z}_A = (\epsilon,\sqrt{1 - \epsilon^2},0), \hspace{20pt} \mathcal{Z}_B = (\epsilon,0,\sqrt{1 - \epsilon^2}),
  \label{eq:badZ}
\end{equation}
where $\epsilon\ll 1$ is a real parameter.
Solving a generalized eigenvalue problem (GEVP) using $2\times 2$ correlation-function matrices with interpolating operators $\{A,B\}$ and times $t_0$ and $t > t_0$ gives eigenvalues
\begin{equation}
  \begin{split}
  \lambda_0^{(AB)} &= e^{-(t-t_0)\eta}\left[ 1 + \epsilon^2  \left(  e^{t \Delta} - e^{t_0 \Delta} \right) + \mathcal{O}(\epsilon^4) \right], \\
    \lambda_1^{(AB)} &= e^{-(t-t_0)(\eta + \delta) }\left[ 1 + \epsilon^2 \left(  e^{t (\Delta + \delta)} - e^{t_0 (\Delta + \delta)} \right) + \mathcal{O}(\epsilon^4) \right].
  \end{split}
\end{equation}
Unless $t$ is large enough such that $e^{t \Delta}$ compensates for the $O(\epsilon^2)$ overlap-factor suppression, the bound obtained from the lowest GEVP eigenvalue will significantly overestimate the true ground-state energy.
However, an asymmetric correlation function of the form $\left< B(t) \overline{A}(0) \right>$ will overlap perfectly with the true ground state with zero excited-state contamination.
This example can be trivially generalized to include more states that have small overlap with the interpolating-operator set $\{A,B\}$ without changing the need for achieving large enough $e^{t \Delta}$ in order to compensate for the smallness of the overlap factors present.
Including additional interpolating operators that have small overlap with the ground state also does not improve GEVP ground-state energy estimates in this example -- it is the inclusion of interpolating operators with sufficiently large overlap with states of interest that is essential for the success of variational methods.

The existence of such models demonstrates that in order to conclusively determine whether two-nucleon systems are bound or unbound with larger than physical values of the quark masses, further variational studies are required to span the subspace of Hilbert space that might be associated with a two-nucleon bound state. While it \emph{is} possible for variational studies to conclusively demonstrate the presence of a bound state, by their very nature they \emph{can not} rule out such a state except if the interpolating operators that are used form a basis for the Hilbert space -- a scenario that cannot be realistically achieved. Similarly, energies extracted from asymmetric correlation functions provide an estimate of the ground state energy which are subject to systematic uncertainties from choices of interpolating operators that may be difficult to estimate.

LQCD studies of multi-nucleon systems will become increasingly refined in the coming years. In future studies, systematic uncertainties in LQCD determinations of nuclear properties associated with lattice spacing and quark-mass extrapolations will be controlled through the use of larger sets of gauge-field ensembles. Excited-state effects will be controlled through variational studies that include more, and more varied, interpolating-operators that better cover the low-energy sector of Hilbert space. The ongoing development of strategies for matching LQCD results to nuclear effective theories and other ways to relate FV and infinite-volume observables will be increasingly essential for extending the reach of robust predictions grounded in the Standard Model up the chart of the nuclides. 

{\bf Acknowledgements:} We are grateful to the former and current members of the NPLQCD Collaboration, especially Martin Savage, for many insightful discussions and valuable collaborations around the topics discussed in this piece.
ZD acknowledges support from the Alfred P. Sloan fellowship, Maryland Center for Fundamental Physics at the University of Maryland, College Park, and the U.S.Department of Energy's (DOE's) Office of Science Early Career Award DE-SC0020271.
WD and PES are supported in part by the U.S. DOE's Office of Science, Office of Nuclear Physics under grant Contract DE-SC0011090.
WD is further supported in part by the SciDAC4 award DE-SC0018121, and within the framework of the TMD Topical Collaboration of the U.S. DOE's Office of Science, Office of Nuclear Physics.
PES is additionally supported by the National Science Foundation under EAGER grant 2035015, by the U.S. DOE's Office of Science Early Career Award DE-SC0021006, by a NEC research award, and by the Carl G and Shirley Sontheimer Research Fund. WD and PES are supported by the National Science Foundation under Cooperative Agreement PHY-2019786 (The NSF AI Institute for Artificial Intelligence and Fundamental Interactions, http://iaifi.org/).
MI is supported in part by the U.S. Department of Energy, Office of Science, National Quantum Information Science Research Centers, Quantum Science Center, and in part by the U.S. Department of Energy, Office of Science, Office of Nuclear Physics, InQubator for Quantum Simulation (IQuS) through the Quantum Horizons: QIS Research and Innovation for Nuclear Science, under Award Number DOE (NP) Award DE-SC0020970.
AP acknowledges financial support from the State Agency for Research of the Spanish Ministry of Science and Innovation through the "Unit of Excellence Mar\'{\i}a de Maeztu 2020-2023" award to the Institute of Cosmos Sciences (CEX2019-000918-M), the European FEDER funds under the contract PID2020-118758GB-I00, and from the EU STRONG-2020 project under the program H2020-INFRAIA-2018-1, grant agreement No. 824093.
This piece has been authored by Fermi Research Alliance, LLC under Contract No. DE-AC02-07CH11359 with the U.S. Department of Energy, Office of Science, Office of High Energy Physics.

\printbibliography[heading=subbibliography]
\end{refsection}
\addtocontents{toc}{\protect\setcounter{tocdepth}{1}}
\begin{refsection}
\chapter{On the reliable lattice-QCD determination of multi-baryon interactions and matrix elements}
\chapterauthor[1,2]{Ra\'ul Brice\~{n}o}
\chapterauthor[3]{Jeremy R. Green}
\chapterauthor[4]{Andrew D. Hanlon}
\chapterauthor[5]{Amy Nicholson}
\chapterauthor[6]{Andr\'e Walker-Loud}

\begin{affils}
\chapteraffil[1]{Thomas Jefferson National Accelerator Facility, Newport News, Virginia 23606,
USA}
\chapteraffil[2]{Department of Physics, Old Dominion University, Norfolk, Virginia 23529, USA}
\chapteraffil[3]{School of Mathematics and Hamilton Mathematics Institute, Trinity College Dublin, Dublin 2, Ireland}
\chapteraffil[4]{Physics Department, Brookhaven National Laboratory,
Upton, New York 11973, USA}
\chapteraffil[5]{Department of Physics and Astronomy, University of North Carolina, Chapel Hill, NC 27516-3255, USA}
\chapteraffil[6]{Nuclear Science Division, Lawrence Berkeley National Laboratory, Berkeley,
CA, 94720, USA}
\end{affils}
\addtocontents{toc}{\protect\setcounter{tocdepth}{-2}}

For about a decade, there has persisted a discrepancy in the literature in which most groups performing calculations of nucleon-nucleon (\NN) systems with lattice QCD (\LQCD) reported the identification of deeply bound di-nucleon systems at pion masses larger than nature~\cite{NPLQCD:2012mex,Yamazaki:2012hi,Orginos:2015aya,Berkowitz:2015eaa,Wagman:2017tmp}, while the HAL QCD Collaboration, utilizing an alternative method known as the HAL QCD potential~\cite{Ishii:2006ec,Aoki:2012tk}, reported 
that the di-nucleon systems do not support bound states at these heavy pion masses~\cite{Inoue:2011ai,Ishii:2012ssm}.
It was initially asserted by many groups that this discrepancy was a sign of unquantified systematic uncertainties in the HAL QCD approach as this method requires additional assumptions beyond those needed for the L\"uscher formalism~\cite{Luscher:1990ux,Rummukainen:1995vs, Briceno:2013lba,Briceno:2014oea}.

However, subsequent work uncovered significant dependence of the extracted spectrum upon the type of local-creation operator used~\cite{Iritani:2016jie}, raising significant concerns on whether or not the previous works correctly determined the \NN spectrum.
The spectrum does not depend upon the creation/annihilation interpolating operators, and so the observation of such dependence is indicative of either a misidentification of the spectrum through ``false plateaus''~\cite{Iritani:2016jie}, or it could be a practical issue that the  operators used couple so poorly to a given state that, at the available finite statistics, one can not numerically resolve the presence of the state through the analysis of the correlation functions.

A shortcoming of all previous works that have identified deeply bound di-nucleons is that they employed asymmetric correlation functions in which the \NN creation and annihilation interpolating operators were not conjugate to each other. In this setup, the overlap factors for the excited states are not guaranteed to have the same sign as for the ground state, making the analysis susceptible to false plateaus.
There are now three independent calculations of two baryon systems which utilize momentum-space creation and annihilation operators leading to Hermitian matrices of correlation functions allowing for a variational approach~\cite{Luscher:1990ck,Blossier:2009kd}:
The Mainz group has computed the $H$ dibaryon and di-neutron systems~\cite{Francis:2018qch} using the distillation method~\cite{HadronSpectrum:2009krc};
The sLapHnn Collaboation has computed the di-nucleon systems~\cite{Horz:2020zvv} using the stochastic Laplacian Heaviside method~\cite{Morningstar:2011ka};
The NPLQCD Collaboation has computed the di-nucleon systems~\cite{Amarasinghe:2021lqa} using a momentum-sparsening method~\cite{Detmold:2019fbk}.
None of these newer works have identified deeply bound di-nucleon systems, including the calculation by NPLQCD which included both momentum-space and local hexaquark interpolating fields in the linearly-independent set of operators.

Resolving the nature of the \NN systems at heavy pion mass is critical for the application of \LQCD to nuclear physics.
It is both a test of the underlying physics, and presently, it is more importantly a test of our ability to perform the calculations with fully quantified theoretical uncertainties.
For example, if it is determined that calculations which utilize local hexaquark creation operators lead to a misidentification of the spectrum, all subsequent calculations of two and more nucleon matrix elements that utilize such a set of creation operators will have unquantified corrections.

A clear picture has emerged: the only calculations which identify deeply bound di-nucleons are those that utilize local hexaquark creation operators and momentum-space annihilation operators.  Can we understand more quantitatively why these sets of correlation functions indicate deeply bound di-nucleons?  HAL QCD has suggested they emerge as a false plateau generated by a linear combination of elastic \NN scattering states, with differing signs for the overlap factors~\cite{Iritani:2018vfn}.  This can be tested with a matrix of correlation functions including both momentum-space and hexaquark operators.

Another recent troubling result is from the Mainz-group calculation of the $H$ dibaryon in the $SU(3)$-flavor symmetric limit, utilizing six lattice spacings, from which they observed very large discretization corrections to the binding energy~\cite{Green:2021qol}.
In contrast, important discretization effects in two-meson systems, which are generally more precisely computed, have so far not been observed.
Therefore, the observation of such corrections in this dibaryon system raises several questions.  Can this be confirmed from independent calculations?  Is it specific to the lattice action used?  Is it unique to the $H$ dibaryon or a general feature of dibaryon interactions?  With such large discretization corrections, does one need to consider discretization corrections to the L\"uscher quantization condition?

In the following, we comment on these and other issues related to \LQCD calculations of two-baryon systems.  We provide some suggestions how to further elucidate some of the perplexing issues that have arisen in the literature, and how the field can make progress.
The first step in making progress is to reliably determine the two-baryon spectrum, prior to moving on to more baryons and/or their matrix elements.

\section{Reliable Spectral Analysis}

For reliable conclusions to be drawn from the L\"uscher finite-volume formalism, it is essential
that the energies input into the quantization condition are accurate.
There are three major challenges to obtaining accurate energies:
\begin{itemize}
\item At early Euclidean time, the correlation functions are contaminated by excited states;

\item The lowest excited state gap in the two-baryon system is given by elastic scattering modes, which have an energy gap corresponding roughly to $p^2/M\approx(2\pi/L)^2/M\approx20-50$~MeV for typical values of $L$ used in present calculations.  The time scale for these excited states to decay is given by the inverse energy gap, which corresponds to $4-10$~fm;

\item Empirically, it is observed that the noise of two-baryon systems overwhelms the signal before $t\approx2$~fm, for calculations at larger-than-physical pion mass.  The exponential degradation of the signal becomes worse as the pion mass is reduced towards the physical point, which for \NN systems scales as $e^{-2(M_N-\frac{3}{2}m_\pi)t}$ at asymptotically large times.

\end{itemize}
Given the results in the literature, the only promising strategy to overcome these challenges is to use a variational method to extract the spectrum~\cite{Luscher:1990ck,Blossier:2009kd} and then use the L\"uscher quantization condition~\cite{Luscher:1990ux,Rummukainen:1995vs, Briceno:2013lba,Briceno:2014oea} to provide important diagnostics on the consistency of the extracted spectrum~\cite{Wilson:2015dqa,Iritani:2017rlk}.
Such a strategy has been used successfully in many studies of two-meson systems, which has become the standard tool there, see the review~\cite{Briceno:2017max} and references therein.

The variational method involves forming a Hermitian matrix of correlation functions with elements defined as
\begin{equation}
  C_{ij} (t) \equiv \langle \mathcal{O}_i (t + t_0) \mathcal{O}_j^\dagger (t_0) \rangle ,
\end{equation}
using a set of $N$ linearly independent operators $\left\{\mathcal{O}_i\right\}$ that, ideally, have
strong overlap with the states that one wants control over.
For example, a baryon-baryon operator could be a linear combination of objects having the form
\begin{equation}
    \mathcal{O}_{BB}(t,\vec P) = \sum_{\vec x,\vec y}
    e^{-i\vec p_1\cdot\vec x}e^{-i(\vec P-\vec p_1)\cdot\vec y}
    (qqq)(t,\vec x)(qqq)(t,\vec y),
\end{equation}
corresponding to the momentum-space operators discussed above,
whereas a hexaquark operator has the structure
\begin{equation}
    \mathcal{O}_H(t,\vec P) = \sum_{\vec x} e^{-i\vec P\cdot\vec x}
    (qqqqqq)(t,\vec x).
\end{equation}
The finite-volume spectrum can be obtained by solving the generalized eigenvalue problem (GEVP) on the resultant correlator matrix
\begin{equation}
  C(t) \upsilon_n (t, \tau_0) = \lambda_n (t, \tau_0) C(\tau_0) \upsilon_n (t, \tau_0),
\end{equation}
whose eigenvalues give
\begin{equation}
  \lambda_n(t, \tau_0) = |A_n|^2 e^{-E_n (t - \tau_0)}\big[1 + O (e^{-\Delta_n t}) \big] ,
\end{equation}
where $E_n$ is the energy of the $n$th eigenstate in the system and $n = 0, \ldots , N - 1$.
Thus, the lowest $N$ eigenstates that have overlap with the chosen set of operators can readily be
extracted from these generalized eigenvalues. But, given the practical limitations on the size of $t$
due to the exponentially bad signal-to-noise ratio, the reliability of this extraction depends strongly
on the size of the gap $\Delta_n$. It has been shown that solving the GEVP with $\tau_0 \ge t/2$
leads to a gap of $\Delta_n = E_N - E_n$~\cite{Blossier:2009kd},
which removes the contribution from all states with $E_m \neq E_n < E_N$ from $\lambda_n(t, \tau_0)$.
This is in contrast to solving for the eigenvalues of $C(t)$ directly, in which case the gap
is in general given by $\Delta_n = \min_{m \neq n} |E_n - E_m|$~\cite{Luscher:1990ck}, and thus does not
help in controlling the contamination from different states.
Hence, by using the GEVP, the gap can be made arbitrarily large by including more operators in the correlator matrix that couple well to the relevant states.

The GEVP is also amenable to self-consistency checks by varying the operators used
in the correlator matrix and observing how the resulting spectrum is affected (e.g. see Ref.~\cite{Amarasinghe:2021lqa} for a recent investigation).
This can help to determine operators that are irrelevant, or, more importantly, essential.
As stated above, further consistency checks can be made by utilizing the quantization condition to look
for any inconsistent behavior in the phase shift coming from the energies extracted from the GEVP.
The resulting phase shift can also be used to predict the energy spectrum and look for any missing energies.
Thus, the GEVP method in combination with the finite-volume L\"uscher formalism is a powerful method
for validating the extracted spectrum.

Pionless EFT indicates that regardless of whether a deep bound state exists in the system or not, a modest variational basis containing only momentum-space operators is sufficient to correctly determine the spectrum, and the inclusion of a hexaquark operator does not improve the convergence~\cite{Amy_proc}, (see also \cite{Francis:2018qch}). 
One lattice calculation found that including a hexaquark operator gave rise to an additional energy level well above threshold, without affecting any other levels~\cite{Amarasinghe:2021lqa}. 
Such a state, if it exists, would have to be a very narrow resonance that is weakly coupled to the \NN system such that it leaves an otherwise imperceptible imprint on the nearby spectrum and the resulting phase shift.
It is important to verify the validity of this state to further understand these strongly interacting systems.

All applications of the variational method to two-baryon systems at larger-than-physical pion masses
either strongly disfavor a bound \NN state~\cite{Horz:2020zvv} or are inconclusive in this regard~\cite{Francis:2018qch,Amarasinghe:2021lqa}.
Direct comparison, on the same ensemble, of variational results to those using the asymmetric correlator setup described above
show inconsistencies in the extracted phase shifts, further providing evidence that these early studies were affected by
uncontrolled excited-state contamination.

The recent switch to utilizing the variational method in two-baryon systems has resulted in great strides
toward resolving the two-baryon controversy in the literature. But, more work is certainly needed.
For example, it will be illuminating to compare the phase shifts determined on the same configurations from the HAL QCD potential method and the variational methods.
Thus a shift to controlling
other sources of systematic error may be the next hurdle for obtaining reliable estimates
of two-baryon observables. We discuss first steps toward this in the next section.

\section{Quantization condition and discretization effects\label{sec:qc}}
Until recently, every lattice calculation of baryon-baryon
interactions was done using a single lattice spacing $a$. This was
based on the assumption that discretization effects mostly cancel when
taking the difference between baryon-baryon energy levels and the sum
of two single-baryon energy levels~\cite{NPLQCD:2010ocs}. In
Ref.~\cite{Green:2021qol}, two of us studied the $H$ dibaryon for a
fixed choice of quark masses using six lattice spacings and found very
large discretization effects: the binding energy in the continuum
limit was 4.6(1.3)~MeV, whereas on the coarsest lattice spacing it was
above 30~MeV. Given this first result in a single physical system, it
will be important to check other systems such as \NN systems to
understand whether large discretization effects are common; work in
this direction is in progress~\cite{Green_Lat21}. In addition, since
discretization effects are not universal, it will be worthwhile to
also perform studies using different lattice actions. Some input from
EFTs or toy models could help in understanding these effects and
whether any lattice action should be preferred.
It will also be interesting and important to understand why they may be relevant for two-baryons but do not seem nearly as relevant for two-meson systems.

If large discretization effects are widespread, this implies that
many previous calculations may also contain 
large systematic errors. In the future, it will be
important to perform calculations using multiple lattice spacings or a
single lattice spacing that is finer than has typically been used in
the past.

\subsection{Applying quantization conditions at nonzero lattice spacing}

A now-standard approach for studying multihadron interactions with
\LQCD is to use finite-volume quantization conditions, which
relate the scattering amplitude to the finite-volume
spectrum~\cite{Luscher:1990ux, Rummukainen:1995vs, Briceno:2013lba,
  Briceno:2014oea}. Given that these conditions have been derived in
the continuum, a natural question is how best to analyze a spectrum
computed at nonzero lattice spacing. Two possible strategies are
illustrated in Fig.~\ref{fig:paths}.

\begin{figure}
  \centering
    \begin{tikzpicture}[
    shorten < =  1mm, shorten > = 1mm,
node distance = 50mm, on grid, auto,
every path/.style = {-Latex,ultra thick},
sx+/.style = {xshift=1 mm},
sy+/.style = {yshift=1 mm},
sx-/.style = {xshift=-1 mm},
sy-/.style = {yshift=-1 mm},
p1/.append style = {draw=red},
state/.append style = {rectangle,draw=none},
                    ]
\node[state] (A) {$E(L,a)$};
\node[state] (B) [right=of A] {$\delta(p^2,a)$};
\node[state] (C) [below=30mm of A] {$E(L)$};
\node[state] (D) [right=of C] {$\delta(p^2)$};
\path[->,blue] %
 (A) edge node[above,black,align=left] {lattice\\quantization?} (B)
 (B) edge node[right,black] {$a\to 0$} (D);
\path[->,red!70!black] %
 (A) edge node[left,black] {$a\to 0$} (C)
 (C) edge node[below,black,align=left] {continuum\\quantization} (D);
\end{tikzpicture}
\caption{Two paths, red and blue, from the lattice finite-volume
  energy levels $E(L,a)$ to the continuum phase shift $\delta(p^2)$.}
  \label{fig:paths}
\end{figure}
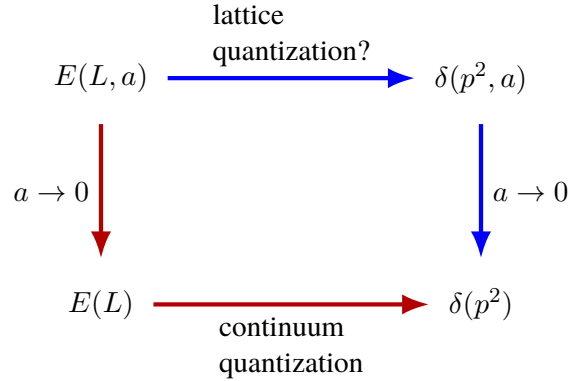

The most theoretically clean approach is to follow the red path, by
first performing continuum extrapolations at fixed volume to obtain
the continuum finite-volume spectrum, then analyzing it using standard
quantization conditions. However, the corresponding lattice
calculations are challenging, since they require matched volumes at
different lattice spacings.

Alternatively, one could follow the blue path, using quantization
conditions to obtain scattering amplitudes at finite lattice spacing
and then extrapolating those to the continuum. Two simplifying
assumptions can be made: the continuum quantization condition can be
applied to data at nonzero lattice spacing, and  the scattering
amplitude at nonzero lattice spacing has the same structure as the
continuum one. This is the main approach used in
Ref.~\cite{Green:2021qol}. Along with a fit ansatz in which
$p\cot\delta_0(p)$ is given by a polynomial in $p^2$ whose
coefficients are affine functions of $a^2$, this approach produced a
good description of the lattice spectra.

A better understanding of discretization effects could help to put the
strategy of following the blue path on theoretically more solid
ground. It would be beneficial to derive a quantization condition that accommodates at least the leading discretization effects. This would, of course, have to incorporate the $\mathrm{O}(a^2)$ discretization effects that reduce the $O(4)$ Euclidean-rotational group down to the hypercubic group.
At this stage it is unclear if these effects could be accounted for using a universal framework or if it is necessary to resort to a specific EFT evaluated to a finite order. First steps in this direction were performed in Ref.~\cite{Korber:2019cuq} for a simple theory.

\subsection{Energy levels on left-hand cuts}

Here we note another important direction of investigation for
the two-particle quantization conditions. It is well known that the standard quantization conditions
break down above certain inelastic thresholds, with considerable
progress being made in recent years to understand quantization
conditions above multi-channel and three-particle thresholds (see Refs.~\cite{Briceno:2017max, Hansen:2019nir} for recent reviews). However, they also fail
for energies that overlap with left-hand cuts that occur below the
lowest threshold. One can see this in the simplest case where the
quantization condition is truncated to $S$-wave:
\begin{equation}
  p\cot\delta_0(p) = \frac{2}{\sqrt{\pi}L\gamma}Z_{00}^{\vec PL/(2\pi)}\left(
    1,\left(\frac{pL}{2\pi}\right)^2\right).
    \label{eq:QC}
\end{equation}
Here, $Z_{00}^{\vec D}$ is a generalized zeta function that is real for
real $p^2$. Below the start of the left-hand cut, $p\cot\delta_0(p)$
is generically complex; as a result, the equation has no solutions.

On the other hand, lattice energy levels below the start of the
left-hand cut have now been observed~\cite{Green:2021qol}.
For SU(3) singlet baryon-baryon
scattering relevant for the $H$ dibaryon, the first left-hand cut is
caused by $t$-channel exchange of a pseudoscalar octet meson. On all
but one of the ensembles in Ref.~\cite{Green:2021qol}, the ground state in the rest
frame lies below the start of the $t$-channel cut; these levels were discarded
from the analysis. For nucleon-nucleon scattering at the physical pion
mass, the $t$-channel cut starts about 5~MeV below threshold; na\"ive
applications of quantization conditions to the scattering amplitude in
the deuteron sector (using models that do not contain a $t$-channel
cut) predict that the ground state in the rest frame will lie this far
below threshold when $L<8$~fm~\cite{Briceno:2013bda, Briceno:2019nns}.

It would be valuable to have quantization conditions that are valid on
left-hand cuts, which could provide subthreshold information on the scattering amplitude. In
fact, the recently-proposed method of Ref.~\cite{Meng:2021uhz} might
already be applicable to some of these challenges. At this point, it is not clear if one could cast such a formalism in a universal form that may be applicable for arbitrary channels.

\section{Two-body matrix elements}

Ultimately, the spectrum of the \NN system serves as a first step towards the determination of more physically interesting quantities, including electroweak elastic and transition form factors and QCD contributions to processes that may provide smoking guns of BSM physics (e.g. neutrinoless double-beta decay). Many of these reactions may be constrained via the evaluation of matrix elements of external currents. With the aim of rigorously determining these matrix elements via \LQCD, there has been significant progress towards providing a non-perturbative connection between finite- and infinite-volume few-body matrix elements~\cite{Briceno:2015tza, Baroni:2018iau}. If we consider the simplest system, where the particles carry no relative angular momentum or intrinsic spin, the relationship between a matrix element for a local scalar current ($\mathcal{J}$) can be compactly written as~\cite{Briceno:2015tza, Baroni:2018iau} 
\begin{align}
L^{3} \langle{P_{f},L}| \mathcal{J}(x=0) |{P_{i},L}\rangle = 
\bigg(
\mathcal{W}_{\rm df }(P_{f}, P_{i}) 
+      \mathcal{M}(P_f^2) \cdot  G(P_{f}, P_{i}, L) \cdot \mathcal{M}(P_i^2)
\bigg)
 \, \sqrt{ \mathcal{R}(P_{f},L) \mathcal{R}(P_{i},L) },
 \label{eq:ME_FV}
\end{align}
where $G$ is a new finite-volume function that is closely related to $Z_{00}$ appearing in Eq.~\eqref{eq:QC}, $\mathcal{R}$ is the so-called Lellouch-L\"uscher factor~\cite{Lellouch:2000pv, Briceno:2014uqa}, $\mathcal{M}$ is the purely hadronic two-body amplitude, and $\mathcal{W}_{\rm df}$ is the desired infinite-volume matrix element~\footnote{The subscript $\rm df$ stands for ``divergence-free'', referring to the fact that simple pole singularities have been removed.}. Although this formalism has not yet been implemented in a \LQCD calculation, important checks have been performed on it, including the consistencies with perturbation theory, the Feynman-Hellmann theorem, and charge conservation~\cite{Briceno:2019nns, Briceno:2020xxs}. 

It is worth emphasizing that $\mathcal{R}$ requires the evaluation of the derivative of the scattering amplitude and the $Z_{00}$ functions with respect to energy. The derivative of the scattering amplitude is not directly accessible via \LQCD, and as a result one must resort to parametrizations of the amplitudes. This, and the additional explicit dependence on $\mathcal{M}$ in Eq.~\eqref{eq:ME_FV}, points to the fact that in order to obtain infinite-volume matrix elements of two-particle states, one requires a tight and accurate constraint on the two-body spectrum and subsequently the two-body scattering amplitudes.

As already mentioned, this formalism has not been implemented in the study of \NN matrix elements. Instead, the published results~\cite{Beane:2015yha, Savage:2016kon, Chang:2017eiq} have restricted their attention to systems that support bound states and relied on the fact that matrix elements of two-body bound states have exponentially suppressed finite-volume effects~\cite{Briceno:2019nns}. If there are no bound states, as indicated by improved spectroscopy calculations, then this approach for avoiding Eq.~\eqref{eq:ME_FV} is not valid. Furthermore, any uncontrolled systematic errors in the spectra will propagate into errors in $\mathcal{M}$ and $\mathcal{R}$, which can be large. In addition, generally the finite-volume matrix element on the left-hand side of Eq.~\eqref{eq:ME_FV} cannot be reliably isolated in a regime where the energy of the corresponding finite-volume state has not been reliably isolated. Although it has not been definitively demonstrated at this point whether the published results for the \NN matrix elements are contaminated by these uncontrolled systematics, this will need further investigation, and these calculations will need to be done using variationally optimized operators as done in, for example, Ref.~\cite{Shultz:2015pfa} for excited mesonic states.

For shallow bound states, like the physical deuteron, finite-volume effects can not be ignored. Instead, one will need to use multiple volumes and/or total momenta to scan the pole region of $\mathcal{W}_{\rm df}$. At the bound-state, this amplitude acquires an energy-dependent pole associated with the initial and final state coupling to the bound state~\cite{Briceno:2020vgp}. And from the residue of this, one can access the form factors of such a state.

To study the matrix elements of \NN states, it will be necessary to generalize the formalism presented in Refs.~\cite{Briceno:2019nns, Briceno:2020xxs} for systems with non-zero intrinsic spin. Finally, as previously emphasized, the formalism discussed only supports currents that are local in time. In other words, it would not be suitable in, for example, the study of Compton scattering or double beta decays. Such observables would need extensions to accommodate the insertion of two currents separated by arbitrary time. Efforts along this track are under way~\cite{Feng:2020nqj,Briceno:2019opb,Davoudi:2020xdv,Davoudi:2021noh, Davoudi:2020gxs}.

 \section{States coupling to three or more nucleons}
 Among the more exciting prospects of the few-nucleon \LQCD program is the possibility of constraining three-nucleon dynamics directly from QCD. The procedure for this follows closely that of the two-body sector. Namely, presently the most rigorous pathway requires the accurate determination of the finite-volume spectra of states with the quantum number of three-nucleons using a large list of operators. The spectrum would then need to be analysed with the extensions of the quantization condition for three particles in order to then constrain the infinite-volume amplitudes. Once the amplitudes have been constrained, these could be analytically continued below threshold to determine the location and residues of possible bound state poles. 
 
 These studies are significantly more challenging than their two-body analogues for multiple reasons, including
 \begin{itemize}
  \item there is a larger density of states, 
  \item for systems with two-body bound states, there can be multiple thresholds, 
  \item the number of allowed contractions generally grows with the number of hadrons,
  \item the numerical cost of evaluating each contraction is generally larger,
  \item the stochastic noise grows with the number of nucleons,
\item the quantization condition is more complicated to derive and implement, 
 \item the quantization condition depends on one, two, and three-body observables, 
 \item the infinite-volume amplitudes have a larger class of singularities.  \end{itemize}
 The first five items point to the fact that determination of finite-volume spectra is generally more complicated, while the last set of items refer to the fact that their analysis is also more challenging.

This program hinges on a non-perturbative formalism to relate the finite-volume spectra and the desired infinite-volume amplitude. There has been significant progress towards this goal~\cite{Hansen:2015zga,Mai:2017bge,Briceno:2012rv,Hammer:2017kms,Hansen:2019nir}. Although these formalisms have not been implemented in the study of nuclear states, they have been successfully implemented in the mesonic sector~\cite{Horz:2019rrn,Blanton:2021llb,Mai:2018djl,Blanton:2019vdk,Culver:2019vvu,Hansen:2020otl}, where calculations are computationally more affordable.

While the formalism for three-nucleon systems in finite volume does not yet exist, there are published results for the three (and four) baryon spectra~\cite{NPLQCD:2012mex, Yamazaki:2012hi}.
There is now evidence that the \NN (and generally, two-baryon) spectra for these results have significant, unaccounted-for systematic errors, as the NPLQCD collaboration first published $B_{NN}\approx20$~MeV binding energies for both di-nucleon systems~\cite{NPLQCD:2012mex,Wagman:2017tmp}, while in their most recent work using momentum-space creation operators, they do not find evidence for deep bound states~\cite{Amarasinghe:2021lqa}.  This is suggestive of an $\mathrm{O}(20 \textrm{ MeV})$ systematic uncertainty on the \NN binding energy.
Ref.~\cite{NPLQCD:2012mex} also quoted a $^3$He binding energy of $B\approx50$~MeV, which is 30 MeV from the quoted proton-deuteron breakup threshold.
Assuming a 20 MeV systematic uncertainty on the deuteron binding energy, it is not unreasonable to also assume a similar or larger uncertainty on the gap from the three-nucleon state to this first open threshold.
This is of the same order of the systematic for the two-body sector, which weakens the claim that a bound ${{}^3\rm He}$ was found for these quark masses. As a result, it will be necessary to do a variational analysis of the spectrum using a larger list of interpolating operators before concluding that ${}^3\rm He$ is indeed bound for these larger pion masses in the range $m_\pi\sim 300-800$~MeV.

In preparation for such studies, the formalism continues being developed~\cite{Blanton:2021mih, Blanton:2020gha, Hansen:2020zhy, Briceno:2017tce} to allow for increasingly complex three-body systems. In parallel to these efforts, toy-model investigation have been continued, exploring nuclear-like theories which support two- and three-body bound states in a finite- and infinite-volume~\cite{Romero-Lopez:2019qrt,Jackura:2020bsk}. Ultimately, these formalisms will need to be extended to accommodate systems with intrinsic spin.

Although calculations of correlation functions with four and more nucleons have been performed, they have used a single local creation operator. 
The challenges discussed above for the \NN sector are expected to be more difficult for four and more nucleons, and the corresponding calculation will be numerically more expensive. Given these two points and the lack of existing formalism to test the validity of the resultant lattice spectra, 
robust investigations of systems composed of four or more nucleons will have to wait.

\section*{Acknowledgements}

JRG acknowledges support from the Simons Foundation through the Simons Bridge for Postdoctoral Fellowships scheme.
ADH is supported by the U.S. Department of Energy, Office of Science,
Office of Nuclear Physics through the Contract No. DE-SC0012704 and
within the framework of Scientific Discovery through Advance Computing
(SciDAC) award ``Computing the Properties of Matter with Leadership
Computing Resources.''
RAB is supported in part by U.S. Department of Energy Contract No. DE-AC05-06OR23177, 
under which Jefferson Science Associates, LLC, manages and operates Jefferson Lab, and is partly supported by the U.S. Department of Energy Contract No. DE-SC0019229. 
ANN is supported by the U.S. National Science Foundation CAREER Award PHY-2047185.  
The work of AWL is supported in part by U.S. Department of Energy, Office of Science, Office of Nuclear Physics under Awards No. DE-AC02-05CH11231.

\printbibliography[heading=subbibliography]
\end{refsection}
\addtocontents{toc}{\protect\setcounter{tocdepth}{1}}
\begin{refsection}
\chapter{Entanglement in nuclear structure}
\chapterauthor[1,2]{Caroline Robin}
\begin{affils}
\chapteraffil[1]{Fakult\"at f\"ur Physik, Universit\"at Bielefeld, D-33615, Bielefeld, Germany}
\chapteraffil[2]{GSI Helmholtzzentrum f\"ur Schwerionenforschung, Planckstra\ss e 1, 64291 Darmstadt, Germany}
\end{affils}
\addtocontents{toc}{\protect\setcounter{tocdepth}{-2}}

As prime examples of quantum many-body systems, atomic nuclei exhibit non-classical correlations, among which, entanglement is certainly the most fascinating.
This peculiar phenomenon, inherent to quantum mechanics, allows particles that have interacted in some way, to remain correlated even when separated by arbitrarily large distances. 
Such non-local correlations play an important role in quantum communication and are the essence of quantum computing.
\\
\\
In order to characterize entanglement in tightly bound many-body systems, where particles are separated by short distances and have overlapping wave functions, one must consider the distinguishable or indistinguishable character of the components. While there appears to be a consensus on the definition of entanglement between distinguishable particles, an extension of this concept to systems of identical particles is difficult and is still subject to debate \cite{Amico:2007ag}.
The issue comes from the fact that, because of their indistinguishability, single components cannot be accessed individually. It is thus unclear how to trace over one subsystem and determine reduced density matrices, which are the key elements to quantify entanglement.
\\
\\
Although possible treatments of two-identical-particle systems have been investigated \cite{ECKERT200288}, how to address particle entanglement in larger systems is not straightforward, and several points of view on the notion of entanglement itself, as well as its characterization, have been developed in the past years \cite{BENATTI20201}.
One possible way around this issue has been to consider the Fock space formulation of the many-body state and evaluate entanglement between modes rather than between particles. In this case, the subsystems are formed by groups of distinguishable single-particle states, or orbitals, so that the total Hilbert space has the required tensor product structure allowing for partial traces and calculation of entanglement measures.
\\
\\
Below we summarize different ways that have been explored to characterize entanglement in the structure of atomic nuclei, and discuss how these studies can not only lead to efficient ways of treating quantum correlations on classical computers, but can also provide valuable insights in order to design quantum computations of nuclei.
Finally we discuss the possible fundamental role of entanglement in the description of nuclei and nuclear forces, and mention some problems and questions to be addressed in the future.
\section{Entanglement in nuclear structure calculations}
The presence of entanglement is the reason why calculating nuclei, and in general quantum many-body systems, on classical computers is so hard. 
As systems with large entanglement cannot be well approximated by separable states they do not possess an efficient classical representation, thus leading to the exponential scaling of the required resources with the number of particles.
In this context, careful investigations and possible manipulations 
of entanglement structures can allow for more efficient calculation schemes. This idea is exploited by methods such as density-matrix renormalization group (DMRG) or tensor networks largely used in other fields, and more recently being developed in nuclear physics \cite{Papenbrock:2003bj,Papenbrock:2003az,White:1992zz,Dukelsky:2002nw,Dukelsky:2004vv,Thakur:2008ct,Rotureau:2006rf,Rotureau:2008rp,Fossez:2021xrx,Zhu:2021pis,TICHAI2021136623}.
As nuclei are composed of two particle species ($Z$ protons and $N$ neutrons), various partitioning of the nuclear state can be defined, for investigation of different forms of entanglement.
For example, Refs. \cite{Papenbrock:2003bj,Papenbrock:2003az,Papenbrock_2005,Gorton:2018tj} made use of the natural bi-partitioning of the nuclear state to study entanglement between neutron and proton subsystems. Analyses of the singular-value decomposition of shell-model ground states showed an exponential fall off of the eigenvalues of the reduced density matrix (singular values), in particular in spherical nuclei with $N>Z$, so that nuclear states could be represented by only a few (correlated) proton and neutron states.
A few investigations of mode correlations and entanglement have now also been performed in the Lipkin model \cite{Faba:2021kop}, two-nucleon \cite{Kruppa:2020rfa} and many-nucleon systems \cite{Legeza:2015fja,Robin:2020aeh}. 
For example, Ref. \cite{Robin:2020aeh} analyzed entanglement properties of several single-particle bases within the ground state of Helium isotopes, and found a clear link between convergence of the ground-state energy (with respect to the size of the model space) and entanglement structures.
In particular, natural orbitals derived from a variational principle, which displayed the fastest convergence of the energy, exhibited much more localized structures of entanglement within the basis, as compared to, for example, harmonic oscillator or Hartree-Fock orbitals, and minimized the total entanglement content of the nuclear ground state.
Measures of entanglement and correlations appeared compressed around the Fermi level, and showed that this basis also effectively decouples the active and inactive single-particle spaces.
Analysis of the two-nucleon mutual information in $^6$He showed that the transformation of single-particle orbitals led to an emergent picture of two interacting $p$-shell neutrons decoupling from an $^4$He core, thus driving the wave function to a core-valence tensor product structure. 
While this study was an {\it a posteriori} investigation of entanglement from no-core configuration-interaction calculations, one could use the structured and minimized entanglement patterns of the variational natural basis to design more efficient calculation schemes.
One intuitive future step in this direction would be to combine such orbital optimization with DMRG, as has already been explored in quantum chemistry \cite{doi:10.1063/1.2883980,MCSCF-DMRG2}.
\section{Entanglement to guide quantum computations of nuclei}
While tremendous progress has been and can still be made in the classical computation of nuclei, quantum computers in principle offer a much more natural way to solve the quantum many-body problem \cite{Feynman:1981tf} and, ultimately, could allow for exact treatments of systems beyond the limits of what could ever be achieved on classical machines. 
Thus, in the past decade a huge effort has been deployed to develop many-body quantum computations, and first proof-of-principle calculations of few-nucleon systems have been performed on quantum devices \cite{Dumitrescu:2018njn,Holland:2019zju,Lu:2018pjk,Stetcu:2021cbj,Cervia:2020fkk}. 
These pioneering studies have so far been limited to very few qubits. Developments of quantum computers are however fast progressing and one can expect that machines with several hundred or thousand qubits will become available in the very near future. To take advantage of these developments, it is of the utmost importance to develop clever algorithms which would limit the error rate to the best extent.
In this respect, understanding the entanglement structure of the system that will be mapped on the quantum machine is crucial, and a careful organization and possible minimization of such structures could possibly allow for smaller numbers of entangling operations on the device, thus potentially leading to an expansion of the reach of quantum calculations.
For example, studies of mode entanglement can be particularly useful to develop algorithms that map modes to qubits. In this context, the natural localization of entanglement into decoupled subspaces that is provided by the natural variational basis could be exploited in order to design hybrid classical-quantum algorithms on present and near-term devices possessing limited connectivity. In particular, the weakly-entangled subspaces could be treated classically while the strongly-entangled part of the Hilbert space would be handled by the quantum device. 
\section{Discussion and questions to address in the future}
The investigation of entanglement in atomic nuclei is overall a rather new and thus exciting line of investigation.
Beyond the computational advantages that could bring an entanglement-based description of nuclear systems, several studies now point out to the fact that entanglement organization and minimization could in fact be fundamental to the description of matter \cite{Klco:2021lap}.
In particular, it was revealed that entanglement suppression is connected to emergent symmetries of the strong interaction at low energy, suggesting that entanglement may be a basic notion related the hierarchy of nuclear forces and could characterize a new power-counting scheme \cite{Beane:2018oxh,Beane:2021zvo}.
Recently, singular-value decompositions of two-nucleon interactions showed that low-rank truncations can be safely applied to non-local potentials, and that the singular-value content is mostly maintained during similarity renormalization group (SRG) evolution of these potentials \cite{Zhu:2021pis,TICHAI2021136623}, suggesting that the entanglement minimization could be preserved in the renormalization flow.
Overall it could be that entanglement minimization is the signal of a 
relevant description for a given energy scale. 
From the aforementioned nuclear structure studies, it seems that this could also be manifest at the many-body level, although more investigations are needed to confirm this statement. 
In particular, it would be interesting to investigate if, similarly to the transition from QCD to nucleons \cite{Beane:2018oxh}, an entanglement suppression also appears when transiting from a description in terms of interacting nucleons to a regime where collective vibrations or rotations become relevant as degrees of freedom. 
\\
\indent Overall, the use of entanglement as driving principle for the development of nuclear forces and many-body methods appears as a promising path to keep exploring. In this context, re-interpretation of existing techniques from an entanglement point of view can also be enlightening. In principle density functional theory tells us that the exact energy of the interacting system can be obtained from a single Slater determinant (SD) \cite{Kohn:1965zzb}, which, by definition, is separable and thus, has no entanglement (beyond anti-symmetrization). The situation is similar in the in-medium SRG method \cite{Tsukiyama:2010rj} which shifts the complexity of the nuclear state to the nuclear Hamiltonian via continuous unitary transformations of the latter. The exact energy (and other observables) can then also be obtained from a SD. On the other hand, these separable states do not characterize the exact wave function of the system. Thus it may not be straightforward to quantify entanglement directly in such approaches and one may need to elaborate different ways to characterize it.
\\
\indent Several other fundamental problems as well as possible applications are to be explored in the future. These include studies of various forms of entanglement, such as bi-partite and multi-partite entanglement, in ground and excited states of diverse types of nuclear systems, from light to mid-mass and heavy nuclei, both near and far from stability. Such works would lead us towards a broader and deeper understanding of entanglement in nuclei and its evolution along the nuclear chart, and could shed light on relations to symmetry breaking and phase transitions, as well as possible links with emergence of new degrees of freedom.
\\
\indent As a more conceptual issue, how to characterize entanglement between individual nucleons (as opposed to modes) in a way that would be independent of the basis should also be clarified. This type of entanglement could potentially provide better insight into physical phenomena such as pairing or clustering, and could perhaps reveal possible experimental signatures of entanglement in nuclei.

\printbibliography[heading=subbibliography]
\end{refsection}
\addtocontents{toc}{\protect\setcounter{tocdepth}{1}}
\setcounter{footnote}{0} 
\begin{refsection}
\chapter{A Perspective on Quantum Information and Quantum Computing for Nuclear Physics}
\chapterauthor[1]{Martin Savage}
\begin{affils}
\chapteraffil[1]{InQubator for Quantum Simulation (IQuS), Department of Physics, University of Washington, Seattle, WA 98195}
\end{affils}
\addtocontents{toc}{\protect\setcounter{tocdepth}{-2}}

This is my contribution to the Panel Discussion on 6 May 2021 in the INT workshop INT 21-1b 
related to quantum information sciences (QIS) for low-energy nuclear physics (NP).   
Each Panelist provided initial comments to start the discussion 
(with topics that we distributed among ourselves to avoid duplication).  
The following text is an approximate transcription
of my remarks, regarding {\it things to keep in mind} when considering
how to go about effectively transferring QIS techniques into and out of NP (theory) research. 
My remarks were mainly reflective in nature, outlining some of the means by which
modern quantum field theory (QFT),
the Standard Model, effective field theory (EFT), quantum chromodynamics (QCD) and lattice QCD 
(\LQCD) techniques and technologies became integrated into the NP community.
\\

Given the multi-disciplinary nature of QIS research, and the many points of connection with NP research, 
it is helpful to understand potential paths for a degree of 
integrating of research in QIS and nuclear theory at the interface to create an effective, robust and mutually beneficial research program.
To provide some insight and guidance about handling the integration of relevant components of QIS research into NP and other domain sciences, and vice versa,
it is worth reflecting on the integration of QFT, the standard model, EFT, QCD and \LQCD techniques and technologies into the NP community.
While QCD was discovered in the early 1970s, and \LQCD soon thereafter, 
its theoretical footprint essentially remained in the domain of high-energy physics (HEP) for more than 15 years.  This was in part because the success of perturbative QCD in systematically describing electroweak processes using EFTs and the RG, and the lack of direct impact on NP at that time beyond hadronic modeling.
Despite a rapidly growing experimental program probing QCD in NP, QFT was not universally considered central to nuclear-theory research even well into the 1990s.  Generally, the existing theoretical tools were integrated into the NP portfolio by 
a modest number of NP theorists re-aligning their research efforts and re-tooling (to some extent) and by 
hiring early-career scientists with PhDs in particle theory, particularly phenomenology, with interest in low-energy problems and electroweak processes.  This was a remarkably successful recruitment, and coincided with one of the swings toward string theory in particle theory providing a significant pool of talent for NP to recruit from,
and has 
contributed in part to present-day cutting-edge nuclear-theory activities.

\LQCD was somewhat delayed in its migration into NP despite its obvious future role, again due to its significant impact on the HEP experimental program, and the challenges faced in computing the properties of even one nucleon with precision. This situation evolved during the 2000s, with the NP community spawning further single-nucleon and multi-nucleon \LQCD efforts and collaborations, and
hiring a number of junior scientists trained in \LQCD by HEP, to utilize the rapidly increasing available classical computing resources.
This was enabled and welcomed by the USQCD collaboration~\footnote{\tt \url{https://www.usqcd.org/}} that represented all \LQCD practitioners in the U.S. and coordinating developments with SciDAC~\footnote{\tt \url{https://www.scidac.gov/}} funding and HPC~\footnote{\tt \url{https://www.usqcd.org/lqcd/WBS/}}.

In addition to increased funding for efforts in new directions in local research groups at universities and national laboratories, national summer schools and conferences, 
major community-driven vehicles to enable a deeper integration of new ideas and concepts into and out of nuclear theory were established around 1990, with the creation of the Institute for Nuclear Theory (INT)~\footnote{\tt \url{https://sites.google.com/uw.edu/int/home}} in Seattle, USA, embedded in a university physics department,
followed a few years later by the European Center for Theoretical Nuclear Physics (ECT*)~\footnote{\tt \url{https://www.ectstar.eu/}} in Trento, Italy connected with their  physics department.
This deliberate co-location provides a  ``low potential barrier'' to engaging graduate students and postdocs in an immersive environment with a large and rotating selection of the world's leading theorists at all career stages.
Currently, ECT* is embracing QIS as an area of importance for future NP research, building upon its prior  efforts in this area and also in HPC.

These discussions and examples of a previously successful integration pathway, provides guidance for considering how to accomplish the analogous 
mutually-beneficial QIS
technique and technology transfer into and out of NP research.   The situation for QIS has important differences, one of them being the major role of technology companies and startups, and the expected growth of the quantum economy to follow a similar path to that of the silicon economy, driven by Moore's Law and financially supported by investors toward a significant fraction of a trillion dollars in the future.  Having said this, there is no obvious reason to ``re-invent the wheel'', and the NP community has the necessary institutional knowledge and infrastructures to rise and meet this challenge.
The {\it InQubator for Quantum Simulation} (IQuS)~\footnote{\tt \url{https://iqus.uw.edu/}} was recently established with the goal of enabling this QIS-NP integration.

\end{refsection}
\addtocontents{toc}{\protect\setcounter{tocdepth}{1}}
\begin{refsection}
\chapter{A perspective on the future of high-performance quantum computing}
\chapterauthor[1]{David J. Dean}
\begin{affils}
\chapteraffil[1]{Thomas Jefferson National Accelerator Facility, Newport News, VA 23606, USA}
\end{affils}

\setcounter{footnote}{0}

\addtocontents{toc}{\protect\setcounter{tocdepth}{-2}}

When contrasting the evolution of quantum computing with the advent of microchip technology, one sees interesting similarities. Let us start with transistors. In 1925, Julian Lilienfeld filed for an early patent for the design of a field-effect transistor\cite{lilienfeld}, marking the conceptual start of the microchip revolution. It took several years of development to move from concept to the first working transistor (its birthdate is given as 23 December 1947) for which the team of John Bardeen, Walter Brattain, and William Shockley won the Nobel Prize in 1956.\footnote{\url{ https://www.nobelprize.org/prizes/physics/1956/summary/}}  Robert Noyce, who received the first integrated circuit (IC) patent in 1961 (filed in 1959~\cite{noyce}), joined with Shockley to found the first microchip production firm, Fairchild Semiconductor, and later founded Intel with Gordon Moore. Gordon Moore is credited with realizing that the number of transistors in a dense integrated circuit doubles about every two years (Moore’s law~\cite{moore}). Indeed, from the Intel-4004 of 1970, with 2,250 transistors on the chip, to today’s chips with billions of transistors, this observation has held true.  The amazing advances made in computer technology have revolutionized every aspect of life and have had tremendous impacts in our approach to the sciences through the development and application of computational science across many domains. 

Quantum computing is following a similar evolutionary path. We often credit Richard Feynman with the idea, see e.g.~\cite{Feynman:1981tf}, of simulating quantum systems with quantum computers. Furthermore, the thermodynamics of computing (and potential reversibility of quantum computing) were being discussed contemporaneously by Toffoli~\cite{toffoli} and Bennett~\cite{bennett}. Influenced by the reversable computing work, Benioff~\cite{benioff} developed universal quantum gate sets for computation. These tremendous leaps in a theoretical understanding of quantum information science and quantum computation mark the start of the current quantum computing era. 

Of course, a theory does not mean a device that can compute. The next steps in the evolution of quantum computing required the hard work of developing and understanding how to make quantum circuits in the laboratory~\cite{popkin}. Many laboratories and researchers from across the world have contributed to the necessary advances required to advance quantum computing in the laboratory~\cite{gyongyosi,motta}. Furthermore, these advances are leading to significant technical investments from different industrial sectors, and generating many new start-up technology companies. For example (and this is not an exhaustive list), progress is being made in generating quantum computers from superconducting qubits (represented by work at IBM\footnote{\url{https://www.ibm.com/quantum-computing/}}, Google\footnote{\url{https://quantumai.google}}, and Rigetti\footnote{\url{https://www.rigetti.com}}), trapped ions (represented by work at IonQ\footnote{\url{https://www.ionq.com}} and Honeywell\footnote{\url{https://www.honeywell.com/us/en/company/quantum}}), and optical computing\footnote{\url{https://psiquantum.com}}.  In each case, the technical difficulty comes with scaling of qubits and with error control. Nevertheless, the field continues to make quick progress in implementation within these technologies. 

The US government has been keenly aware of the need to invest in quantum computing for several years. Starting a completely new industry plays into this need, as does the threat of competition from across the world~\cite{dean}. The US Department of Energy began to formulate plans~\cite{carter,moore2,awschalom,beck} for research in the area that would directly impact both coherence times and quantum computing scalability. At the same time, the DOE began funding research in quantum algorithm development and in use cases that are germane to the DOE’s mission in scientific R\&D. The reports (and others that I have not listed),  coupled with the initial base programmatic research, formed the basis for DOE’s role in the National Quantum Information Act\footnote{\url{https://www.congress.gov/bill/115th-congress/house-bill/6227}}  which was signed into law on December 21, 2018. The NQI Act establishes federal coordination among various US government agencies pursuing quantum R\&D\footnote{\url{https://www.quantum.gov}} and provides funding for NQI Research Centers to be established by NSF, DOE, and the DOD. The DOE Office of Science funds 5 National Quantum Information Science Research Centers, each operating at \$25M/year. 

The Quantum Science Center (QSC)\footnote{For further details see \url{https://qscience.org} and follow QSC at @QuantumSciCtr (on twitter).},  for which I was the PI until January 2022 when I transitioned to JLab, is dedicated to overcoming key roadblocks in quantum-state resilience, quantum-state controllability, and ultimately the scalability of quantum technologies. This mission is being achieved by integrating the discovery, design, and demonstration of revolutionary topological quantum materials, algorithms, and sensors, catalyzing the development of disruptive technologies. The QSC also develops the next generation of scientists and engineers through the active engagement of students and postdoctoral associates in research and professional development activities. Furthermore, by closely coordinating with industry, the QSC is strongly coupling its basic science foundation and technology development pathways to transition new applications to the private sector to make quantum technologies a reality.
Specifically, the QSC is organized into three scientific thrusts. First, the QSC is addressing the fragility of quantum states through the design of new topological materials for quantum information science (QIS). QSC researchers focus on the design, synthesis~\cite{rosa}, and characterization of topological superconductors and quantum spin liquids, both of which are candidate materials for discovering non-abelian quasiparticle states (anyons) that promise to yield quantum computing gates that are protected from environmental noise, and thus increasing the robustness of quantum computation, see e.g.~\cite{lahtinen}. Ultimately, the QSC will demonstrate controlled interactions between these topological states, or Majorana zero modes, to realize scalable topological quantum computation. Recent work~\cite{scheie} includes the observation of quantum entanglement phenomena in a triangular antiferromagnet using neutron scattering. The material, KYbSe2, is a quantum spin liquid candidate which should be exhibit anyonic behavior. 

Second, the QSC is developing scalable algorithms and software to exploit the new physics enabled by topological systems. The QSC develops and tests these algorithms on several noisy intermediate-scale quantum platforms to characterize their behaviors and to devise algorithms that mitigate the noise~\cite{wang}. In other computational work, QSC researchers presented and demonstrated entanglement-enhanced methods that can be used to learn an entire unitary rather than just its action on a low-lying subspace. The result is a framework for quantum machine learning of continuous variable (for instance, photonic) quantum systems capable of exponentially reducing input-output state training resources~\cite{Volkoff:2021aat}.

Third, the QSC is designing new quantum devices and sensors to unambiguously detect topological quasiparticles and explore meV regions of dark matter phase space. To manipulate Majorana states, one must first unambiguously identify them. To that end, the QSC develops new capabilities in detecting the ultralow-noise nondestructive local sensing of electromagnetic fields. These techniques are also being developed to detect for the first time, “light” dark matter, one of the theoretically favored candidates for dark matter, which constitutes 85\% of the matter in the universe. Recent results include the discovery of room-temperature single-photon emitters in SiN, which have the potential to enable direct, scalable and low-loss integration of quantum light sources with well-established photonic on-chip platforms~\cite{senichev}.

The QSC will transition discoveries in fundamental QIS through a progressive series of capability demonstrations that assess the readiness of quantum technology. The QSC method of co-design actively engages a broad range of researchers directly involved in pursuing an end goal to design and implement solutions. Thus, co-design processes generate the scientific integration of QSC-wide research projects toward a common programmatic outcome, and each co-design process provides feedback across research projects to focus innovations on new quantum science and technologies. During Year 1, the leads for the co-design processes identified common interfaces between projects and adopted quarterly milestones for Year 1 activities. The co-design process for topologically protected quantum information in Year 1 focused on the co-design of materials that host anyon physics. The co-design process for quantum simulations for scientific applications demonstrates quantum simulation for scientific applications by using quantum computing hardware with tailored quantum algorithms. The co-design process for quantum sensing in real-world applications demonstrates quantum sensors developed for material science characterization and dark matter detection.

Nuclear theorists, working with quantum computing colleagues, pursued several early calculations on quantum computing platforms. For example, early cloud-based quantum computers were used to calculate the binding energy of the deuteron~\cite{Dumitrescu:2018njn}. Furthermore, in the early explorations, researchers also studied the dynamics of the Schwinger model using quantum computers~\cite{Klco:2018kyo}. These early works laid the groundwork for collaborations of nuclear theorists and QIS theorists to work at the interface of the two fields. Indeed, one recent example of progress coming from QSC involves describing neutrino oscillations at high neutrino density~\cite{Hall:2021rbv}. These examples indicate how theoretical nuclear physics research is incorporating quantum computing technology to solve interesting problems. It is still early days, and continuing collaborations coupled with efficient utilization of increasingly powerful quantum computers, will enable considerable progress in the coming decade. 

I can draw conclusions from the progress made during the first year of the QSC, and from the exciting work being performed across the quantum computing community. This could be considered a prediction of the future. As Yogi Berra said, ‘It’s tough to make predictions, especially about the future’. Nevertheless, I will end this brief discussion with some predictions for the next decade of QIS research, particularly as it pertains to quantum computing.
\begin{itemize}
    \item 
    Errors will be addressed in quantum computing. Materials that make up quantum computers (particularly superconducting materials) will be engineered to the point where the materials are not the primary sources of loss of coherence. Longer coherence times will enable gate depth to increase, thus opening possibilities for tackling larger problems. Furthermore, algorithms that mitigate error will continue to develop and improve. We should remember that HPC derives its power to solve increasingly difficult scientific problems from both hardware and algorithm improvement. A similar story is emerging in quantum computing. 
    \item 
    Quantum ‘supremacy’ will be claimed several times before it truly happens. Evidence suggests that the Google supremacy claim~\cite{Arute:2019ts} rested on the use of a less than effective algorithm for the HPC comparison~\cite{liu}.  
    \item 
    While leadership computing may hit a plateau in implementation, staying at the exascale for several years and likely moving away the power-law progression that has characterized the top500 list for decades\footnote{\url{ https://www.top500.org/statistics/perfdevel/}}, quantum computers will show significant advances in scientific reach over the next decade. The scale of R\&D funding in quantum computing today is enormous, with roughly \$24.4B being spent in globally thus far\footnote{See, for example \url{https://www.qureca.com/overview-on-quantum-initiatives-worldwide-update-mid-2021/}}. Trends suggest that the rate of R\&D expenditures will continue to increase over the next decade.
    \item 
    Anyons will be unambiguously detected in two-dimensional superconductors or quantum spin liquids and manipulation of several anyons and their braiding to produce two-qubit gates will occur within the next decade. The QSC is built on this scientific goal, and I look forward to the day when we can say that we succeeded. 
    \item
    Nuclear theorists, working with QIS experts, will develop research problems that only a quantum computer can solve, and that are relevant for advances in nuclear physics. 
    \item 
    Quantum computing technology will eventually fold into HPC as accelerator technology. Recent press releases indicate that this development is already being pursued in Europe\footnote{\url{https://thequantuminsider.com/2021/11/17/q-exa-collaborative-iqm-quantum-computer-will-be-first-quantum-system-to-be-integrated-into-a-hpc-supercomputer/}}.
\end{itemize}

Without a doubt, scientists can tackle the challenges that remain in quantum computing. It will take years of sustained R\&D and multidisciplinary cooperation to get there, and that is as exciting as the future that quantum computing will usher in.  

{\bf Acknowledgements:} This work was supported by the Quantum Science Center (QSC), a National Quantum Information Science Research Center of the United States Department of Energy (DOE) under Contract No. DE-AC05-00OR22725 with UT-Battelle, LLC. 

\printbibliography[heading=subbibliography]
\end{refsection}

\end{document}